\documentclass[11pt]{article}
\usepackage[a4paper, total={7.5 in, 10.7 in}]{geometry}
\usepackage{blindtext}
\usepackage{titlesec}

\usepackage{authblk}
\usepackage{amsmath}
\usepackage{amssymb}
\usepackage{graphicx}
\usepackage{enumitem}
\usepackage{mathtools}
\usepackage{multicol}
\usepackage{titlesec}

\usepackage[
backend=biber,
doi=false,isbn=false,url=false,eprint=false,
style=apa,
]{biblatex}
\title{Biomass transfer on autocatalytic reaction networks: \\
a delay-differential equation formulation}
\author[1,2,3,4,$*$]{Wei-Hsiang Lin}
\affil[1]{Institute of Molecular Biology, Academia Sinica, Taipei, Taiwan}
\affil[2]{National Center for Theoretical Sciences, Physics Division, Taiwan}
\affil[3]{Sarafan Chemistry, Engineering, and Medicine for Human Health Institute, Stanford University, Stanford, CA 94305, USA}
\affil[4]{Howard Hughes Medical Institute, Stanford University, Stanford, CA 94305, USA}
\affil[$*$]{Corresponding author (whl243@as.edu.tw)}

\begin{document}
\maketitle

\graphicspath{{/Users/wei-hsiang/Desktop/biomass_transfer/2025Mar_maintext/figures/}}

\setlength{\columnsep}{0.8cm}


\begin{multicols}{2} 

\textbf{Abstract}. For a biological system to grow, mass must be transferred from environment to the system and be assimilated into its reaction network. Here, I characterize biomass transferring process for a large class of exponential-growing systems including linear and nonlinear dynamics. By tracking biomass along reaction pathways, we can characterize asymptotic growth trend of an n-dimensional ordinary differential equation (ODE) by reformulating it as a one-dimensional delay differential equation (DDE). The kernel function of this DDE represents the overall amplification and transfer delay of the system biomass and summarizes the autocatalysis dynamics. The DDE formulation allows us to compare reaction networks with various topologies and complexities, and provides a rigorous scheme for estimating system growth rate under dimensional reduction. \\

\textbf{Significances}. Living systems which grow and replicate are often out-of-equilibrium, and the speed of system expansion depends on their autocatalytic efficiency. In this work, a mathematical framework is developed for analyzing autocatalysis in general exponential- growing systems. This approach focuses on calculating the biophysical quantity $\alpha(\tau)$, which represents the waiting time distribution for biomass to self-amplify within a reaction network. This catalytic kernel $\alpha(\tau)$ allows us to compare autocatalysis dynamics across different growth models, and reveals general principles on how evolution can shape reaction networks to optimize their long-term growth rate.  \\

\textbf{0. Introduction}. An essential feature of biological systems is their ability to grow and expand, which involves incorporating external resources into internal biomass. This conversion process often involves a reaction network that performs biosynthesis and energy production. In many natural systems, reaction networks are autocatalytic in the broad sense that the system synthesize its own components (\cite{blokhuis_universal_2020}, \cite{hordijk_structure_2012}, \cite{letelier_organizational_2006},\cite{vassena_unstable_2024},
	\cite{despons_nonequilibrium_2025},
	\cite{despons_structural_2024},
	\cite{marehalli_srinivas_characterizing_2024},
	\cite{gagrani_polyhedral_2024},
	\cite{muller_elementary_2022}). Understanding the general principles in metabolic conversion and biomass synthesis is fundamentally important for all living systems.  \\

To obtain useful insights, we consider the analogy between material transfer in physical systems and biomass transfer in biochemical systems. In fluid mechanics, Eulerian description inspects the influx and efflux in terms of absolute coordinates, whereas Lagrangian description inspects movement of a fluid parcel along a streamline. In reaction networks, one could focus on the influxes and effluxes of each nodes and calculate the material change rate of each node, which corresponds to the Eulerian view. Alternatively, a small cohort of biomass may be tracked and transfer along the reaction pathways, reflecting a Lagrangian view. \\

For growing reaction networks, the majority of theoretical studies adopt the Eulerian description (\cite{kondo_growth_2011},\cite{maitra_bacterial_2015}, \cite{mb_synthetic_2000}, \cite{scott_interdependence_2010}). The Lagrangian view has been used to develop metabolic models of pulse-chase experiments with isotope labeling (\cite{antoniewicz_elementary_2007}, \cite{thommen_stochastic_2023}), but it has not been formulated for network growth and autocatalysis. In this work, I demonstrate that analyzing biomass transfer along reaction pathways is an intuitive and powerful way of studying autocatalytic growth dynamics. The analysis focuses on nodes that regulate boundary influxes of the entire system. These nodes, termed as \textit{gatekeepers}, limit the entry of the biomass into the system. By adopting a Lagrangian view, biomass transfer from boundary influxes to gatekeepers can be analyzed as an autocatalytic process, enabling formulation of delay-differential equations (DDEs) for gatekeeper biomass (denoted as $Z$). \\

Specifically, the long-term growth trend of a reaction network can be accessed via equation $\frac{dZ}{dt} = -\beta Z + \int \alpha(\tau)\,Z(t-\tau)\,d\tau$ , whereby the catalytic kernel $\alpha(\tau)$ summarizes the amplification magnitude and waiting time distribution for gatekeeper biomass $Z$ to catalyze itself. Note that the DDE here is distinct from the classical Volterra integral equation or renewal equation $x(t) = f(t) + \int K(\tau)\,x(t-\tau)\,d\tau$, which is used in age-structured population growth (\cite{britton_essential_2003}). In this work, $\alpha(\tau)$ is calculated explicitly for scalable reaction networks (SRNs), which is a general class of reaction networks that includes linear and nonlinear flux functions and allows long-term exponential growth (\cite{lin_origin_2020}). \\

The catalytic kernel $\alpha(\tau)$ is not merely a mathematical tool but a concrete biophysical quantity. It can be measured experimentally in the future, for example, by tracking isotope labeling of metabolites. It is a general phenomenological concept for autocatalysis and does not depend on a specific reaction network. By adopting the DDE formulation, one can compare reaction networks of differing dimensions and topologies and perform coarse-graining on complex reaction networks. Overall, using DDEs to analyze the dynamics of biomass transfer would provide a deeper and systems-level understanding of autocatalysis. \\


\vspace{10 pt}

\textbf{1. Reaction networks, flux function, and gatekeepers.} Let us consider an open system in which nutrients and waste can be exchanged with the environment (Sughiyama et al., 2022). The reaction network structure (Figure 1A) includes system nodes $\{x_1,\cdots,x_n\}:=\mathcal{X}$, environmental nodes $\{E_1,\cdots,E_{n^{'}} \}:=\mathcal{E}$, and reactions between nodes $\{\phi_1,\cdots,\phi_m \}$. Each node represents one chemical species. A reaction $\phi_a$ is represented by
\begin{equation} \label{eqM1}
	\phi_a: \sum_{i=1}^n c_{ia}x_i + \sum_{j=1}^{n'}c_{ja}'E_j
	\rightarrow
	\sum_{i=1}^n d_{ia}x_i + \sum_{j=1}^{n'}d_{ja}'E_j,
\end{equation}
where $c_{ia},c_{ja}',d_{ia},d_{ja}' \geq 0$ are stoichiometric coefficients. We say $x_i$ is an upstream (or downstream) node of $\phi_a$ if $d_{ia}-c_{ia}$ is greater (or less) than 0, and $\phi_a$ is the influx (or efflux) of node $x_i$ if $x_i$ is a downstream (or upstream) node of $\phi_a$. A node can have multiple influxes and effluxes, and each reaction has at least one upstream node and one downstream node. We denote $up(\phi_a)$, $dw(\phi_a)$ as the collections of upstream and downstream nodes of $\phi_a$, respectively, and $in(x_k), out(x_k)$ as the collections of influxes and effluxes of node $x_k$, respectively. \\

To describe biomass transfer, a nonnegative number $\mathfrak{m}(x_k)$ is associated to each node $x_k$. In practice, this number could be molecular weight, or carbon atom count of a biochemical molecules. The biomass-weighted stoichiometry matrix is defined as $S_{ka}:=(d_{ka}-c_{ka})\,\mathfrak{m}(x_k)$. In this way, stoichiometry can be interpreted by the unit of biomass. Mass conservation is required to avoid anomalous behaviors of system growth (i.e., autocatalysis from void). In the framework of this study, mass conservation can always be achieved by introducing additional environmental nodes. \\

Our discussion so far focus on reaction network structure. To construct a dynamical system, we need to define variables and flux functions. We define $X:=(X_1,\cdots,X_n)^T$ as the biomass vector, where $X_k := \mathfrak{m}(x_k) \times$ (number of $x_k$-type objects in the system). We define $J:=(J[\phi_1],\cdots,J[\phi_m])^T$ as the flux function vector, equal to the number of reaction events that happen per unit time. Note that the units for $X$ and $J$ are [biomass] and [1/time], respectively, and the unit for the biomass-weighted stoichiometry matrix (described in the previous section) is [biomass]. Together, the biomass flux of reaction $\phi_a$ associated with node $x_k$ is $S_{ka} J_a$, with unit [biomass/time]. \\

We assume that substances in the environment are unlimited and are maintained at constant densities surrounding the system (i.e., similar to a particle reservoir of a grand canonical ensemble in statistical mechanics). For the system, by assuming large number of objects and reaction events, $X$ and $J$ are approximated as continuous variables (Figure 1D) and follow the systems of ordinary differential equation (ODE):
\begin{equation}
	\frac{dX_k}{dt} = \sum_{a=1}^m
	S_{ka} J_a,
\end{equation}	
for $k=1,\cdots,n$. Coefficients for environmental nodes (i.e., $c_{ja}', d_{ja}'$ do not appear in the equation since it is assumed that their amounts are unlimited. \\

In general, each flux function $J_a(X)$ can be multivariate nonlinear function from $\mathbb{R}^n$ to $\mathbb{R}$. Let $N=X_1+\cdots+X_n$ denote the total biomass, define the long-term growth rate $\lambda$ (if exists) as
\begin{equation}
\lambda := \lim_{t\rightarrow\infty} \frac{1}{t}\,\log N(t).
\end{equation}
Appropriate conditions will be imposed later such that the system grows exponentially in the long term. \\

\textit{Boundary influx reactions} are reactions whose upstream has environmental nodes. In this work, each boundary influx reaction $\phi_a$ (red arrows in Figure 1A) is assumed to be catalyzed by one system node (labeled by the encircled nodes next to the red arrow). This node is called the \textit{gatekeeper} of $\phi_a$. The collection of all gatekeepers is denoted $\mathcal{G}$. Intuitively, since gatekeepers catalyze all of the boundary influxes collectively, the rates of gatekeeper growth limits the rate of system growth. The long-term dynamics of gatekeepers will be the key concept in our following analysis to formulate DDE in Lagrangian view. \\

\textbf{2. Reaction pathways and biomass transfer.} To adopt the Lagrangian view and tracking biomass, we assume the biomass can be discretized conceptually in microscopic level. The discretized \textit{biomass units} can be tracked conceptually from an upstream node to a downstream node during a reaction. In this way, a Markov process which is consistent with ODE flux model is established (Figure 1B). Specifically, we define the instantaneous transition rate from node $x_k$ via reaction $\phi_a$ to node $x_j$, denoted as $r_{kaj}$, as 
\begin{equation}
	r_{kaj}(t)= \frac{|S_{ka}|J_a(t)}{X_k(t)}
    \times 
	\xi[x_j, \phi_a],
\end{equation}
where $\xi[x_j, \phi_a] \in [0,1]$ is the fraction of biomass flux of $\phi_a$ that transfer into node $x_j$. The formulation details are described in Supplementary Information (SI).\\

Reaction pathways are natural structures to model biochemical reactions. To achieve sustainable growth, an open system must convert environmental materials into internal biomass, with reaction pathways being indispensable for these interconversion processes. Here, a reaction pathway is defined by
\begin{equation}
	\pi(u,\omega)
	= u_0\,\omega_0\,
	\cdots
	\omega_L\,
	u_{L+1},
\end{equation}
where nodes $\{u_0,\cdots,u_{L+1} \}=:\{u\}$ and reaction $\{\omega_0,\cdots,\omega_L \}:=\{\omega\}$ are ordered sets. For each reaction $\omega_j$, the nodes $u_{j-1}$ and $u_j$ are its upstream and downstream nodes, respectively (Figure 1C). We require all nodes in $\{u\}$ to be system nodes, except that the first or last nodes can be environmental nodes. We allow $\{u\}$ and $\{\omega\}$ to have repeated members, and the length $L$ can be infinite (Figure 1D). Where there is no confusion, $\{\omega\}$ can be omitted and the reaction pathway can be written as $\pi:u_0 u_1 \cdots u_{L+1}$. \\


\end{multicols} 



\vspace{10 pt}
\begin{center}
	\includegraphics[scale = 0.45]{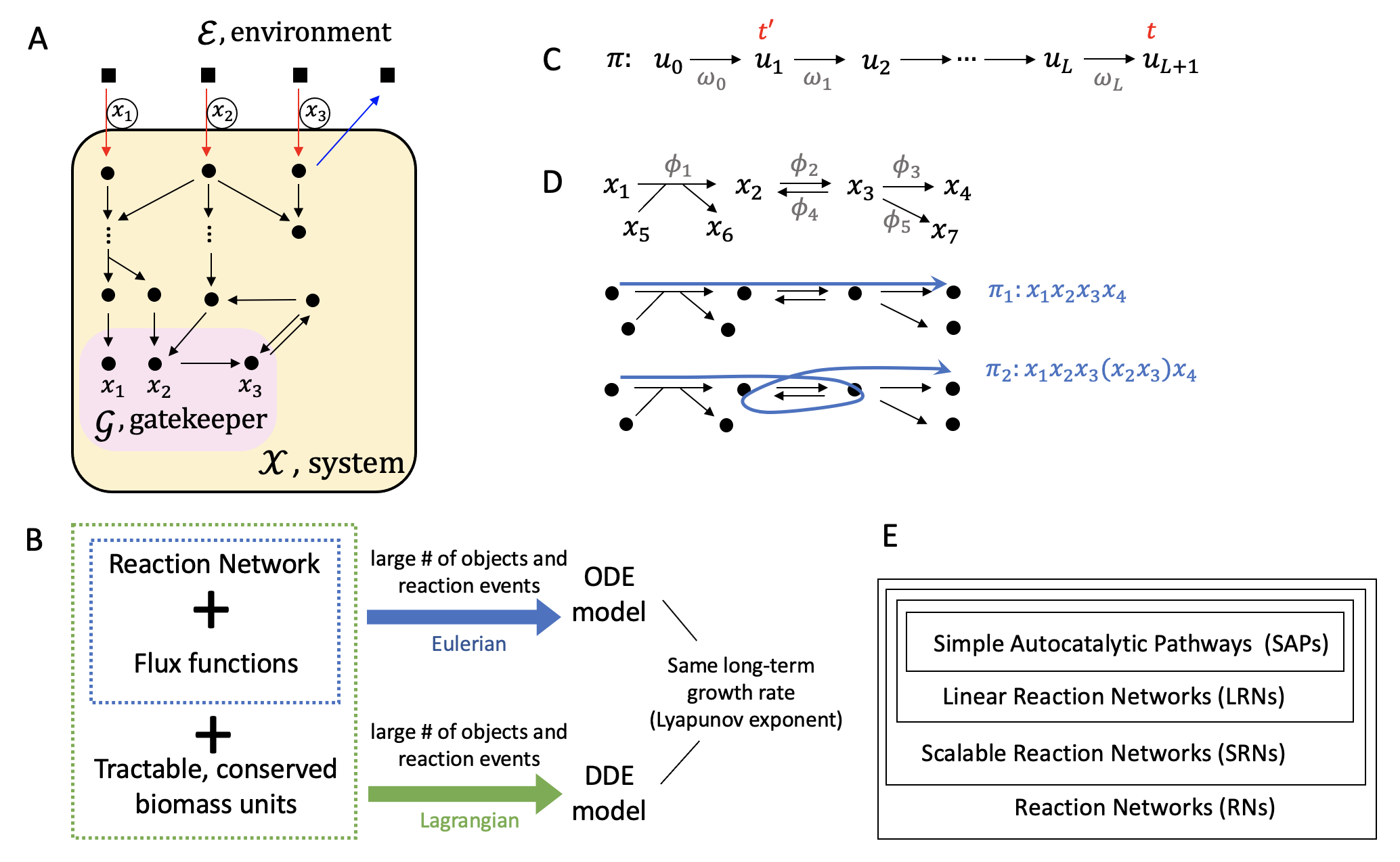}
\end{center}
\small
Figure 1. Growing reaction networks and the modeling frameworks.  (A) A diagram of a reaction network, with system nodes represented in circles and environmental nodes represented in squares. The gatekeeper set $\mathcal{G}$ is a subset of the system node $\mathcal{X}$. Gatekeeper nodes control the boundary reaction influxes (red arrows). (B) Eulerian and Lagrangian views on reaction network and biomass. (C) A reaction pathway, with $t',t$ represent the time points where biomass arrive node $u_1$ and $u_{L+1}$, respectively. (D) Schematics of reaction pathways in a network. For example, $\pi_1$ is a reaction pathway directly from $x_1$ to $x_4$, and $\pi_2$ is another reaction pathway from $x_1$ to $x_3$, back to $x_2$, and ultimately reaching $x_4$. Blue paths indicate the trajectory of biomass movement. (E) Reaction networks with different flux function classes (defined in later sections). \\
\normalsize


There are two important quantities for reaction pathways $\pi$ involving autocatalysis dynamics. The first quantity is the amplification rate $\kappa_{\pi}$, which measures the efficiency of the last node $u_{L+1}$ to promote the first reaction flux $\omega_0$ in reaction $\pi$ (Figure 1C):  
\begin{equation}
	\kappa_{\pi}(t') := 
	\frac{\text{total mass flux arrived } u_1 }
	{\text{total mass on }u_{L+1} }.
\end{equation}

The second important quantity is called the arrival function $H_{\pi}(t',t)$, which represents the transition probability for biomass to travel from $u_1$ to $u_{L+1}$ via $\pi$ (Figure 1C):
\begin{equation}
	H_{\pi}(t',t):= 
	\frac{\text{mass arrived $u_1$ at time  $t^{'}$},
	\text{transferred along $\pi$, and arrived $u_{L+1}$ at time $t$} }
    {\text{mass arrived  $u_{L+1}$ at time  $t'$}}.
\end{equation}

The definition of $\kappa_{\pi}(t')$ and $H_{\pi}(t',t)$ (both with unit $\sim [time]^{-1}$) would be clear when we analyzing biomass transfer on reaction pathway; they are important for constructing the DDE in Lagrangian perspective. Our focus will be investigating how $\lambda$ is determined by the catalytic efficiency and biomass transfer time of the system. This paper is structured in three parts. First, I analyze biomass transfer in simple autocatalytic pathways (SAPs) (\cite{fiedler_dynamics_2013}) for developing intuition and concrete formulae. Then, I generalize the analysis to linear reaction networks (LRNs) (\cite{nandori_growth_2022}, \cite{unterberger_stoechiometric_2022}), before extending to scalable reaction networks (SRNs), which enable us to study many nonlinear systems that grow exponentially (Figure 1E). \\


\vspace{1 pt}


\begin{multicols}{2} 

\textbf{3. Biomass transfer along simple autocatalytic pathways (SAPs).} To illustrate the basic idea, let us consider a reaction network with $n$ system nodes and $n$ reactions, including $\phi_k:x_k \rightarrow x_{k+1}$ for $k=1,\cdots,n-1$ and $\phi_n: E_1\rightarrow x_1$ (Figure 2A). We assume internal fluxes from $x_k$ to $x_{k+1}$ follow linear functions $J[\phi_k]=a_k X_k$, while the boundary influx $\phi_n$ catalyzed by node $x_n$ has flux function $J[\phi_n]=bX_n$. We also assume $\mathfrak{m}(x_j)=1$ for all nodes. This type of network is termed a simple autocatalytic pathway (SAP); it grows exponentially and represents a primitive autocatalytic structure. In the following, we analyze the long-term growth rate $\lambda$ using two different perspectives: \\

\textit{Eulerian view}\: By inspecting the influx and efflux of each node, we have a linear ODE: $\frac{dX}{dt}=MX$ with  
\begin{equation}
M=
\begin{pmatrix}
-a_1 & 0 & \cdots &\cdots & b  \\
a_1 & -a_2 & 0 &\cdots & 0 \\
0 & a_2 & \ddots &\ddots & \vdots \\
\vdots & \ddots & \ddots & -a_{n-1} & 0\\
0 & \cdots & 0 & a_{n-1} & 0\\
\end{pmatrix}.
\end{equation}
The solution is $X(t)=e^{Mt} X(0)$. The characteristic polynomial of $M$ is $p_M(t)=(-1)^{n}[t(t-a_1)\dots(t-a_{n-1})-ba_1\dots a_{n-1}]$. Let $\sigma_{max}$ represent the principal eigenvalue of $M$ (the eigenvalue with the largest real part). In this case, it can be shown (by the Perron-Frobenius Theorem) that $\sigma_{max} \in \mathbb{R}$. For large $t$, $N(t)\approx e^{\sigma_{max} t}$ and $\lambda =\lim_{t\rightarrow \infty} \frac{1}{t} \log N(t) = \sigma_{max}$ (see Supplementary Information (SI) for details). \\

\textit{Lagrangian view}: In this perspective, the biomass is conceptually discretized at microscopic level and can be “tracked” in the system. The SAP behaves as an autocatalytic structure, with the last node $x_n$ promoting the entry of environmental material $E_1$ into the system and this material later becoming biomass on $x_n$. An SAP is a reaction pathway on its own. The amplification rate $\kappa$, which measures the efficiency by which $x_n$ brings in external resources, is
\begin{equation}
	\kappa := 
	\frac{\text{total mass flux arrived } x_1}
	{\text{total mass on }x_n }
	=\frac{bX_n}{X_n} =b.
\end{equation}

To calculate $H(t',t)$, each step of the internal reactions $\phi_k, k \geq 1$ is considered. During a small time interval $\Delta t$, the total biomass transferred across $\phi_k$ is $a_k X_k \Delta t$, which corresponds to a Poisson process with a “reaction event” of rate $a_k$. For biomass on node $x_k$, the waiting time $\tau$ for the “reaction event” to happen follows an exponential distribution $T_k \sim a_k e^{-a_k \tau}$. Therefore, the total waiting time for biomass from $x_1$ to $x_n$ is $T=T_1+\cdots+T_{n-1}$. Define $\tau=t-t'$, $H(t',t):=h(\tau)$ is the convolution of exponential waiting time, i.e.
\begin{equation}
    h(\tau) = (a_1 e^{-a_1 \tau})
    \ast \cdots \ast 
    (a_{n-1} e^{-a_{n-1} \tau}). 
\end{equation}
Intuitively, the greater value of $\kappa$, the more effective node $x_n$ is at taking in biomass, and so the faster the system grows. The shorter the waiting time $T\sim h(\tau)$, the faster biomass can be incorporated into node $x_n$, so the system grows more rapidly. \\

We notice that $h(\tau)$ also acts as a delay kernel between the first and last fluxes in the SAP, namely  
\begin{equation} \label{E3A}
	J_{n-1}(t)
	= \int_{0}^t
	h(\tau)\,J_0(t-\tau)\,d\tau + \epsilon(t).
\end{equation}
Here, $J_0, J_{n-1}$ represent the fluxes of $\phi_n, \phi_{n-1}$, respectively, and the error term $\epsilon(t)$ is negligible for large $X_n$. Note that $J_{n-1}(t) = \frac{dX_n}{dt}$ and 
and $J_0(t)=bX_n(t)= \kappa X_n (t)$, so Eq (\ref{E3A}) can be written as a delayed differential equation (DDE) of $X_n$:
\begin{equation}
	\frac{dX_n}{dt}
	= \int_{0}^t
	\kappa h(\tau)\,X_n(t-\tau)\,d\tau + \epsilon(t).
\end{equation}
The long-term behavior of $X_n$ is determined by $\kappa h(\tau) := \alpha(\tau)$, defined as the \textit{catalytic kernel}. The Laplace transform of catalytic kernel, denoted by $\tilde{\alpha}(s)$, is called the \textit{catalytic spectrum}. \\

For SAP with $n=1$, biomass is transferred from $E_1$ to $x_1$ directly without time delay, and the arrival function is a delta function, i.e., $h(\tau)=\delta(\tau)$. Hence $\alpha(\tau) = b\delta(\tau)$ and $\tilde{\alpha}(s) = b$. For SAPs with $n>1$,
\begin{equation} \label{E3C}
\tilde{\alpha}(s)=b \frac{a_1}{s+a_1}
\times \cdots \times
\frac{a_{n-1}}{s+a_{n-1}}.	
\end{equation} 
Importantly, there is a algebraic relation (see SI for derivation).  
\begin{equation} \label{E3B}
	\lambda = \tilde{\alpha}(\lambda).
\end{equation}
The Eulerian and Lagrangian view yields consistent results, since from the characteristic polynomial $p_M(t)$ we can also recover the formula (\ref{E3B}). \\


\textbf{Examples 3.1.} We consider SAPs with $n=1,2,3$ (Figure 2B). Note that $\tilde{\alpha}(s)$ is non-increasing on positive real line. Increasing the number of nodes causes $\tilde{\alpha}(s)$ to have more factors in the form $\frac{a_k}{s+a_k}$ in Eq.(\ref{E3C}), and makes it decay faster. Interestingly, $\lambda$ can be determined geometrically as the intersection between $y=s$ and $y=\tilde{\alpha}(s)$ by Eq. (\ref{E3B}). This allows us to study $\lambda$ from qualitative analysis of the catalytic spectra. \\

\textbf{Example 3.2.} The exact formula for an SAP with $n=2$ can be expressed as $\lambda = \lambda_2(a_1,b) := \frac{a_1}{2} \big( \sqrt{1+\frac{4b}{a_1}}-1 \big )$, whereas for SAPs with $n>2$ the formulae of $\lambda$ become cumbersome. An interesting question is whether multiple reactions can be “coarse-grained” into one reaction to obtain an “equivalent reaction pathway”, in a fashion similar to the law of resistors in an electric series circuit (Figure 2C). This possibility can be assessed by comparing the catalytic spectra between multiple-reaction SAPs and a single-reaction SAP. \\




\end{multicols} 

\begin{center}
	\includegraphics[scale = 0.5]{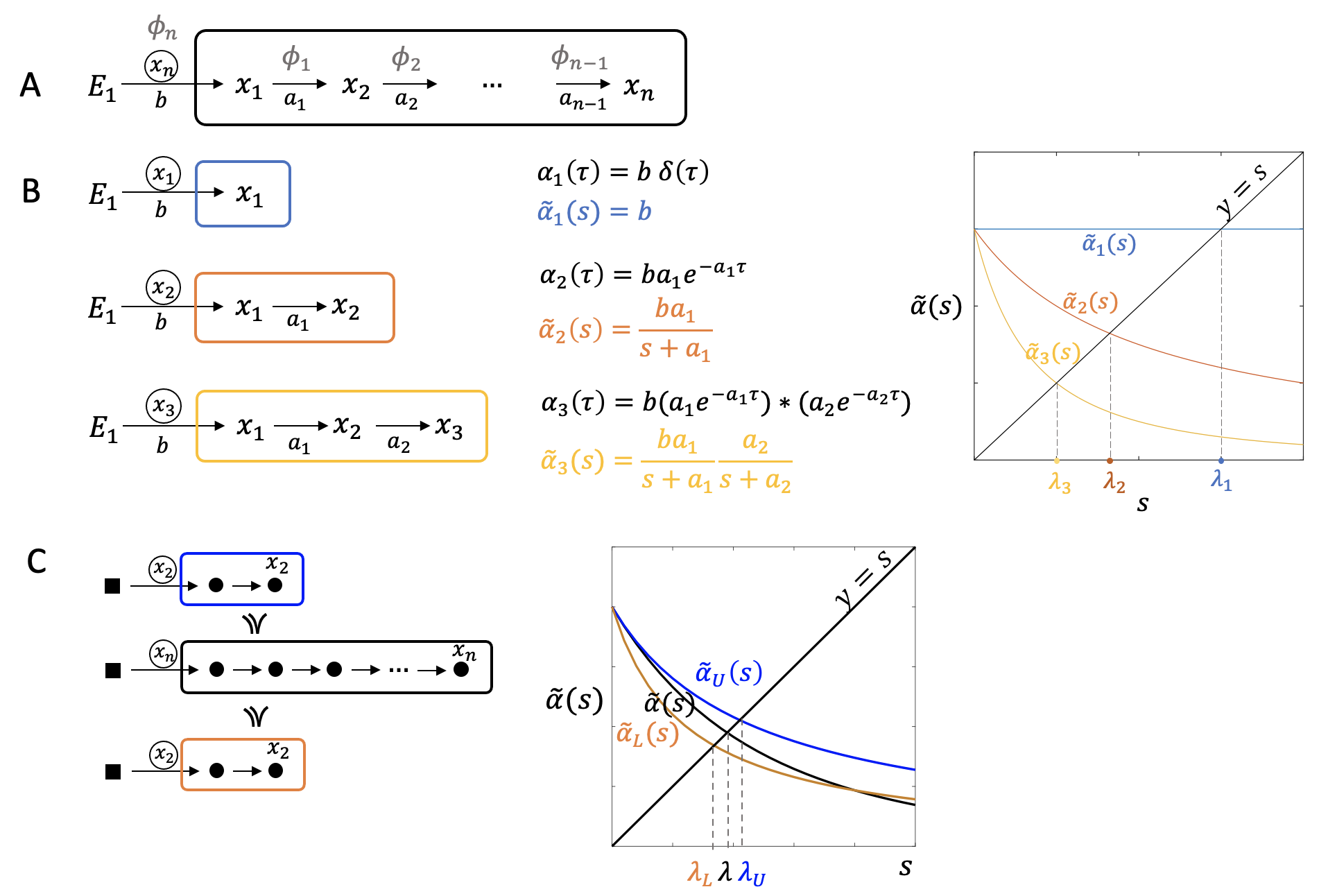}
\end{center}
\small
Figure 2. Simple autocatalytic pathways (SAPs).  (A) Illustration of a SAP. The rate constants ($a_k$ and $b$) for each reaction flux are labeled under the arrows. The encircled $x_n$ for boundary reactions represents the gatekeeper node of this influx. (B) Left: SAPs with $n$=1, 2 or 3. Middle: the decay kernels and decay spectra of the three SAPs at left. Right: The decay spectra $\tilde{\alpha}_n(s)$ of the three SAPs at far left. Note that the intersections between $y=\tilde{\alpha}_n(s)$ and $y=s$ correspond to the value of $\lambda_n$. (C) An SAP of length $n$ can be bounded by two SAPs of length two; one as an upper bound (blue) and one as a lower bound (green). The ordered relation $\preceq$ is used to represent that one SAP has a higher growth rate than the other one, i.e., $\tilde{\alpha}_1\preceq \tilde{\alpha}_2$ if $\lambda_2 \preceq \lambda_1$. \\
\normalsize




\begin{multicols}{2} 

Mathematically, no exact equivalency can be established between a SAP of length 2 and of length $n>2$. Nevertheless, a coarse-grained pathway can be used to deduct upper and lower bounds for $\lambda$. Specifically, define 
\begin{equation}
a_U^{-1} := a_1^{-1}+\cdots+ a_{n-1}^{-1},
\qquad    
a_L^{-1} :=\frac{\tilde{\alpha}(b)-1}{b},
\end{equation}
and then the growth rate $\lambda$ can be estimated by inequality $ f(a_L,b) = \lambda_L \leq \lambda \leq \lambda_U = f(a_U,b)$ (see SI). Note that the coefficient of upper estimation $a_U$ is the harmonic mean of $a_j$, indicating that the mean coarse-grained waiting time is the sum of mean waiting times of each of the reaction steps. However, doing so only provides the upper bound and not the exact formula, since the system is growing and hence not in detail balance. \\



\vspace{10 pt}

\textbf{4. Biomass transfer on growing linear reaction networks (LRNs).} In previous section, the idea of biomass transfer is illustrated by simple autocatalytic pathway. Now we study its generalization, the linear reaction networks (LRNs). Let $J_a(X)$ denote the linear flux function of reaction $\phi_a$, we assume the networks follow \textit{LRN conditions}:

\begin{enumerate}[label=(\roman*)]
	\item Each reaction $\phi_a$ has exactly one upstream node, denoted as $up(\phi_a)$. 
	\item If $up(\phi_a)=x_k$ is a system node, then its flux function is $J_a = R_a X_k$. 
	\item If $up(\phi_a)=E_k$ is an environmental node, then the flux function follows $J_a = R_a X_g$ for some node $x_g$ in the system. In this case, we say $x_g$ \textit{gatekeeps} $\phi_a$.
\end{enumerate}

Condition (ii) requires the reaction to be upstream-limited, that is, $J_a (X)\rightarrow 0$ as $X_k\rightarrow 0$. For condition (iii), the node $x_g$ is called the gatekeeper of reaction $\phi_a$. Next, let us analyze long-term growth rate for LRNs. \\

\textit{Eulerian view.} The LRN system can be represented in matrix form. By defining matrix $P$ as
\begin{equation}
	P_{ak} = 
	\begin{cases}
		R_a,  & \text{ if } J_a(X) = R_a X_k, \\
		0,    & \text{ otherwise. }
	\end{cases}
\end{equation}
Then, the LRN follows a linear ODE $\frac{dX}{dt} = MX$ with $M:=SP\in \mathcal{M}^{n,n}$. Note that from LRN condition (ii), the nonnegative orthant is invariant for LRNs. Denote $N:=X_1+\cdots+X_n$ as the system biomass and $Y:=X/N$ as the rescaled trajectory. For a general linear ODE with trajectory $X(t)$, the rescaled trajectory $Y(t)$ can potentially converge to a limit torus. However, for LRNs the long-term behavior of each $Y(t)$ can only be a fixed point as shown by Theorem A (see SI for the proof): 
\begin{center}
	\fbox{\parbox{{3.3in}}{
		\textbf{Theorem A.} Consider a linear reaction network $\frac{dX}{dt}=MX$ with a nonnegative initial condition. The long-term growth rate $\lambda$ is equal to one of the eigenvalues of $M$, and the rescaled trajectory $Y(t)$ converges to an eigenvector $Y^*$ affiliated with $\lambda$. 
	}}
\end{center}
 
Note that $\lambda$ may depend on the initial condition $X(0)$. Moreover, $Y^{*}$ may depend on initial condition $X(0)$ if the eigenspace of $\lambda$ displays multiplicity. When $M$ is irreducible, $\lambda$ is the dominant eigenvalue (the eigenvalue of largest real part, see SI, \cite{horn_matrix_2012}).  \\

For high dimensionality, the analytical formula for eigenvalues is generally intractable. It is difficult to analyze the relationship between $\lambda$ and the kinetic constants or topology of the network. Lagrangian perspective, on the other hand, could provide insights for analyzing $\lambda$. \\ 

\textit{Lagrangian view.} Let us generalize the SAP-scheme in the previous section. The last node $x_n$ of SAPs is critical since it acts as the gatekeeper for the reaction pathway, so the DDE is formulated on $X_n$. Let $\mathcal{G}$ denotes the collection of all gatekeeper nodes, and $Z$ be the \textit{gatekeeper biomass} of these node, i.e.
\begin{equation}
\begin{aligned}
Z &:= \sum_{x_g \in \mathcal{G}} X_g,
\end{aligned}
\end{equation}
When $\lambda>0$, the growth rate of $Z$ must be the same as for the entire system. This is because gatekeepers control all influx from the environment into the system, so if $Z$ grows more slowly than system size $N$, then the ratio $Z/N$ gradually decays to zero, and the boundary influx will also decay to zero, contradicting the assumption that $\lambda >0$ (see Proposition 2.2 in SI for details). In the following steps, I assume that $\lambda>0$ and demonstrate how the DDE of $Z(t)$ is constructed. \\

\textbf{Step 1: Decomposition of reaction pathways.} Let $\mathcal{A}, \mathcal{B}$ be collection of nodes. We say a reaction pathway is “first hitting” from $\mathcal{A}$ to $\mathcal{B}$, if in the pathway the first node belongs to $\mathcal{A}$, the last node belongs to $\mathcal{B}$, and the intermediate nodes are not in either $\mathcal{A}$ or $\mathcal{B}$. Let $\mathcal{F}(\mathcal{A},\mathcal{B})$ denoting the collection of first hitting pathways from $\mathcal{A}$ to $\mathcal{B}$. \\

\end{multicols} 


\newpage

\begin{multicols}{2} 
	
\textbf{Step 2: Amplification rates and arrival functions for first hitting pathways.} The following concepts are generalized from SAPs. Consider a reaction pathway $\pi(u,\omega): u_0\, \omega_0\,u_1\, \cdots \, \omega_L\, u_{L+1}$ (see Figure 1B). Assume $\pi$ in $\mathcal{F}(\mathcal{E},\mathcal{G}) \cup \mathcal{F}(\mathcal{G},\mathcal{G})$, by definition, the first reaction $\omega_0$ has its upstream node either in $\mathcal{E}$ or $\mathcal{G}$. Let $J[\omega_0]$ denote the flux function of the reaction $\omega_0$, and $S[u_1,\omega_0]$ denote the stoichiometric coefficient of node $u_1$ and reaction $\omega_0$. We express $J[\omega_0]:= R_a X_g$ and define the amplification rate $\kappa_{\pi}$ as 
\begin{equation}
	\kappa_{\pi}(t'):= \frac{\text{mass flux arrived } u_1}
	{\text{gatekeeper mass}} = 
	\frac{S[u_1,\omega_0]\,J[\omega_0](t')}{Z(t')}.
\end{equation}
For $t \rightarrow \infty$, we have $\kappa_{\pi}(t') \rightarrow S[u_1,\omega_0] R_a W_g^*$ where $W_g^*:=X_g/Z$ is the long-term gatekeeper sub-fraction of $x_g$. \\

Calculating the arrival function $h_{\pi}(\tau)$ in LRNs requires more effort than SAPs. In general, reaction pathways can have branches, so each pathway has a pathway probability $q_{\pi}$ of biomass being transferred. Furthermore, nodes in LRNs can have multiple effluxes, with the biomass being partitioned among those effluxes. In this case, only a fraction of biomass is transferred via $\pi$, which is denoted as the transmission efficiency $\theta_{\pi}$. Both $q_{\pi}$ and $\theta_{\pi}$ are constants with values of between 0 and 1, and they represent “biomass transfer dissipation” along the reaction pathway (see SI for explicit formula). \\

In addition to the transfer dissipation, the pathway branches also affect the waiting time distribution. From a stochastic viewpoint, the biomass undergoes the earliest reaction event, so this waiting time is conditional for the reaction in $\pi$ to be the earliest. It follows an exponential distribution of the total reaction rate (as described in classical Gillespie algorithm, see SI). Denoting $f_{T_{\pi}}(\tau)$ as the conditional waiting time distribution for $\pi$, the arrival function then is the product
\begin{equation} \label{E4A0}
	h_{\pi}(\tau) = q_{\pi}\theta_{\pi}f_{T_{\pi}}(\tau).
\end{equation}

\textbf{Step 3. Summation of all relevant reaction pathways.} The final task is to sum the function $\kappa_{\pi} h_{\pi} (\tau)$ for all reaction pathways in $\mathcal{F}(\mathcal{E},\mathcal{G})$ and $\mathcal{F}(\mathcal{G},\mathcal{G})$, which generates the catalytic kernel $\alpha(\tau)$:
\begin{equation} \label{E4A}
	\alpha(\tau) := 
	\sum_{\pi \in \mathcal{F}(\mathcal{E}, \mathcal{G}) 
	\cup  \mathcal{F}(\mathcal{G}, \mathcal{G})}
	\kappa^*_{\pi}\, h_{\pi}(\tau)
\end{equation}
where $\kappa^*_{\pi} := \lim_{t'\rightarrow\infty}\kappa_{\pi}(t')$.

\end{multicols}

\vspace{1 pt}
\begin{center}
	\includegraphics[scale = 0.5]{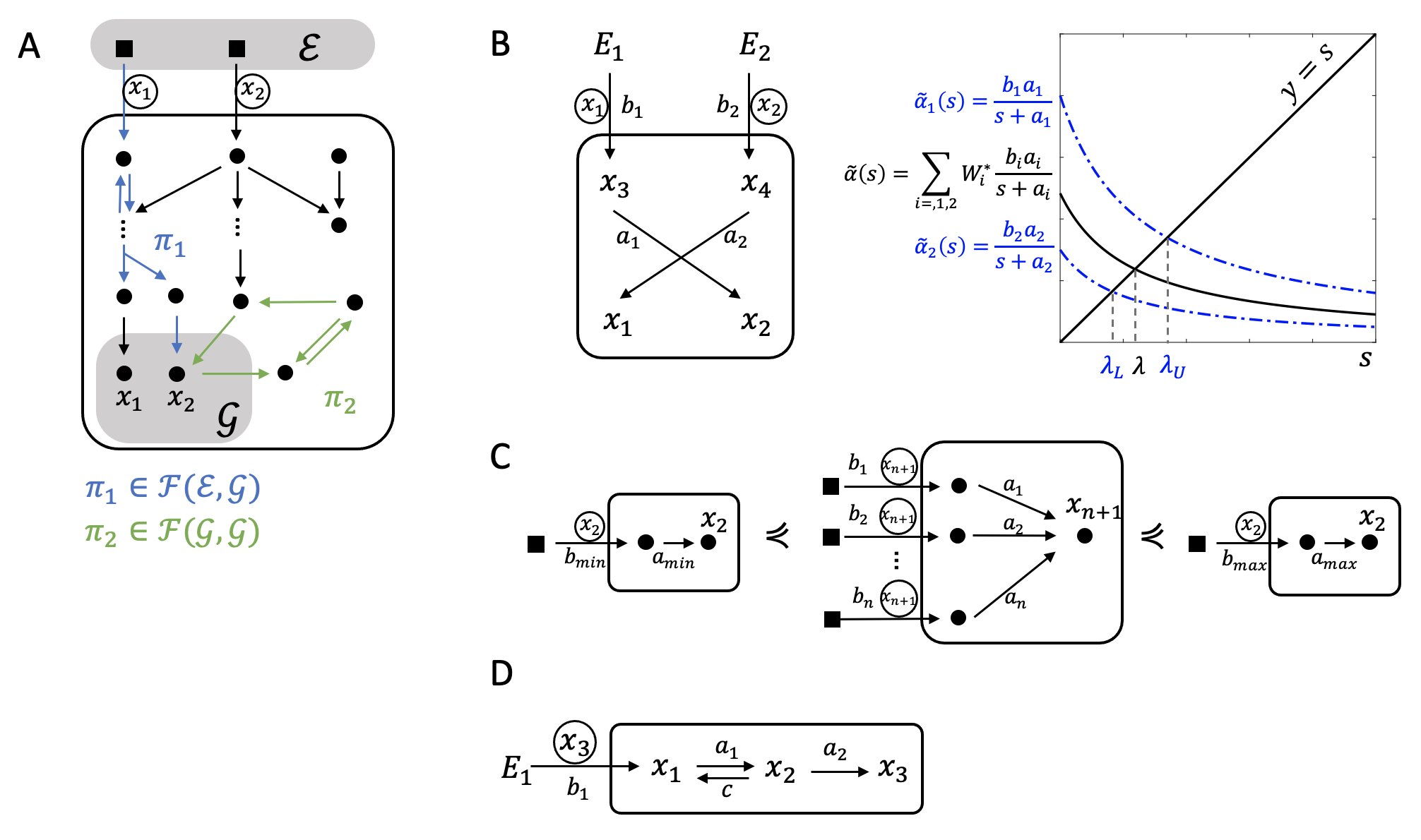}
\end{center}
\small
Figure 3: Linear reaction networks (LRNs). (A) Examples for the first-hitting pathways from the environment ($\mathcal{E}$) to gatekeepers ($\mathcal{G}$) ($\pi_1$ as one example), and from gatekeepers back to gatekeepers ($\pi_2$ as one example). (B) Left: the LRN of Example 4-1 with two boundary reaction influxes and two gatekeepers. Right: $\tilde{\alpha}_1(s)$ and $\tilde{\alpha}_2(s)$ as the upper and lower bounds of $\tilde{\alpha}(s)$, used to estimate $\lambda$. (C) Illustration of the LRNs with parallel reaction pathways from Example 4-2. The ordered relation $\preceq$ is used to represent that one LRN has a higher growth rate then the other, i.e., $\tilde{\alpha}_1 \preceq \tilde{\alpha}_2$ if $\lambda_2 \geq \lambda_1$. The rate constants are described in Example 4-2. (D) An LRN from Example 4-3 with reversible reactions between $x_1$ and $x_2$. \\ 
\normalsize


\begin{multicols}{2} 

In addition, for effluxes from gatekeeper biomass, the gatekeeper degradation rate $\beta$ is defined by
\begin{equation} \label{E4B}
\beta := 
\sum_{x_g \in \mathcal{G}} \;
\sum_{\phi_c \in out(x_g)}
(1-\xi[\mathcal{G}, \phi_c])\;
|S[x_g,\phi_c]|\;
R_c W_g^*,
\end{equation}
where $S[x_g, \phi_c]$ is the stoichiometry coefficient of node $x_g$ and reaction $\phi_c$, and $1-\xi[\mathcal{G}, \phi_c]$ is the downstream fraction that gatekeeper biomass goes to non-gatekeeper nodes (see SI). Note that $\alpha(\tau)$ is a nonnegative function and $\beta \geq 0$. Together, these attributes provide the following theorem (see SI for the proof):

\begin{center}
	\fbox{\parbox{{3.3in}}{
		\textbf{Theorem B.} Consider an LRN and assume given an initial condition with $\lambda>0$. Then, the gatekeeper biomass follows the long-term dynamics
		\begin{equation}
			\frac{dZ}{dt}(t)= -\beta Z(t)
			+ \int_0^t 
			\alpha(\tau)\,Z(t-\tau)\,d\tau
			+ \gamma(t), 
		\end{equation}
with $\lim_{t\rightarrow\infty} \frac{\gamma(t)}{Z(t)}=0$. Furthermore, define $\tilde{\alpha}(s):=\mathcal{L}[\alpha(\tau)]$ as the catalytic spectrum, we have 
\begin{equation}
	\lambda + \beta = \tilde{\alpha}(\lambda).
\end{equation} 
}}
\end{center}


In the following examples, we assume all $\mathfrak{m}(x_j)=1$ for all nodes in all LRNs. \\

\textbf{Example 4.1.} Consider an LRN with four nodes and gatekeepers $\mathcal{G}=\{x_1,x_2\}$ that collectively import each other’s precursors $\{x_3,x_4\}$ (see Figure 3C). The ODE is $\frac{dX}{dt}=MX$ with 
\begin{equation}
	M = 
	\begin{pmatrix}
	0 & 0 & 0 & a_2 \\
	0 & 0 & a_1 & 0 \\
	b_1 & 0 & -a_1 & 0 \\
	0 & b_2 & 0 & -a_2 \\
	\end{pmatrix},
\end{equation}
and the formula of the top eigenvalue (from a nontrivial quartic equation) is quite complicated. Nevertheless, a Lagrangian view provides additional insights. By inspecting the network topology, there are two first-hitting reaction pathways in $\mathcal{F}(\mathcal{E}, \mathcal{G}): \pi_1 = E_1 x_3 x_2$ and $\pi_2 = E_2 x_4 x_1$. The gatekeeper fraction of $x_1$, $x_2$ is denoted by $W_1^*$, $W_2^*$. From Eq. (\ref{E4A}), $\kappa_{\pi_1} = b_1 W_1^*$ and $\kappa_{\pi_2} = b_2 W_2^*$. From Eq. (\ref{E4B}), $h_{\pi_1}(\tau)=a_1 e^{-a_1 \tau}$ and $h_{\pi_2}(\tau)=a_2 e^{-a_2 \tau}$. Therefore, direct calculation gives 
\begin{equation} \label{E4C}
	\tilde{\alpha}(\lambda)= W_1^* \frac{a_1b_1}{\lambda+a_1} 
+ W_2^* \frac{a_2b_2}{\lambda+a_2} = \lambda,
\end{equation}
and $\beta = 0$. To fully solve $\lambda + \beta = \tilde{\alpha}(\lambda)$, 
the algebraic expressions of $W_1^*$ and $W_2^*$ are needed, which are quite intractable. However, since $W_1^*+W_2^*=1$, we can compare $\tilde{\alpha}(s)$ with the upper and lower bounds $\tilde{\alpha}_U(s), \tilde{\alpha}_L(s)$ (Figure 3B, right panel, see SI for details). The range of $\lambda$ can be estimated as being between $\lambda_U = \lambda_2(a_1,b_1)$ and $\lambda_L = \lambda_2(a_2,b_2)$, where $\lambda_2(a,b):= \frac{a}{2} \big(\sqrt{1+\frac{4b}{a}}-1 \big)$, similar to Example 3.2.
  \\


\textbf{Example 4.2.} We are interested in determining if reaction pathways in parallel could be coarse-grained into a single reaction pathway (see Figure 3D), i.e., similar to the law of resistance of an electric circuit. This scenario can be analyzed rigorously by comparing the catalytic spectra of parallel reaction pathways with that of a single coarse-grained pathway. Mathematically, there is no exact “equivalent reaction pathway”, but upper and lower bounds for the system growth rate can be obtained. For the single gatekeeper $\mathcal{G}=\{x_{n+1} \}$ in Figure 3D, $\tilde{\alpha}(s)=\frac{a_1 b_1}{s+a_1} + \cdots +\frac{a_n b_n}{s+a_n}$. Hence, define 
\begin{equation}
\begin{aligned}
a_{\min} = \min_{k} \{a_k\}, 
\qquad
a_{\max} = \max_{k} \{a_k\}, \\
b_{\min} = \min_{k} \{b_k\}, 
\qquad
b_{\max} = \max_{k} \{b_k\},
\end{aligned}
\end{equation}
the catalytic spectrum is bounded as: 
\begin{equation}
	\tilde{\alpha}_L(s):= \frac{n a_{\min} b_{\min} }{s + a_{\min} }
	\leq \tilde{\alpha}(s)
	\leq \frac{n a_{\max} b_{\max} }{s + a_{\max} }
	=: \tilde{\alpha}_U(s).
\end{equation}
This provides the estimation $\lambda \in [\lambda_L, \lambda_U]$, where $\lambda_L = \lambda_2(a_{min},nb_{min})$, $\lambda_U = \lambda_2(a_{max},nb_{max})$. \\


\textbf{Example 4.3.} For general LRNs, it is possible to have reversible reactions (i.e., both reaction $x_j \rightarrow x_k$ and $x_k \rightarrow x_j$ exist in the system), so there are infinite reaction pathways. To illustrate this, consider the simple example shown in Figure 3D. In that case, the gatekeeper set $\mathcal{G}=\{x_3\}$ and the infinite first hitting pathways $\mathcal{F}(\mathcal{E}, \mathcal{G}) = \{\pi_0,\pi_1, \cdots \}$, with 
\begin{equation}
	\pi_k: \; E_1\,x_1\,x_2\,
	(x_1\,x_2)^k\, x_3,
	\qquad k=0,1, \cdots
\end{equation}

Intuitively, pathway $\pi_k$ makes $k$ extra “loops” between nodes $\{x_1,x_2 \}$ before first hitting $x_3$. The path probability to making extra loop decreases geometrically with $k$, namely, $q_{\pi_k} = (\frac{c_2}{c_2+a_2})^k(\frac{a_2}{c_2+a_2})$. Also, $\kappa_{\pi_k}=b$ and $\theta_{\pi_k}=1$ for all $\pi_k$. Let $f_k(\tau)$ denote the conditional waiting time distribution of $T_{\pi_k}$ in (\ref{E4A0}), its Laplace transform is 
\begin{equation}
\mathcal{L}[f_k] =
\frac{a_1}{s+a_1} 
\bigg( \frac{a_2+c_2}{s+a_2+c_2} 
	   \frac{a_1}{s+a_1} 
\bigg)^k
\frac{a_2}{s+a_2+c_2},
\end{equation}
for $k=0,1,\cdots$. Using the formula (\ref{E4A0}),(\ref{E4A}) and perform simplification, we obtain
\begin{equation}
	\tilde{\alpha}(s) 
	= \frac{a_1 a_2 b}{s^2+ (a_1+a_2+c)s + a_1a_2}\, 
\end{equation}
By means of the formula $\tilde{\alpha}(\lambda)=\lambda$, we can obtain an equation that is identical from the ODE method. Furthermore, the Lagrangian view provides insights. Intuitively, reverse reactions delay the autocatalytic process, which can be rigorously showed by noticing $\tilde{\alpha}(s;c)$ decreases as $c$ increases and hence growth rate decreases with $c$ (see SI, Proposition 6.1). \\


\vspace{10 pt}

\textbf{5. Biomass transfer on scalable reaction networks (SRNs).} In the previous section, we discussed biomass transfer on LRNs. However, modeling with LRNs has two major drawbacks. First, all flux functions must be linear functions of X. Second, to satisfy linear flux and the upstream-limited condition simultaneously, each reaction can only have one upstream node. Consequently, to study a larger class of reaction networks, we can consider \textit{scalable reaction networks} (SRNs), whereby each flux function follows the three conditions that:
\begin{enumerate}[label=(\roman*)]
	\item $J_a(X)$ is positive on $\mathbb{R}^n_{>0}$, and is continuously differentiable on $\mathbb{R}^n_{\geq0}$ except the origin.
	\item $J_a(X)$ is upstream-limited, i.e. if $S_{ka}<0$ then $J_a(X) = 0$ whenever $X_k =0$.
	\item $J(cX) = cJ(X)$ for all $c \geq 0$.
\end{enumerate}

The properties of SRNs have been studied previously (\cite{lin_origin_2020}). Condition (ii) is upstream-limited condition which guarantees the system trajectory $X(t)$ is nonnegative. Condition (iii) above requires that $J_a(X)$ be a homogeneous function of degree one, i.e., it scales as $J_a (cX_1,\cdots,cX_n )=cJ_a (X_1,\cdots,X_n)$ for all $c>0$. This scenario encompasses linear fluxes, as well as many nonlinear fluxes such as $J_a(X_1,X_2)= \frac{X_1 X_2}{X_1+X_2}$. Nonlinear fluxes that are scalable can be used to describe the reactions for multiple upstream nodes. Furthermore, by scaling, the dynamics of original ODE: $\frac{dX}{dt}=SJ(X):=F(X)$ can be studied by decomposing $X(t)$ into $N(t)$ and $Y(t)$, where $N := \sum_{k=1}^n X_k$ is the system size and $Y := X/N$ is the rescaled system. Specifically, 
\begin{equation}
	\begin{aligned}
		\frac{1}{N}\frac{dN}{dt} 
		&= \sum_{k=1}^n F_k(Y) 
		:= \mu(Y),\\
		\frac{dY_k}{dt} 
		&= F_k(Y)=\mu(Y)Y_k. \\
	\end{aligned}
\end{equation}
Note that the rescaled trajectory $Y(t)$ is an autonomous ODE on the unit simplex space $\Delta^{n-1} := \{X:X_1+\cdots +X_n=1\}$. The dynamics of the system SRNs are hence decoupled into \textit{simplex dynamics} of $Y(t)$ and \textit{radial dynamics} of $N(t)$. Unlike for LRNs where $Y(t)$ always converges to a fixed point in the long term, the rescaled trajectory $Y(t)$ of SRNs can converge to various omega-limit sets in $\Delta^{n-1}$, including fixed point, limit cycle and torus, heteroclinic cycle, and chaotic attractors, among others. Consequently, it is possible to study unbounded dynamic systems with various growth modalities. \\

To proceed, we assume $X(t)$ in the SRNs also satisfies the following auxiliary conditions: 
\begin{enumerate}[label=(\roman*), start=4]
	\item If a reaction $\phi_a$ has an upstream node in the environment, then exist a system node $x_g$ such that $J[\phi_a] =0$ whenever $X_g = 0$. 
	\item The trajectory $Y(t)$ is $\rho$-\textit{regular} with  an ergodic measure $\rho$ on $\Delta^{n-1}$.
	\item The system has long-term growth rate $\lambda > 0$. 
\end{enumerate}	
	
		
Condition (iv) requires that each boundary influx (which has upstream environmental nodes) is controlled by at least one system node, representing a natural generalization of the gatekeeper concept from LRNs. Together, conditions (i) - (iv) imply that each flux function in SRNs can be expressed as 
\begin{equation}
J_a (X)=R_a (Y) X_k,	
\end{equation}
where $R_a (Y)\geq 0$ is continuous and bounded on $\Delta^{n-1}$, and $x_k$ represent a upstream node or a gatekeeper node of reaction $\phi_a$. Note that the above expression may not be unique. For example, a reaction $x_1+x_2 \rightarrow x_3$ with $J_a(X)=(aX_1 X_2)/N$ can be expressed as $J_a(X)=(aY_2) X_1$ when $x_1$ is regarded as the upstream node, or $J_a (X)=(aY_1) X_2$ when $x_2$ is considered as the upstream node. \\

Condition (v) allows us to perform phase averaging of $Y(t)$ on the unit simplex. We say a trajectory $Y(t)$ is $\rho$-regular if for every Borel set $B \subseteq \mathbb{R}^n$ the following limit always exists
\begin{equation}
	\rho(B)	:= \lim_{T\rightarrow\infty} 
	\frac{1}{T} \int_{0}^T
	\chi_B(Y(t)) \,dt.	
\end{equation}
Here, $\rho$ is called the \textit{occurrence frequency measure} (see SI for details), and $\chi_B(Y)$ be the characteristic function of a set $B$,  with $\chi_B(Y)=1$ if $Y\in B$ and $\chi_B(Y)=0$ otherwise. For $\rho$-regular trajectory, time average of a continuous function $f(Y)$ can be replaced by phase average with respect to measure $\rho$, i.e. \\
\begin{equation}
 \lim_{T\rightarrow\infty} 
	\frac{1}{T}
	\int_{0}^T
	f(Y(t)) \,dt
	= \int_{\Delta^{n-1}}
	f(Y)\,\rho(dY).
\end{equation}
The time average is denoted by $\langle f \rangle_t$ and the phase average with respect to $\rho$ is denoted by $\langle f \rangle_{\rho}$. With this notation, the long-term growth rate can be calculated by $\lambda = \langle \mu(Y)\rangle_{\rho}$. \\

We adopt condition (vi) ($\lambda > 0$) to focus on systems with autocatalytic growth. Note that any SRNs with boundary influxes can be modified as $\lambda > 0$ by removing the boundary effluxes. Consequently, Condition (vi) is not a strong restriction and can be extended in the future. \\

Analyzing system growth from a Eulerian perspective has been discussed previously (\cite{lin_origin_2020}). Accordingly, here I focus on the Lagrangian perspective. First, I introduce a systematic program to decompose an SRN into an equivalent SRN in which each reaction only has one upstream node. Then, the amplification rate and arrival function for first hitting pathways can be calculated. Ultimately, a formula for the effective catalytic spectrum is calculated. \\

\textbf{Step 1: Mono-upstream decomposition.} Consider an SRN satisfying conditions (i)-(vi), denoted as $\mathcal{N}(x, S, J)$. Consider a general reaction $\phi_a$ as in equation (\ref{eqM1}). Let $up(\phi_a)$, $dw(\phi_a)$ represents upstream and downstream nodes, respectively. We can define another reaction $\phi_a^*$ by
\begin{equation} 
\begin{aligned}
	\phi_a^*: 
	&\sum_{x_i \in up(\phi_a)} |d_{ia} - c_{ia}|\, x_i 
	+ \sum_{E_j \in up(\phi_a)} |d_{ja}' - c_{ja}'|\, E_j \\
	\rightarrow
	&\sum_{x_k \in dw(\phi_a)} |d_{ka} - c_{ka}|\, x_k
	+ \sum_{E_{\ell} \in dw(\phi_a)} 
	|d_{\ell a}' - c_{\ell a}'|\,E_{\ell},
\end{aligned}
\end{equation}
One can verify that $\phi_a$ and $\phi_a^*$ yield the same ODE in the deterministic setting. Now, let $\nu[x_i, \phi_a^*] \in [0,1]$ represent the biomass fraction from upstream node $x_i$ in reaction $\phi_a^*$. We can split the reaction $\phi_a^*$ into multiple reactions, where each reaction has exactly one upstream nodes from $\phi_a^*$. Namely, for each node $x_i \in up(\phi_a^*)$, define
\begin{equation}
	\begin{aligned}	
		\phi_a^{x,i}
		: |d_{ia}-c_{ia}|\,x_i
		\rightarrow 
		&\sum_{x_k \in dw(\phi_a)}
		\bigg(
		\nu[x_i,\phi_a^*] \cdot
		|d_{ka}-c_{ka}| 
		\bigg)\,x_k \\
		+ &\sum_{E_\ell \in dw(\phi_a)}
		\bigg(
		\nu[x_i,\phi_a^*] \cdot
		|d'_{\ell a}-c'_{\ell a}| 
		\bigg)\,E_\ell,
	\end{aligned}	 
\end{equation}
and define similarly for each node $E_j \in up(\phi_a^*)$ (see SI, Definition 8.4). For all reactions in above, define their flux functions as $J[\phi_{a}^{x,i}] = J[\phi_{a}^{E,j}] = J[\phi_a]$. Perform this decomposition for all reactions, we arrive a new SRN where each reaction has one upstream node and also satisfies condition (i)-(vi), denoted as $\mathcal{N}(x,S^*,J^*)$. This equivalent SRN follows the same ODE as the original SRN. \\


\textbf{Step 2: Amplification rates and arrival functions for first-hitting pathways.} After obtaining the equivalent SRN $\mathcal{N}(x,S^*,J^*)$, we next look for its first-hitting pathways. Consider a reaction pathway $\pi$ in $\mathcal{F}(\mathcal{E}, \mathcal{G}) \cup \mathcal{F}(\mathcal{G}, \mathcal{G})$, $\pi(u,\omega) = u_0\,\omega_0\,	\cdots \omega_L\, u_{L+1}$, and assume that biomass arrives at node $u_1$ at time $t'$ and arrives at $u_{L+1}$ at time $t$ (shown in Figure 3B). \\

The \textit{state-dependent amplification rate} is defined as  
\begin{equation}
\begin{aligned}
\kappa_{\pi}^{dep}(Y(t'))
:&= \frac{\text{mass flux arrived } u_1}{Z} \\
&= \frac{S[u_1,\omega_0]\,J[\omega_0](t')}{Z(t')}.
\end{aligned}
\end{equation}
Note that for SRNs, the ratio $\frac{J[\omega_0](t')}{Z(t')}$ is a function of $Y(t')$, with superscript ‘dep’ emphasizing this the amplification rate should be evaluated at the biomass \textit{departure time} $t'$. \\

\end{multicols}


\vspace{10 pt}

\begin{center}
	\includegraphics[scale = 0.5]{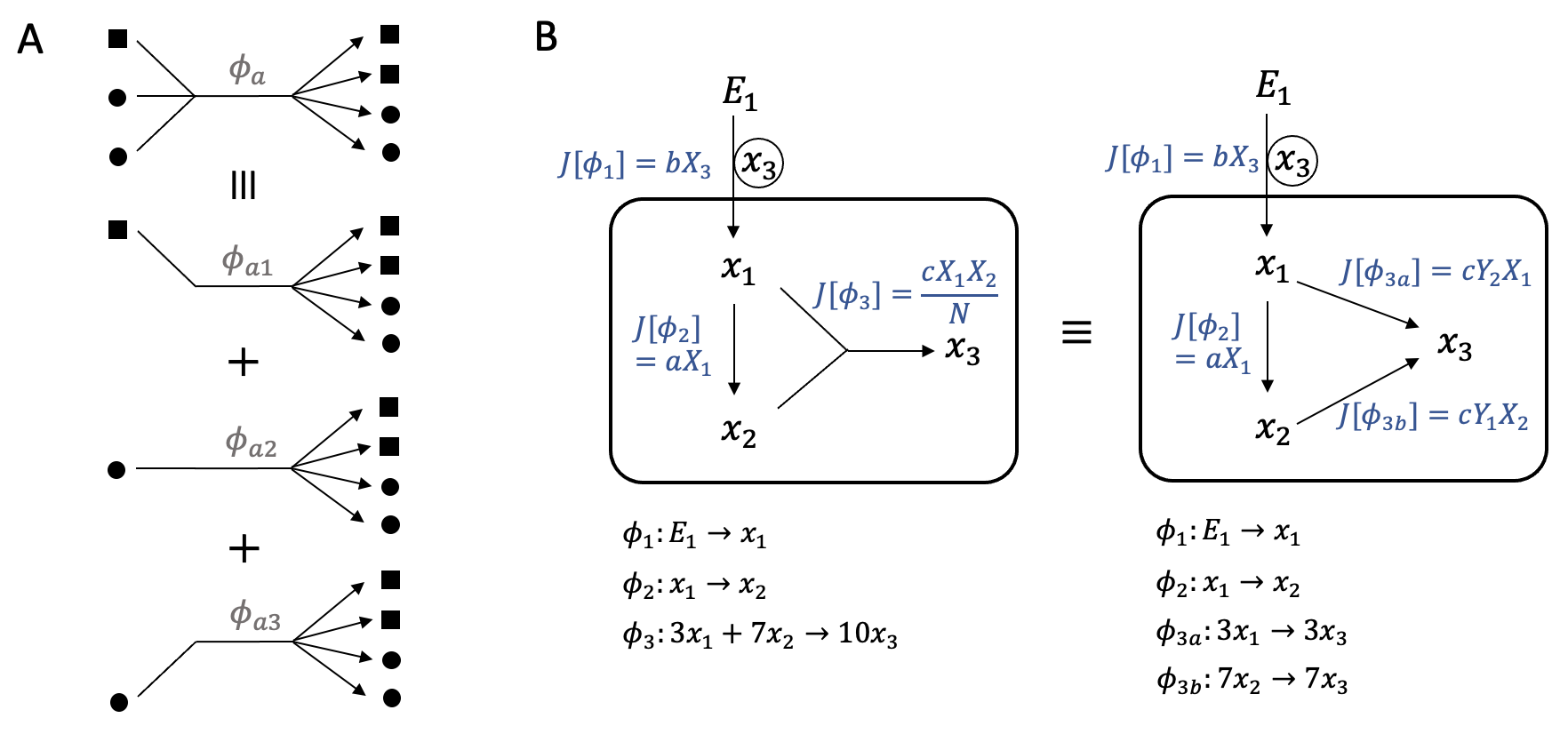}
\end{center}
\small
Figure 4: Mono-upstream decomposition for SRNs. (A) Each reaction (environmental nodes as squares, system nodes as circles) with multiple upstream nodes is decomposed into multiple reactions having single upstream nodes. (B) An example of decomposition. Here, reaction $\phi_3$ is decomposed into $\phi_{3a}$ and $\phi_{3b}$. \\ 
\normalsize

\newpage

Next, we define the \textit{state-dependent arrival function} as:
\begin{equation}
	h_{\pi}^{arr}(\tau;Y(t)) := H(t',t).
\end{equation}
We put superscript ‘arr’ to emphasize that this arrival function should be evaluated at the biomass \textit{arrival time} $t$. The function $h_{\pi}^{arr}$ can be calculated by according to the Chapman-Kolmogorov equation (\cite{klenke_probability_2013}), 
\begin{equation}
\begin{aligned}
H(t',t) = \int_{t'\leq t_2 \leq\cdots\leq t_L \leq t}
 P(t',t_2|\;u_1,\omega_1,u_2)\;P(t_2,t_3|\; u_2,\omega_2,u_3) 
\cdots P(t_L,t|\;u_L,\omega_L,u_{L+1})\; 
dt_2\cdots dt_L, 
\end{aligned}
\end{equation}
with
\begin{equation}
\begin{aligned}
	P(t_0,t_1|\,x_j,\phi_a,x_k) 
    := 
	r_{ja}(t_1)\,\exp\bigg( 
	-\int_{t_0}^{t_1} r_{tot}(z) \,dz 
	\bigg )
     \times\, \xi[x_j,\phi_a], \\
    r_{ja}(t) := 
	\frac{1}{X_j} |S_{ja}|\, J_a(t),
    \qquad
    r_{tot}(t) := 
	\frac{1}{X_j}\sum_{\phi_c \in out{(x_j)} } |S_{jc}|\, J_c(t).
\end{aligned}
\end{equation}\


\textbf{Step 3. Summation of all relevant reaction pathways.} Similar to the case of LRNs, we sum up all reaction pathways in $\mathcal{F}(\mathcal{E}, \mathcal{G})$ and $\mathcal{F}(\mathcal{G}, \mathcal{G})$. This gives a state-dependent catalytic kernel $\alpha(\tau)$ with $\tau :=t-t'$:
\begin{equation} 
\begin{aligned}
\alpha(\tau; Y(t)) :=
\sum_{\pi \in \mathcal{F}(\mathcal{E},\mathcal{G}) \;\cup\; \mathcal{F}(\mathcal{G},\mathcal{G})}
\kappa_{\pi}^{dep}(Y(t'))\, h_{\pi}^{arr}(Y(t)).
\end{aligned}
\end{equation} 
For effluxes from gatekeeper nodes, the state-dependent gatekeeper degradation rate $\beta$ is defined as:
\begin{equation} 
\beta(Y(t)) :=
\sum_{x_g \in \mathcal{G}} \;
\sum_{\phi_c\in out(x_g)}
(1-\xi[\mathcal{G}, \phi_c])\,
|S[x_g,\phi_c]|\;
\frac{J_c(X(t))}{Z(t)}.
\end{equation}


\begin{center}
	\hrulefill
\end{center}

\begin{multicols}{2} 
Under SRN conditions (i)-(vi), the gatekeeper biomass $Z(t)$ grow as fast as the entire system. If $Y(t)\rightarrow Y^*$, then $Z(t)$ grows exponentially on the delta measure $\rho(dY)=\delta(Y-Y^*)$ and it is clear that $Z(t)/Z(t+\tau) \rightarrow e^{-\lambda \tau}$. For general attractor with ergodic measure measure $\rho$, similar result holds and is summarized by Theorem C (see SI for proof). 

\begin{center}
\fbox{\parbox{{3.3in}}{		
\textbf{Theorem C}: Consider an SRN satisfying condition (i)-(vi) with long-term growth rate $\lambda$. Then, the gatekeeper biomass $Z(t)$ satisfies 
\begin{equation}
\lim_{T\rightarrow\infty}\frac{1}{T} \int_0^T \; 
\frac{Z(t)}{Z(t+\tau)}\;dt
= e^{-\lambda \tau},
\end{equation}
for every $\tau \geq 0$. \\
}}
\end{center}

Note that for general SRNs, both $\alpha(\tau; Y(t))$ and $\beta(Y(t))$ are functions of $Y \in \Delta^{n-1}$. If $Y(t)\rightarrow Y^*$, then the relevant quantities in the long term are $\alpha(\tau, Y^*)$ and $\beta(Y^*)$ and we obtain a DDE expression similar to that of LRNs. In general, SRNs may converge to other types of attractors, such as limit cycles. For these nontrivial cases, we found that the appropriate average is \textit{not} $\langle \alpha(\tau;Y) \rangle_{\rho}$. Denote $\Gamma^{\tau}: X(t) \mapsto X(t+\tau)$ as the semigroup operator for the ODE solution trajectory $X(t)$. The appropriate average which gives the \textit{effective catalytic kernel} is described in Theorem D. \\

\begin{center}
\fbox{\parbox{{3.3in}}{

\textbf{Theorem D}: Consider an SRN satisfying condition (i)-(vi), with a trajectory having long-term growth rate $\lambda$ and occurrence frequency measure $\rho$. Define 
\begin{equation}
\begin{aligned}
\alpha_{eff}(\tau) 
&:= e^{\lambda \tau} 
\bigg \langle \alpha(\tau;Y(t)) \frac{Z(t)}{Z(t+\tau)}
\bigg \rangle_{t} \\
&= e^{\lambda \tau} 
\bigg \langle \alpha(\tau;Y) \frac{Z(X)}{Z(\Gamma^{\tau}X)}
\bigg \rangle_\rho \\
\beta_{eff} 
&:=  
\big \langle \beta(Y(t))\big \rangle_t
= \big \langle \beta(Y) 
\big \rangle_\rho
\end{aligned}
\end{equation} 
Then the long-term growth rate satisfies 
\begin{equation} \label{E-D2}
\lambda + \beta_{eff} = \tilde{\alpha}_{eff}(\lambda).
\end{equation} 
}}
\end{center}

In the following, we use three examples to illustrate the analyses of biomass transfer in SRNs.  \\


\textbf{Example 5.1.} Consider an SRN (Figure 4B, left) that has a reaction $\phi_3: 3 x_1 + 7 x_2 \rightarrow 10 x_3$ with two upstream nodes $x_1,x_2$. The scalable flux function for this reaction is $J[\phi_3 ]=(cX_1 X_2)/(X_1+X_2+X_3)$. With parameter $a=b=c=1$ and a positive initial condition, $Y(t)$ converges to a fixed point $Y^*$. To analyze this system from the Eulerian view, we solve $dY_k/dt=0$ for $k=1,2,3$ numerically, yielding $Y^* \approx (0.3041,0.1121,0.5838)$. Therefore, $\lambda = \mu(Y^* )=bY_3^* \approx 0.5838$. \\ 

To analyze the SRN from a Lagrangian perspective, mono-upstream decomposition is performed to obtain a new SRN with a different topology but follows the same ODE (Figure 4B, right). There are four reactions in this new system, all of which have a single upstream node. There are two first hitting pathways in $\mathcal{F}(\mathcal{E},\mathcal{G})$, i.e., $\pi_1:E_1 x_1 x_3$ and $\pi_2:E_1 x_1 x_2 x_3$. Analyzing their catalytic spectra gives (see SI for details)
\begin{equation}
	\tilde{\alpha}(s) = \frac{1}{s+1+3Y_2^*} \;
	\frac{3Y_2^*+7Y_1^*}{s+7Y_1^*}. 	
\end{equation}
It can be verified numerically that $\tilde{\alpha}(\lambda)=\lambda$ is satisfied. \\


\textbf{Example 5.2.} To illustrate how to calculate an effective catalytic kernel, consider an SRN with rescaled trajectory $Y(t)$ displaying oscillatory behavior (see Figure 5A). This model is inspired by the classical repressilator (\cite{mb_synthetic_2000}), where node $x_2$ is the gatekeeper and nodes $x_2,x_3,x_4$ repress each others' synthesis fluxes. Since the system is oscillatory (Figure 5B, left), the occurrence frequency measure $\rho$ is supported by a limit cycle. We denote $\tau_{cyc}$ as the “phase parameter" of the limit cycle. The length of the oscillatory period is $\tau_{cyc}^{max} \approx 35$ time units. \\

From a Eulerian viewpoint, the long-term growth rate is obtained by 
\begin{equation}
	\lambda  = \int_0^{\tau_{cyc}^{max}}\,
	\mu(Y(\tau_{cyc}))\,d\tau_{cyc}.
\end{equation} 
For the Lagrangian view, note that the catalytic kernel $\alpha(\tau; Y(\tau_{cyc}))$ is state-dependent, which can be calculated numerically for different $\tau_{cyc}$ (see Figure 5B, right). Numerically, we calculate the effective kernel $\alpha_{eff}(\tau)$ as well as the mean kernel $\langle \alpha(\tau;Y) \rangle_{\rho}$. The former, but not the later gives the correct growth rate formula in (\ref{E-D2}) (see SI for details). \\

\textbf{Example 5.3.} To illustrate how SRNs can be applied to a cell-like system, a simple proteome partition model inspired by the earlier work (\cite{scott_interdependence_2010}) is constructed. This  SRN has five nodes (labeled as $S, U, P, Q$ and $R$ in Figure 5C). Nodes $S$ and $U$ represent as small metabolites, and proteome sectors are represented by transporters ($P$), housekeeping enzymes for metabolic conversion ($Q$), and ribosomal proteins ($R$), respectively. The transporter $P$ acts as a gatekeeper to control system influx. The housekeeping enzyme $Q$ controls metabolic conversion of $S$ into $U$. The ribosomal proteins ($R$) control the synthesis of $P,Q,R$ from $U$. We consider the parameter set $a_1=50,a_2=100,b=K_1=K_2=10$, where the three proteome sectors are synthesized according to the fractions $c_P,c_Q,c_R$, respectively, with $c_P+c_Q+c_R=1$. The rescaled system converges to a fixed point $Y^*$. \\ 

From the Eulerian view, the long-term growth rate follows $\lambda=bP^*$, where $\lambda$ is determined by differing allocation strategies of $c_P,c_Q,c_R$. The optimal growth is achieved when $(c_P^*,c_Q^*,c_R^*) \approx (18\%,46\%,36\%)$ (Figure 5D, left). \\

From a Lagrangian perspective, the only reaction pathway in $\mathcal{F}(\mathcal{E},\mathcal{G})$ is $\pi:E_1 SUP$, representing gatekeeper $P$ (the transporter) catalyzing itself. Clearly, this reaction pathway is dependent on $P$ (required for $E_1 \rightarrow S$), $Q$ (required for $S \rightarrow U$), and $R$ (required for $U \rightarrow R$). The amplification rate is $\kappa_{\pi}=b$, and the pathway probability $q_{\pi}$ equals $c_P$. Note that there is a clear trade-off among sectors: increasing $c_P$ increases the chance for biomass goes to $P$ from $U$, while increasing $c_Q$ and $c_R$ accelerates the arrival time of $\pi$. The arrival function is
\begin{equation}
\begin{aligned}
	h_{\pi} (\tau;Y^*)  
	  = c_P (e^{-r_1\tau} )*(e^{-r_2\tau}), \\
	r_1 = \frac{a_1 Q^*}{K_1+S^*}, \quad
 	r_2 = \frac{a_2 R^*}{K_2+U^*}. 
\end{aligned}
\end{equation}

Using the parameter $r_{avg}:=( \frac{1}{r_1} + \frac{1}{r_2})^{-1}$ as the \textit{overall transfer rate} of $\pi$, the trade-off among $c_P,c_Q,c_R$ can be approximated by a trade-off between $\kappa = c_P$ (the amplification rate) and $r_{avg}$ (the transfer rate), see Figure 5D. Thus, the Lagrangian viewpoint provides additional insight into how optimal biomass synthesis being constrained by multiple factors in biochemical pathways. \\


\textbf{6. Discussion.} Delay differential equations (DDEs) have been used prevalently to model biological systems (\cite{macdonald_biological_1989}, \cite{smith_introduction_2010}), including cell signaling (\cite{heltberg_inferring_2019}), disease propagation (\cite{culshaw_mathematical_2003}, \cite{young_consequences_2019}), and population dynamics (\cite{kuang_delay_1993}), among others. Reaction networks belong to a special type of system wherein the delay originates from the stochastic waiting time of biomass during the reactions. This type of dynamics is distinct from other DDEs, such as age-structured populations, wherein the delay is attributable to the age of individuals (\cite{britton_essential_2003}), or for general reaction networks with a single, specific delay time (\cite{craciun_delay_2020}). \\

In our framework, simple reaction pathways (SAPs) are used as the basic building block for more complicated autocatalytic systems, such as LRNs and SRNs. Networks with topology similar to SAPs have  been explored with different types of flux functions (\cite{buzi_analysis_2011}). In our work, we focused on scalable fluxes which allow the expression $J_a (X)=R_a (Y) X_k$, where the pre-factor $R_a(Y)$ can act as the Markov transition rate of biomass. \\

\end{multicols}


\begin{center}
	\includegraphics[scale = 0.5]{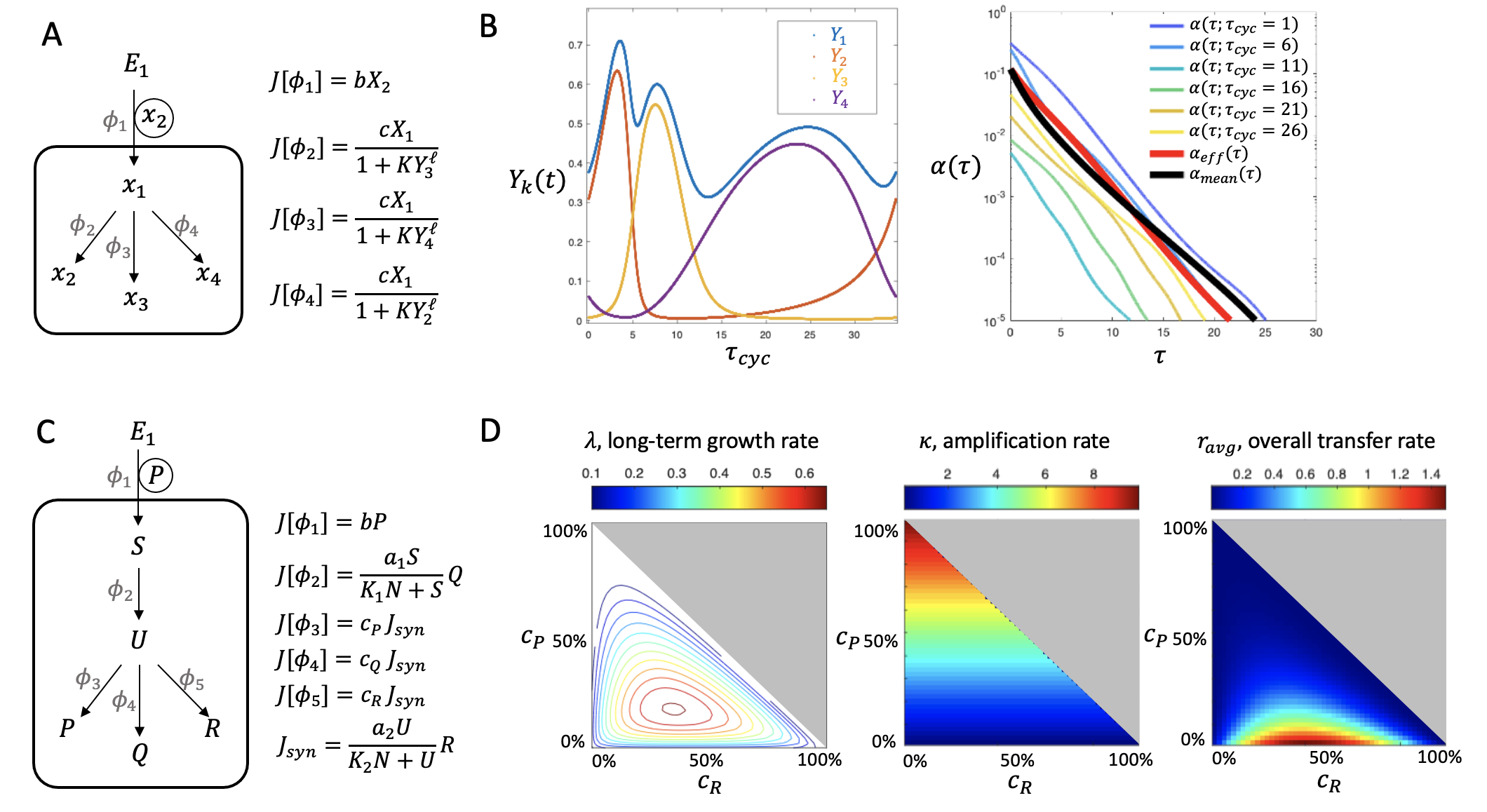}
\end{center}
\small
Figure 5: Examples of SRNs. (A,B) A growing repressilator system with oscillatory dynamics. (A) Flux diagram of the SRN with three nonlinear fluxes repressed by each other’s reactant. (B) Left: Oscillatory pattern of the rescaled trajectory $Y(t)$, where $\tau_{cyc}$ is the phase parameter for the limit cycle. Parameters used: $b=c=1, K=1000, \ell = 2.5$. Right: State-dependent catalytic kernel at different phases, as well as the effective (red) and the mean (black) catalytic kernels. (C,D) A proteome partition system with two metabolites ($S, U$) and three proteome sectors ($P,Q,R$). Parameter used are $a_1=50, a_2 = 100, b=10, K_1 = K_2 = 10$. (C) Flux diagram for biosynthesis and proteome allocation. (D) Dependency of $\lambda$ (left), $\kappa=c_P$ (middle) and $r_{avg}$ (right) on various allocations of $c_P,c_Q,c_R$ across the parameter space subjected to $c_P+c_Q+c_R=1$. \
\normalsize
\begin{center}
	\hrulefill
\end{center}


\begin{multicols}{2} 

For DDEs, it is common practice to analyze system stability in the frequency domain (\cite{gu_stability_2003}). Here, the “instability” or “Lyapunov exponent” of the growing DDE is analyzed, yielding the relation $\lambda + \beta = \tilde{\alpha}(\lambda)$ for the long-term growth rate. Characterizing the positive Lyapunov exponent in the frequency domain represents an intriguing future research direction. The catalytic spectrum $\tilde{\alpha}(s)$ monotonically decreases in the positive axis, hence facilitating geometric intuition and interpretation of $\lambda$ (Figure 2B). It is plausible to regard $\tilde{\alpha}(s)$ as a “fingerprint” for network autocatalysis, and the information contained within it allows further exploration. \\

It is worth emphasizing that the catalytic kernel $\alpha(\tau)$ is not merely a mathematical tool, but also possesses a real biophysical quantity in nature. It represents the overall biomass transfer delay and amplification under gatekeeper control. For cellular or ecological systems, $\alpha(\tau)$ could be measured experimentally by means of isotope labeling (\cite{wang_metabolic_2020}). In simulated flux network models (e.g., \cite{yamagishi_adaptation_2021}), catalytic kernel can also be computed during the mass transferred processes. With help from ergodic theory, I characterized the effective kernel $\alpha_{eff}(\tau)$ as the suitable generalization for this biophysical quantity when the system has more complex growth dynamics. \\

Reaction networks in the real world can be extremely complex. To describe cellular systems, models range in complexity from a few coarse-grained nodes (\cite{scott_interdependence_2010}) through intermediate-level (e.g. central metabolism, see  \cite{lao-martil_kinetic_2022}), to a comprehensive simulated cell comprising thousands of nodes (\cite{karr_whole-cell_2012}). It is difficult to compare among different reaction network models since their network topologies and complexities differ by orders of magnitude. However, in terms of the growing scalable reaction networks proposed in the current study, different models can be compared via their catalytic kernels $\alpha(\tau)$. Doing so enables us to rigorously estimate how parameter and dimensional reductions affect the range of growth rate (Figure 2C, 3C). \\

In summary, analyzing biomass transfer through the lens of the Lagrangian perspective raises many interesting directions for future studies. The purpose for DDE approach is not to replace the ODE analysis, but to provide additional insight on mass transfer dynamics and the network structure. Mathematically, the coarse-graining process for reaction network can use to define a partial-order relationship among SRNs. Biologically, applying the concept of the catalytic kernel and estimating this quantity for the biochemical networks of cells could help unify and reconcile the divergent models currently in existence, providing a deeper understanding of autocatalysis. \\

\textbf{Acknowledgements.}  The author thanks Bernold Fiedler, Lai-Sang Young, Edo Kussell for useful discussions and comments. The author thanks Cheng-Hung Chang, Jia-Yuan Dai, Wei-chung Liu for reviewing the earlier version of this manuscript. The author thanks Christine Jacobs-Wagner for supporting this research while this work was initiated as an independent project during the author's postdoc training period. The author thanks colleagues in reaction network community and colleagues in Academia Sinica for discussions. This work was supported by the Howard Huges Medical Institute (via Christine Jacobs-Wagner) and the Institute of Molecular Biology, Academia Sinica (to Wei-Hsiang Lin). \\

\end{multicols}


%
%
%
%
%
%


\newpage


\Large
\begin{center}
	Supplementary Information. \\
	Biomass transfer on autocatalytic reaction networks: \\
	a delay-differential equation formulation
\end{center}
\normalsize

\tableofcontents

\newpage

\begin{center}
\textbf{Notation}	
\end{center}

\begin{center}
\begin{tabular}{|p{2.5cm}*{1}{|l|l|}r}
         
\hline
Variable
& Unit
& Definition \\
\hline
$\alpha^{\bigstar}(\tau;t)$ 
& $[time]^{-2}$ 
& Catalytic kernel for general RNs \\
$\alpha(\tau)$ 
& $[time]^{-2}$ 
& Catalytic kernel for LRNs \\
$\alpha(\tau; Y(t))$
& $[time]^{-2}$ 
& State-dependent catalytic kernel for SRNs \\
$\alpha_{eff}(\tau)$
& $[time]^{-2}$ 
& Effective catalytic kernel for SRNs \\
$\tilde{\alpha}(s)$
& $[time]^{-1}$ 
& Catalytic spectrum, Laplace transform of $\alpha(\tau)$ or $\alpha_{eff}(\tau)$  \\
\hline
$\beta^{\bigstar}(t)$ 
& $[time]^{-1}$ 
& Gatekeeper degradation rate for general RNs \\
$\beta$ 
& $[time]^{-1}$ 
& Gatekeeper degradation rate for LRNs \\
$\beta(Y(t))$ 
& $[time]^{-1}$ 
& Gatekeeper degradation rate for SRNs \\
$\beta_{eff}$ 
& $[time]^{-1}$ 
& Effective gatekeeper degradation rate for SRNs \\
\hline
$\gamma(t)$ 
& $[mass][time]^{-1}$ 
& Residual terms in delay differential equation \\
$\theta_{\pi}$
& $1$ 
& transmission efficiency for reaction pathway $\pi$ \\
$\kappa_{\pi}$
& $[time]^{-1}$ 
& Amplification rate for reaction pathway $\pi$ \\
$\kappa_{\pi}^*$
& $[time]^{-1}$ 
& Amplification rates of reaction pathway $\pi$ in LRNs \\
$\kappa^{dep}_{\pi}(Y(t'))$
& $[time]^{-1}$ 
& State-dependent amplification rates of reaction pathway $\pi$ in SRNs\\
$\lambda$
& $[time]^{-1}$ 
& Long-term growth rate of the system \\
$\mu(t)$
& $[time]^{-1}$ 
& Instantaneous growth rate of the system \\
$\nu[x_i, \phi_c]$
& $1$ 
& Upstream biomass fraction of $\phi_c$ from node $x_i$ \\
$\xi[x_j, \phi_c]$
& $1$
& Downstream biomass fraction of $\phi_c$ into node $x_j$ \\
$\xi[A, \phi_c]$
& $1$
& Downstream biomass fraction of $\phi_c$ into node set $A$ \\
$\pi_k$
& $-$
& Reaction pathways \\
$\rho(dY)$
& $1$
& Occurrence frequency measure of a trajectory $Y(t)$ \\
$\phi_c, \omega_c$
& $-$
& Reactions in general reaction networks \\

\hline
$\Gamma^t$
& $-$
& Semigroup operator with parameter $t$ \\
$\Delta^{n-1}$
& $-$
& Unit simplex of dimension $n-1$ \\
$\Omega$
& $-$
& Omega-limit set of a trajectory $Y(t)$ \\

\hline
$h_{\pi}(\tau)$
& $[time]^{-1}$
& Arrival function of reaction pathway $\pi$ in LRNs \\ 
$h^{arr}_{\pi}(\tau, Y(t))$
& $[time]^{-1}$ 
& Arrival functions of reaction pathway $\pi$ in SRNs \\
$m$
& $1$ 
& Number of reactions in system \\
$m_A^+(t), m_A^-(t)$
& $[time]^{-1}$ 
& Biomass flux that enters or leaves node collection $A$ \\
$\mathfrak{m}(x_k)$
& $[mass]$ 
& Biomass of node $x_k$ \\
$\mathfrak{m}_0$
& $[mass]$ 
& Biomass unit \\
$n$
& $1$ 
& Number of nodes in system \\
$q_{\pi}$
& $1$
& Pathway probability for reaction pathway $\pi$ \\
$x_k, u_k$
& $-$
& System nodes \\

\hline
$E_k$
& $-$ 
& Environmental nodes \\
$H_{\pi}(t',t)$
& $[time]^{-1}$ 
& Arrival function of reaction pathway $\pi$ in general RNs  \\
$J_c(X), J[\phi_c]$
& $[time]^{-1}$ 
& Flux function of reaction $\phi_c$ \\
$M$
& $[time]^{-1}$ 
& Matrix of LRN \\
$N$
& $[mass]$ 
& Total biomass in the system \\
$S_{ka}, S[x_k,\phi_a]$
& $[mass]$ 
& Stoichiometry matrix (biomass-weighted) 
  for general RNs  \\ 
&& \;\;of node $x_k$ and reaction $\phi_a$\\
$W_g$
& $[mass]$ 
& Gatekeeper sub-fraction of node $x_g$; defined as $X_g/Z$ \\
$X_k(t)$
& $[mass]$ 
& Biomass of system node $x_g$ \\
$Y_k(t)$
& $1$ 
& Relative biomass of system node $x_g$; defined as $X_k/N$ \\
$Z(t)$
& $[mass]$ 
& Gatekeeper biomass \\
\hline
$\mathcal{E}$
& $-$ 
& Collection of environmental nodes \\
$\mathcal{F}(\mathcal{A}, \mathcal{B}$)
& $-$ 
& Collection of first hitting pathways 
from $\mathcal{A}$ to $\mathcal{B}$  \\
$\mathcal{G}$
& $-$ 
& Collection of gatekeeper nodes \\
$\mathcal{X}$
& $-$ 
& Collection of system nodes \\
\hline
\end{tabular} \\
\end{center}



\newpage

\section{Eulerian and Lagrangian perspectives for reaction networks (RNs)}

In this section, we consider growing systems surrounded by environments and compare the dynamics of reaction networks between discrete and continuous frameworks. We introduce concept of "biomass unit" to facilitate the Lagrangian perspective for tracking biomass. \\

\textbf{Definition 1.1 (Discrete framework).} We consider a model with \textit{system} and surrounding \textit{environments}, where the reaction networks have system nodes $x_1, \cdots, x_n$ and environmental nodes $E_1, \cdots, E_{n'}$. Let $\#(x_i) $ and $\#(E_j)$ represent the numbers of objects of $x_i$-- or $E_j$-- types, respectively. In this work we assume environmental objects are unlimited, so $\#(E_j)$ can be regarded as arbitrary large numbers. \\

A reaction $\phi_a$ is represented by 
\begin{equation}
	\phi_a: \sum_{i=1}^n c_{ia}x_i + \sum_{j=1}^{n'}c_{ja}'E_j
	\rightarrow
	\sum_{i=1}^n d_{ia}x_i + \sum_{j=1}^{n'}d_{ja}'E_j,
\end{equation}
where  $c_{ia}, c'_{ja}, d_{ia}, d'_{ja}$ are nonnegative integers. While we have not specifiy the kinetic rules for reactions, every proper reaction kinetics would require that reaction $\phi_a$ does not happen when $\#(x_i) < c_{ia}$ or $\#(E_j) < c'_{ja}$. After a $\phi_a$--reaction event, $\#(x_i)$ and $\#(E_j)$ change to $\#(x_i) + d_{ia} - c_{ia}$ and $\#(E_j) + d'_{ja} - c'_{ja}$, respectively for all $i, j$. \\


\textbf{Note.} The above description is \textit{discrete} in the sense that the number of objects are nonnegative integers and reaction events happen at discrete time points. When object numbers are large, $\#(x_j)$ can be modeled by continuous variables. Also, when reaction events are frequent on the timescale of interest, we can use reaction frequency (a variable continuous in time) to replace individual reaction event in the model. This leads to our next definition. \\


\textbf{Definition 1.2 (Continuous framework).} Consider a reaction network in Definition 1.1, we construct a systems ODE to describe its dynamics.  Let $\mathfrak{m}(x_i)>0$ and  $\mathfrak{m}(E_j)>0$ represents the \textit{biomass} for object on node $x_i$ and $E_j$, respectively. In this way, the total biomass on node $x_i$ is equal to 
\begin{equation}
	X_i := \mathfrak{m}(x_i)\cdot \#(x_i).
\end{equation}
We call $X =(X_1,\dots,X_n)^T$ as the \textit{biomass vector} of the system. Next, we assume there are $m$ reactions in the system and denote them as $\phi_1,\cdots, \phi_m$. The \textit{biomass-weighted stoichiometry matrix} $S$ (with dimension $n$-by-$m$) is defined by
\begin{equation}
	S_{ia} := (d_{ia}-c_{ia})\,\mathfrak{m}(x_i).
\end{equation}
We also define $J_a(t)\,\delta t$ as the frequency of $\phi_a$-type reaction in time $[t, t+\delta t]$, and $J = (J_1,\dots,J_m)^T$ as the \textit{flux functions} for reactions. Flux function of specific reaction $\phi_a$ is also denoted by $J[\phi_a]$. The stoichiometry coefficient corresponds to $x_i$ and $\phi_a$ is also denoted by $S[x_i,\phi_a]$. \\

Note that here $X_i(t)$ and $J_a(t)$ are nonnegative real variables, defined for continuous time $t \geq 0$. With these definition, the biomass dynamics of system nodes can be described by 
\begin{equation}
	\frac{dX_i}{dt} = \sum_{a=1}^m
	S_{ia} J_a.
\end{equation}


\textbf{Note.} We introduce the concept of "biomass" as a weighting factor for different kinds of objects. In chemical or biological systems, different objects can represent  different types of molecules, and biomass can be molecular weight or carbon contents of the molecule. \\ 


\textbf{Definition 1.3.} Given a node $x_k$ in a reaction network, we denote $in(x_k)$ and $out(x_k)$ as the collection of reactions $\phi_a$ that has $S_{ka}>0$ and $S_{ka}<0$, respectively; i.e. reactions that provide net influx and efflux of $x_k$. Given a reaction $\phi_a$, the  \textit{upstream nodes of} $\phi_a$, denoted by $up(\phi_a)$,  are the nodes with $d_{ia}>c_{ia}$ or $d'_{ia}>c'_{ia}$. Similarly, the \textit{downstream nodes of} $\phi_a$, denoted by $dw(\phi_a)$, are nodes with $d_{ia}<c_{ia}$ or $d'_{ia}<c'_{ia}$. \\

\textbf{Note.} A node can appear as both reactant and product in a reaction. In our framework, we only consider \textit{net biomass change} of a reaction, and use the term "upstream" and "downstream" nodes to represent source and sink of biomass for this reaction. \\


The ODE framework is an Eulerian perspective, since we focus on mass influxes and effluxes of each node. To adopt Lagrangian perspective, we need to track biomass transition on reaction network. This motivates us to include additional assumptions: \\

\textbf{Definition 1.4 (Discrete framework with tractable biomass units).} Following Definition 1.1 and 1.2., we assume additionally that 

\begin{enumerate} [label=(\roman*)] 
	\item All biomass are composed by discrete \textit{biomass units}. Each unit has the same biomass $\mathfrak{m_0}$, and these units cannot be created, divided or destroyed. 
	\item Biomass units are distinguishable from each others. 
	\item All object biomass $\mathfrak{m}(x_j)$ are multiples of $\mathfrak{m_0}$, i.e. $\mathfrak{m}(x_j)/\mathfrak{m_0}$ are integers. 
	\item When a reaction event happens, biomass units are draw randomly from upstream nodes and distributed randomly into downstream nodes according to biomass stoichiometry.
\end{enumerate}

\textbf{Note.} These additional assumptions are motivated from physical and chemical systems. Here, biomass units behave as atoms, which are indivisible for typical chemical reactions, and the objects are behaved as molecules. By assuming biomass units cannot be created or destroyed, we have \textit{mass conservation}, i.e. for each reaction $\phi_a$ we have 
\begin{equation}
	\sum_{i=1}^{n} (d_{ia}-c_{ia})
	\,\mathfrak{m}(x_i)
	+ \sum_{j=1}^{n'} (d'_{ja}-c'_{ja})
	\,\mathfrak{m}(E_j) 
	= 0.
\end{equation}

We assume that the biomass units to be randomly distributed from upstream nodes to downstream nodes after reaction occurs. In typical chemical reactions, atoms from reactants can be allocated into specific positions in the products in a non-random way. We adopt this minimal assumption here given that the molecular configuration of objects are unspecified. With more information on reactions, this assumption can be relaxed and modified in the future. Since biomass is conserved, we can define the fraction of biomass in and out of each node during a reaction, stated in the following.  \\  


\textbf{Definition 1.5. (Upstream and downstream fractions of reaction fluxes).} We define the biomass-weighted stoichiometry constants by 
\begin{equation}
\begin{aligned}
	S_{ia} &:= (d_{ia} - c_{ia}) \,\mathfrak{m}(x_i), \\
	S_{ja}' &:= (d_{ja}' - c_{ja}') \,\mathfrak{m}(E_j). \\
\end{aligned}
\end{equation}
 Note that if $x_i \in up(\phi_a)$, then $S_{ia}<0$. Similarly, if $E_j \in up(\phi_a)$, then $S_{ja}' < 0$. Given a reaction $\phi_a$, the total biomass being transferred via $\phi_a$ in single reaction event is denoted by 
\begin{equation}
	\mathfrak{m}_{tot}(\phi_a) := 
	\sum_{x_i \in up(\phi_a)} |S_{ia}| +
	\sum_{E_j \in up(\phi_a)} |S_{ja}'| > 0.
\end{equation}
By mass conservation, we also have
\begin{equation}
	\mathfrak{m}_{tot}(\phi_a) = 
	\sum_{x_k \in dw(\phi_a)} |S_{ka}| +
	\sum_{E_\ell \in dw(\phi_a)} |S_{\ell a}'|.
\end{equation}
 
For reaction $\phi_a$, we define the \textit{upstream fractions} (denoted by $\nu[x_i,\phi_a]$ and $\nu[E_j, \phi_a]$) as the biomass fraction from $x_i \in up(\phi_a)$ and $E_j \in up(\phi_a)$. Namely,
\begin{equation}
	\nu[x_i,\phi_a] := 
	\frac{ |S_{ia}| }{ \mathfrak{m}_{tot}(\phi_a)},  \qquad
	\nu[E_j,\phi_a] := 
	\frac{ |S_{ja}'| }{ \mathfrak{m}_{tot}(\phi_a)}.
\end{equation}

For reaction $\phi_a$, we define the \textit{downstream fractions} (denoted by $\xi[x_k, \phi_a]$ and $\xi[E_\ell, \phi_a]$) as the biomass fraction into from $x_k \in dw(\phi_a)$ and $E_j \in dw(\phi_a)$. Namely,
\begin{equation}
	\xi[x_k,\phi_a] := 
	\frac{ |S_{ka}| }{ \mathfrak{m}_{tot}(\phi_a)},  \qquad
	\xi[E_\ell,\phi_a] := 
	\frac{ |S_{\ell a}'| }{ \mathfrak{m}_{tot}(\phi_a)}.
\end{equation}
By definition, $\nu[x_i,\phi_a], \,\nu[E_j,\phi_a]$, $\xi[x_k, \phi_a],\, \xi[E_\ell, \phi_a]$ are positive numbers between 0 and 1. \\

Finally, sometimes we are interested in upstream/downstream fraction for a collection of system nodes. Let $\mathcal{C}_1, \mathcal{C}_2$ be collections of upstream and downstream system nodes of $\phi_a$, respectively. We define 
\begin{equation}
	\nu[\mathcal{C}_1, \phi_a] := 
	\sum_{x_i \in \mathcal{C}_1} \nu[x_i,\phi_a],\qquad
	\xi[\mathcal{C}_2, \phi_a] := 
	\sum_{x_k \in \mathcal{C}_2} \xi[x_k,\phi_a].
\end{equation}


\textbf{Example}: The following example shows how to calculate upstream and downstream fractions of reaction $\phi_a$. \\

\begin{center}
	\includegraphics[scale = 0.6]{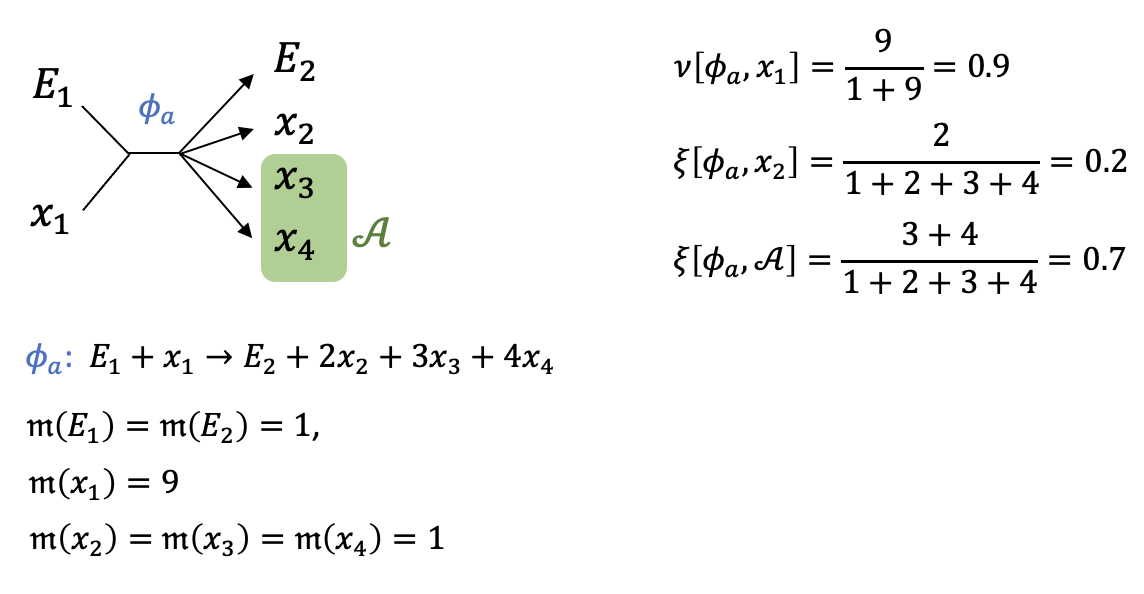}
\end{center}


Under the framework of Definition 1.1 to Definition 1.4, we can construct a stochastic process based on discrete biomass units, with kinetics to be consistent to the ODE framework. This stochastic process will be used to track the "trajectory" of individual biomass unit in the reaction network. \\


\begin{center}
	\includegraphics[scale = 0.4]{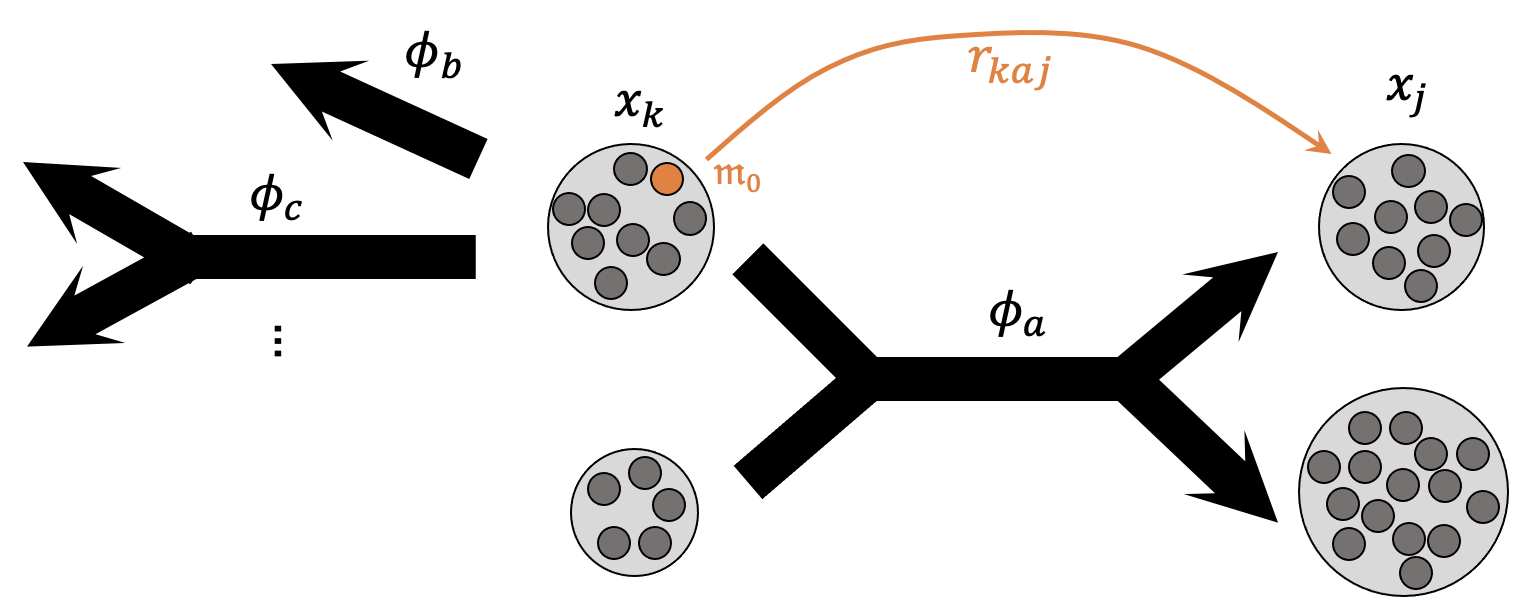}
\end{center}


Consider a reaction $\phi_a$ with $x_k \in up(\phi_a)$ and $x_j\in dw(\phi_a)$. Based on ODE, the biomass flux from $x_k$ entering reaction $\phi_a$ within time $[t,t+\delta t]$ is $|S_{ka}|J_a(t)\,\delta t$ (note that here $S_{ka}<0$ since $\phi_a$ is an efflux of node $x_k$). Among these biomass, only a fraction of them arrive $x_j$ since there are other downstream nodes. This is the downstream fraction $\xi[x_j,\phi_a]$ in Definition 1.5. Therefore, the biomass flux from $x_k$ transfer to $x_j$ via reaction $\phi_a$ within infinitesimal time $\delta t$ is 
\begin{equation}
    |S_{ka}| J_a(t) \times 
    \xi[x_j,\phi_a] \,\delta t.
\end{equation}
To compare with the discrete framework, the flux in above equation is equivalent to $N_{kaj} \delta t$ biomass units, where
\begin{equation}
    N_{kaj} = 
    \frac{|S_{ka}|J_a(t)}{\mathfrak{m_0}} \times 
    \xi[x_j,\phi_a].
\end{equation}
Now, if we focus on a single biomass unit located on $x_j$ (labeled as an orange circle in above figure), then this specific unit may be one of the $N_{kaj}$ unit that goes to $x_j$. Alternatively, it could remains at $x_k$ or goes to other reactions $\phi_b, \phi_c$, etc. Since at time $t$ the number of biomass units on node $x_k$ is $X_k/\mathfrak{m_0}$, and every biomass unit has equal chance to enter $\phi_a$, the probability for a specific biomass unit to enter $\phi_a$ and arrive $x_j$ within time $[t,t+\delta t]$ is 
\begin{equation}
    N_{kaj}\, \delta t 
    \bigg \slash
    \frac{X_k(t)}{\mathfrak{m_0}}.
\end{equation}
We summarize this result in Proposition 1.6. \\


\textbf{Proposition 1.6.} Under the framework of Definition 1.1 to Definition 1.4. Let $r_{kaj}(t)\,\delta t$ denote the transition probability within time $[t,t+\delta t]$ for a single biomass unit at node $x_k$ to transit to node $x_j$ via reaction $\phi_a$. Assume the ratio $\frac{J_a}{X_k}$ exists for all fluxes. We have 	
\begin{equation}
	r_{kaj}(t)= \frac{|S_{ka}|J_a(t)}{X_k(t)}
    \times 
	\xi[x_j, \phi_a].
\end{equation}

Interestingly, the transition rate $r_{jak}$ is independent of the value of $\mathfrak{m_0}$. Conceptually, we could discretize our system for any $\mathfrak{m_0} >0$ as long as conditions in Definition 1.4 are satisfied. Therefore, we do not need to specify an explicit value for $\mathfrak{m_0}$, while still able to formulate the transition process for biomass unit to move on the reaction networks. \\

Note that in Definition 1.4 the reaction dynamics is still discrete in time, that is, reaction event happened at certain time points, and biomass units were transferred at those time points. In contrast, in Proposition 1.6, we defined $r_{kaj}$ as the \textit{instantaneous transition rate} for a continuous-time Markov process. In this work, growing systems are considered, and hence the reaction event number per unit of time increases as system grows. The continuous-time approximation is reasonable since reaction event numbers are large. \\

As a final note, so far we have not specifies the flux function $J(X)$. To ensure the ratio $J_a(X)/X_k$ is bounded, flux functions must be  upstream-limited and scales as $J_a(X) \propto X_k ^\alpha$ with $\alpha \geq 1$. These condition will be introduce in the latter sections. To analyze the growth behavior of the system, we have the following definition. \\


\textbf{Definition 1.7 (System growth).} Consider the systems ODE in Definition 1.2. The \textit{system size} is defined by $N := X_1+\dots+X_n$. The \textit{relative fraction of biomass} $Y$ is defined as $Y := X/N$, also called as the \textit{rescaled system}. \\

We denote $\mathbb{R}^{n}_{\geq0}$ and $\mathbb{R}^n_{>0}$ as the nonnegative and positive orthants, and 
\begin{equation}
	\Delta^{n-1} := \{X: X_1+\dots+X_n = 1\}
\end{equation}
as the unit simplex. We consider system dynamics $X(t) \in \mathbb{R}^n_{\geq 0}$ and $Y(t) \in \Delta^{n-1}$. The \textit{instantaneous growth rate} of the system is defined as 
\begin{equation}
\mu := \frac{1}{N}\frac{dN}{dt}.
\end{equation}
The trajectory $X(t),N(t),Y(t)$ are governed by the ODEs \cite{lin_origin_2020}:

\begin{equation}
\begin{aligned}
	\frac{dX_k}{dt} &= \sum_{a=1}^m S_{ka}J_a(X) =: F_k(X),
	\;\;k=1,\dots,n, \\
	\frac{dN}{dt} &= \sum_{k=1}^n F_k(X), \\
	\frac{dY_k}{dt} &= \sum_{a=1}^m S_{ka}
	\frac{J_a(NY)}{Y}
	 - \mu(Y) Y_k,
	\;\;\;k=1,\dots,n. \\
\end{aligned}
\end{equation}\

The \textit{long-term growth rate} of the system (if exists) is defined as 
\begin{equation}
\lambda := \lim_{t\rightarrow\infty} \frac{1}{t}\,\log N(t).	
\end{equation}

For a solution trajectory $Y(t), \;t \in [0,\infty)$, we denote \textit{time average} of a function $f(Y(t))$ (if exists) as 
\begin{equation}
	\langle f \rangle_t := \lim_{T\rightarrow\infty} \frac{1}{T} \int_0^T f(Y(t))\,dt.
\end{equation}


\textbf{Note.} The time average may not exist (for example, if $X(t)$ converge to a heteroclinic cycle \cite{gaunersdorfer_time_1992}). If the above average exists for $\mu(t)$, we have $\lambda = \langle \mu\rangle_t$. Other than time average, we will also consider phase average on probability measures (see section 6). \\


\textbf{Definition 1.8 (Reaction pathway).} We define a reaction pathway by
\begin{equation}
	\pi(u,\omega) = u_0 \,\omega_0 \,u_1 \cdots \omega_{L} \,u_{L+1}
\end{equation}
where nodes $\{u_0,\cdots ,u_{L+1} \} =: \{u\}$ and reaction $\{\omega_0, \cdots, \omega_L \} =: \{\omega\}$ are ordered sets. For each reaction $\omega_j$, the nodes $u_{j-1}$ and $u_j$ must be its upstream and downstream nodes, respectively. We require all nodes in $\{u\}$ to be system nodes, except for the first or last nodes (which can be environmental node). We allow $\{u\}$ and $\{\omega\}$ to have repeated members and hence the length $L$ can be infinite. When there is no confusion, we omit the $\{\omega\}$ and write $\pi: u_0 \,u_1 \cdots \,u_{L+1}$. \\ 


\begin{center}
	\includegraphics[scale = 0.45]{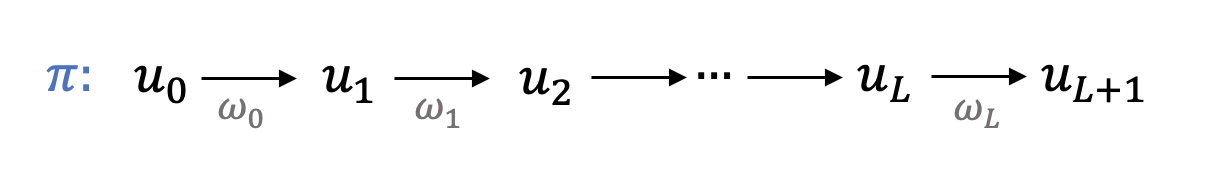}
\end{center}


\textbf{Note.} In the figure example above, the $\pi$ only includes the nodes and reactions in the pathway and each reactions has exact one upstream and one downstream node. In general, it is possible that these reactions have multiple upstream/downstream nodes. It is also possible that a node has multiple influx/efflux reactions. \\



In summary, our goal is to construct Lagrangian perspective and analyze how discrete biomass units move on a reaction network. The relation between different frameworks is depicted in below. 

\begin{center}
	\includegraphics[scale = 0.6]{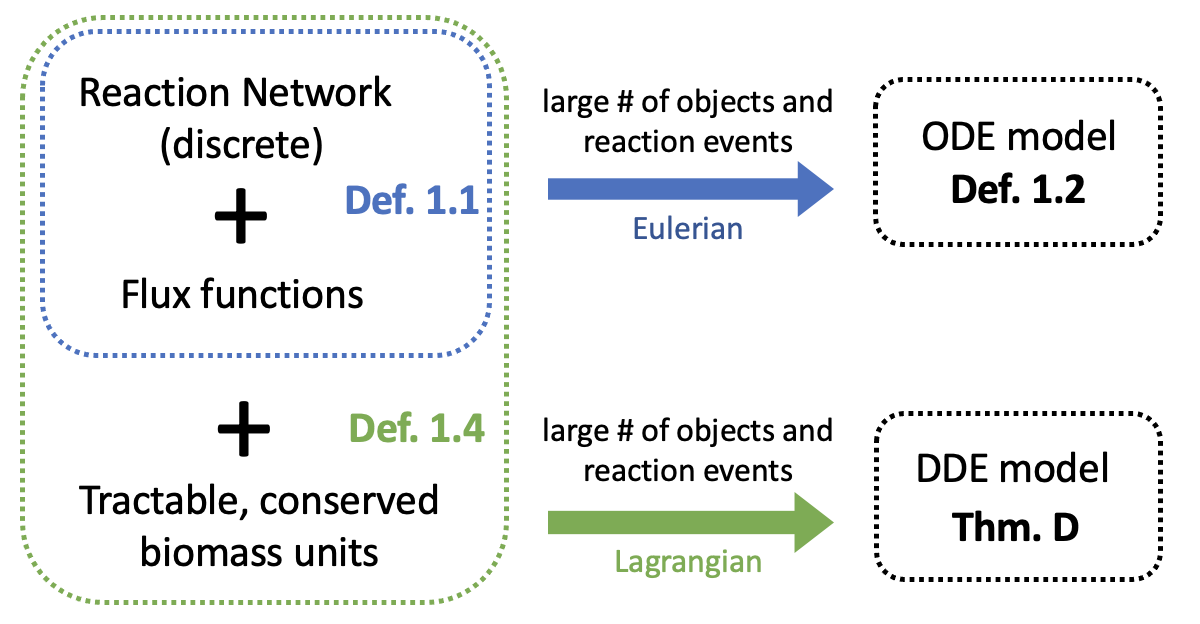}
\end{center}



\newpage
\section{Gatekeepers and DDE formulation for growing systems}

In this section, we describe how to analyze biomass transfer using the Lagrangian perspective for growing system. Our main idea is to focus on biomass of "gatekeeper nodes" and formulate the delay differential equations of the biomass of these nodes. Throughout this work, we assume the reaction networks and biomass satisfies the assumptions in section 1. \\

\textbf{Definition 2.1.} Consider a reaction network with system nodes $\mathcal{X}$ and environmental node $\mathcal{E}$. We call those reactions with upstream node in $\mathcal{E}$ as \textit{boundary influx reactions}. A boundary influx reaction $\phi_a$ is said to have a \textit{gatekeeper node} $x_g$ if its flux function can be expressed as 
\begin{equation}
J[\phi_a](X) = R_a(Y)X_g	,
\end{equation}
where $R_a(Y) \geq 0$ is bounded on the unit simplex space $\Delta^{n-1}$. \\

\textbf{Note.} The definition of gatekeeper nodes depends on the flux functions $J[\phi_a](X)$. It is possible that for a boundary reaction $\phi_a$ the choice of $x_g$ is not unique; in this case, we will assign one $x_g$ as the gatekeeper for $\phi_a$ and fixed this assignment throughout the analysis. The figure below shows some examples of gatekeeper $x_g$ and the flux functions of boundary influx reactions. 
\begin{center}
	\includegraphics[scale=0.45]{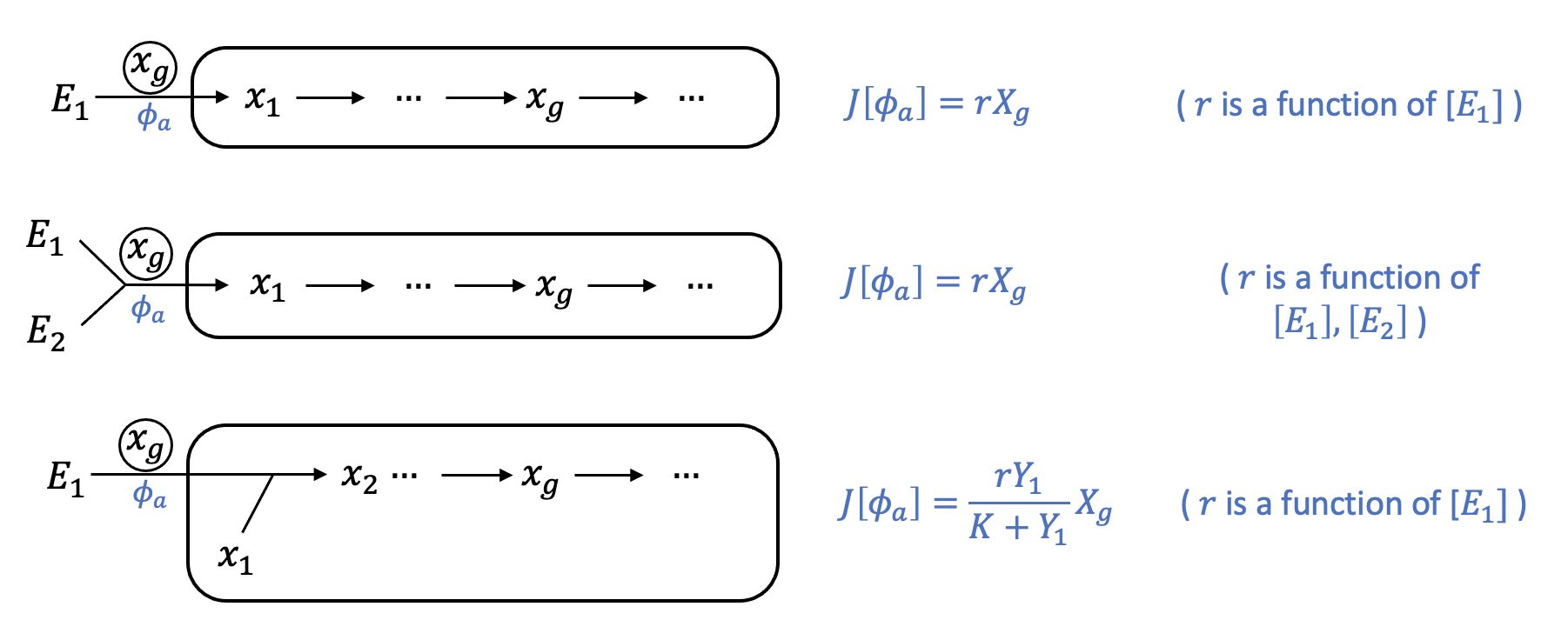} \\
\end{center}

Note that the coefficient $r$ can be potentially a function of $[E_1], [E_2]$, i.e. depend on the concentration of environmental nodes. In this work, we assume environmental nodes has constant concentration, and hence $r$ is regarded as a constant. \\

We denote $\mathcal{G} \subseteq \mathcal{X}$ as the collection of gatekeeper nodes. The \textit{gatekeeper biomass} $Z$ and \textit{gatekeeper sub-fractions} $W_g$ are defined as  
\begin{equation}
\begin{aligned}
Z &:= \sum_{x_g \in \mathcal{G}} X_g, \\
W_g &:= X_g / Z,
\qquad \text{for }\; x_g \in \mathcal{G}.
\end{aligned}
\end{equation}
Intuitively, the existence of gatekeepers indicates that the system growth is controlled and rate-limited by gatekeeper nodes. This is shown in the next proposition. \\


\textbf{Proposition 2.2.} Consider a reaction network whose boundary influxes all have gatekeepers. We have the following properties: 

\begin{enumerate}[label=(\roman*)]
	\item If $\lim_{t\rightarrow\infty} Z(t) = 0$, then $\lim_{t\rightarrow\infty} \frac{dN(t)}{dt} \leq 0$. 
	\item If $\lambda$ exists, then $\lim_{t\rightarrow\infty} \frac{Z}{N} = 0$ implies $\lambda \leq 0$. 
	\item If $\lambda$ exists and is positive, and the limit 
\begin{equation}
\mu_Z:= 
\lim_{t\rightarrow\infty}\frac{1}{t} \log Z(t) 
= \bigg \langle\frac{1}{Z}\frac{dZ}{dt} \bigg \rangle_t
\end{equation}
also exists, then $\mu_Z = \lambda$.
\end{enumerate}

\textbf{Proof.} Let $\mathcal{I}$ denote the collection of boundary reactions. Since all boundary influxes have gatekeepers, we have $J[\phi_a](X) = R_a(Y)X_g \leq M_a Z$ for $\phi_a \in \mathcal{I}$, where $M_a := \max\{R_a(Y):\,Y\in \Delta^{n-1}\} $. Define $M_Z := \max_{\phi_a \in \mathcal{I}}\{M_a\}$. We have \\
\begin{equation} \label{E2.1}
	\frac{dN}{dt} = 
	\sum_{a=1}^m S_{ka}J_a(X) \leq 
	\sum_{\phi_a \in \mathcal{I}} S_{ka}J_a(X) \leq
	\bigg(\sum_{\phi_a \in \mathcal{I}} S_{ka}\bigg)M_Z Z.
\end{equation}
This implies the statement (i). Now, dividing the left and right hand sides in (\ref{E2.1}) by $N$ gives 
\begin{equation}
	\frac{1}{N}\frac{dN}{dt} \leq C'\frac{Z}{N},
\end{equation} 
where $C'$ is a positive constant. If $Z(t)/N(t) \rightarrow 0$ then the long-term time average $\langle Z/N \rangle_t = 0$. This implies $\langle \frac{1}{N}\frac{dN}{dt}\rangle_t = \lambda \leq 0$, which is the statement (ii). \\

To show (iii), we suppose the contrary that $\lambda -\mu_Z > \varepsilon > 0$ for some $\varepsilon$. This implies $\lim_{t\rightarrow\infty} \frac{1}{t} \log(N(t)/Z(t)) > \varepsilon$; hence 
\begin{equation}
0 \leq \lim_{t\rightarrow\infty} \frac{Z(t)}{N(t)} \leq 
\lim_{t\rightarrow\infty} e^{-\varepsilon t} = 0
\end{equation}
and we have $\lim_{t\rightarrow\infty} Z(t)/N(t) = 0$. By the property (ii), this implies $\lambda \leq 0$, which contradicts to our assumption that $\lambda > 0$. Therefore, we must have $\lambda \leq \mu_Z$. \\

Next, we show that $\lambda < \mu_Z$ leads to a contradiction. Suppose $\mu_Z -\lambda > \varepsilon > 0$ for some $\varepsilon$. This implies  
\begin{equation}
\log Z(t) - \log N(t) > \varepsilon t 
\end{equation}
for large enough $t$. However, this is impossible since $Z\leq N$ for all $t$ by definition. Therefore, we  must have $\mu_Z = \lambda$. $\blacksquare$ \\   


%
%

Proposition 2.2 tells us that if $\lambda > 0$ then gatekeeper biomass $Z(t)$ has the same long-term growth rate as the system. This allows us to study growing system via a DDE of $Z(t)$. We use the following example to illustrate the idea. For simplicity, all reactions has linear flux functions in form of $J_a(X)=R_a X_k$, all stoichiometry coefficients to be $\pm1$ for the reactions, and $\mathfrak{m}(x_j)=1$ for all nodes. \\

\textbf{Example 2.3.} Consider a reaction network with three pathways $\pi_1,\pi_2,\pi_3$ as shown in below figure. This system has only one gatekeeper node $\mathcal{G}=\{x_1\}$. All biomass enters the system by $\phi_1$ and using the pathway $\pi_1$ to reach the node $x_1$. Another pathway $\pi_2$ start from $x_1$ and circulated via other nodes and eventually back to $x_1$. The pathway $\pi_3$ start from $x_1$ and export the biomass into the environment. 

\begin{center}
\includegraphics[scale=0.5]{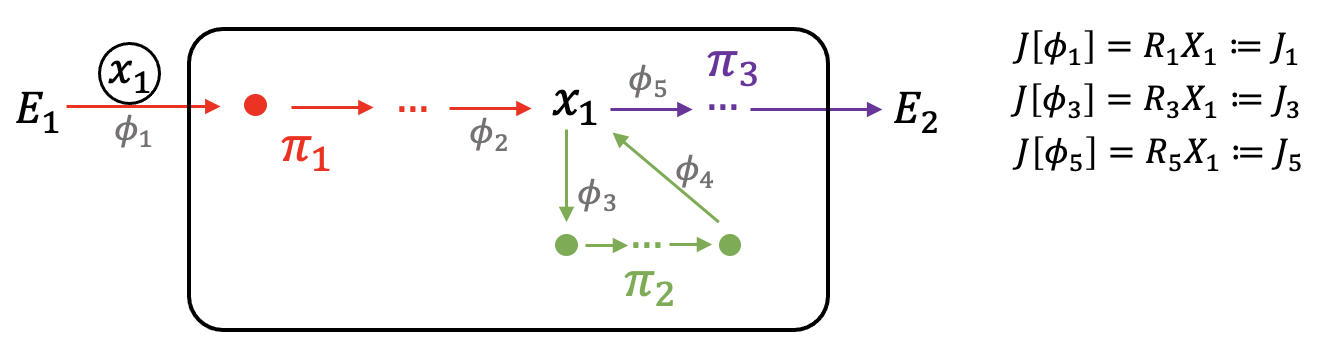} \\	
\end{center}

We assume all stoichiometry coefficients to be $\pm1$ for these reactions. Let $J_k$ denote $J[\phi_k]$, the systems ODE follows
\begin{equation}\label{ExampleLRN}
	\frac{dX_1}{dt} = J_2-J_3+J_4-J_5.
\end{equation}

Our next goal is to write down a Markov kernel function $m_{\pi}(t',t), \; t'\leq t$ that correlates the upstream biomass flux $J_1(t')$ and the downstream biomass flux $J_2(t)$. In this way, we can express flux $J_2(t)$ in terms of an integral on $J_1(t')$ with this kernel function, i.e.
\begin{equation}
J_2(t) = \int_0^t m_{\pi_1}(t',t)\, J_1(t')\,dt'.\\
\end{equation}

Since $x_1$ gatekeeps $\phi_1$, we can write $J_1(t') := R_1X_1(t')$. As we will show later, the Markov function $m_{\pi_1}(t',t)$ for an linear reaction network is a function of $t'-t:=\tau$ and we can define  $\alpha_1(t-t') := m_{\pi_1}(t',t)$. This allow us to express 
\begin{equation}
J_2(t) = \int_0^t \alpha_1(\tau)\, R_1 X_1(t-\tau)\,d\tau.
\end{equation}

Similarly, for the pathway $\pi_2$, there is another Markov function that relates the earlier biomass flux $J_3(t')$ and the latter biomass flux $J_4(t)$, by defining $\alpha_2(t-t') := m_{\pi_2}(t',t)$. Note that since $x_1$ is the upstream of $J_3$, we can write $J_3 = R_3X_1$. Therefore,
\begin{equation}
J_4(t) = \int_0^t \alpha_2(\tau)\, R_3 X_1(t-\tau)\,d\tau.
\end{equation}

For the effluxes from $x_1$, we note that $J_3=R_3X_1$ and $J_5=R_5X_1$. Using the equation (\ref{ExampleLRN}), we derive an DDE for $X_1$, 
\begin{equation}
	\frac{dX_1}{dt}(t) = -\beta X_1(t)
		+ \int_0^t \alpha(\tau)\, X_1(t-\tau)\, d\tau
\end{equation}\
with $\alpha(\tau) := \alpha_1(\tau) + \alpha_2(\tau)$ and $\beta := R_3 + R_5$. \\

This example illustrates how the kernel function $\alpha$ is derived from the Markov relation between fluxes. Our goal is to develop a systematic way to characterize the kernel function. \\


\textbf{Definition 2.4.} Let $\mathcal{A}, \mathcal{B}$ denote collections of nodes and consider $\pi: u_0\,\omega_0\, u_1\cdots\omega_L u_{L+1}$. We say $\pi$ is a \textit{first hitting pathway from $\mathcal{A}$ to $\mathcal{B}$} if $u_0 \in \mathcal{A}$, $u_{L+1} \in \mathcal{B}$, and other nodes of $\pi$ are not in $\mathcal{A}$ or $\mathcal{B}$. We denote 
\begin{equation}
	\mathcal{F}(\mathcal{A}, \mathcal{B}):= 
	\text{collection of first hitting reaction pathways from }
	\mathcal{A}\text{ to }\mathcal{B} .
\end{equation}
In addition, we define nonnegative quantities below to describe the amount of mass being transferred:
\begin{equation}
\begin{aligned}
m^+_{\mathcal{A}}(t) 
& := 
\{\text{mass flux arrived $\mathcal{A}$ at time $t$}\}, \\
m^-_{\mathcal{A}}(t) 
& := 
\{\text{mass flux left $\mathcal{A}$ at time $t$}\}, \\
m_{\pi}(t',t) 
& := \{
\text{mass flux arrived $u_1$ at time $t'$, transferred via $\pi$, and arrived $u_{L+1}$ at time $t$}\}.
\end{aligned}
\end{equation} 
Note that the unit for $\int m^+_{\mathcal{A}}(t) \,dt$ and $\int m^-_{\mathcal{A}}(t) \,dt$ are both [mass]. Also, the unit for $\int \int m_{\pi}(t',t) \,dt dt'$ is [mass]. \\   



\textbf{Lemma 2.5.} Consider a reaction network with $\lambda > 0$ and assume each boundary influx has a gatekeeper node. Then the  gatekeeper biomass $Z$ follows
\begin{equation}
\frac{dZ}{dt} =\sum_{x_g \in \mathcal{G}} \frac{dX_g}{dt}
= m^+_{\mathcal{G}} - m^-_{\mathcal{G}}.
\end{equation}
The quantities $m^+_{\mathcal{G}}, m^-_{\mathcal{G}}$ are given by the formula

\begin{equation}
\begin{aligned}
    m^+_{\mathcal{G}}(t) &=
    \sum_{\pi \in \mathcal{F}(\mathcal{E}, \mathcal{G})
    \cup \mathcal{F}(\mathcal{G}, \mathcal{G}) } 
    \int_{0}^{t}
    m_{\pi}(t',t)\; dt' \
    + C_{ini}(t), \\
    m_{\mathcal{G}}^-(t)
	&= (-1)
	\sum_{x_g \in \mathcal{G}} \;
	\sum_{\phi_c \in out(x_g)} 
	(1-\xi[\mathcal{G}, \phi_c]) \, S[x_k,\phi_c] \, J[\phi_c](t),
\end{aligned}
\end{equation} 
with $\int_0^\infty C_{ini}(t) dt \leq N(0)$ bounded by initial biomass. \\

\textbf{Proof.} Recall our definition about biomass units (see Definition 1.4), which are indivisible and tractable. The discretization on microscopic scale allows us to adopt the "Lagrangian view" on reaction network. Since each biomass unit can be tracked, we can record its  transition time of every reaction event. This is called the \textit{history} of a biomass unit on the reaction network. For example, in a reaction pathway $\pi: u_0 \; u_1 \cdots \; u_L \; u_{L+1}$, we use the time sequence $\{t_0 \leq t_1<\cdots<t_L \leq t_{L+1}\}$ to represent a biomass unit started at node $u_0$ at time $t_0$, transferred along $\pi$ and arrived $u_k$ at time $t_k$, $k=1,\cdots,L+1$ (see figure below). \\

\begin{center}
\includegraphics[scale=0.4]{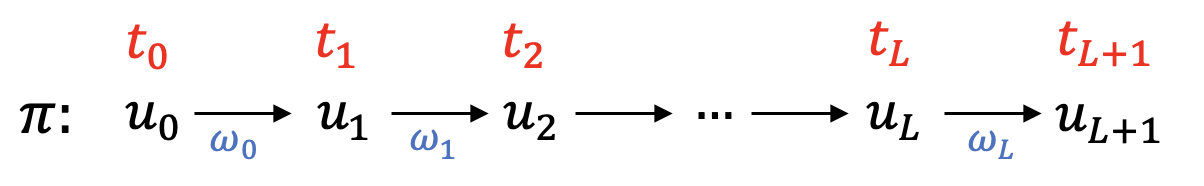} \\
\end{center}

It is understood that if the biomass arrives $x_j$ at time $t_j$, then the corresponded reaction event $\omega_{j-1}$ also happens at time $t_j$. In the following, we denote $t':=t_1$ and $t:=t_{L+1}$. Consider a biomass unit arrived gatekeeper node $x_g \in \mathcal{G}$ at time $t$. We could classify the history of this biomass unit into one of the three categories:\\
\begin{center}
\includegraphics[scale=0.4]{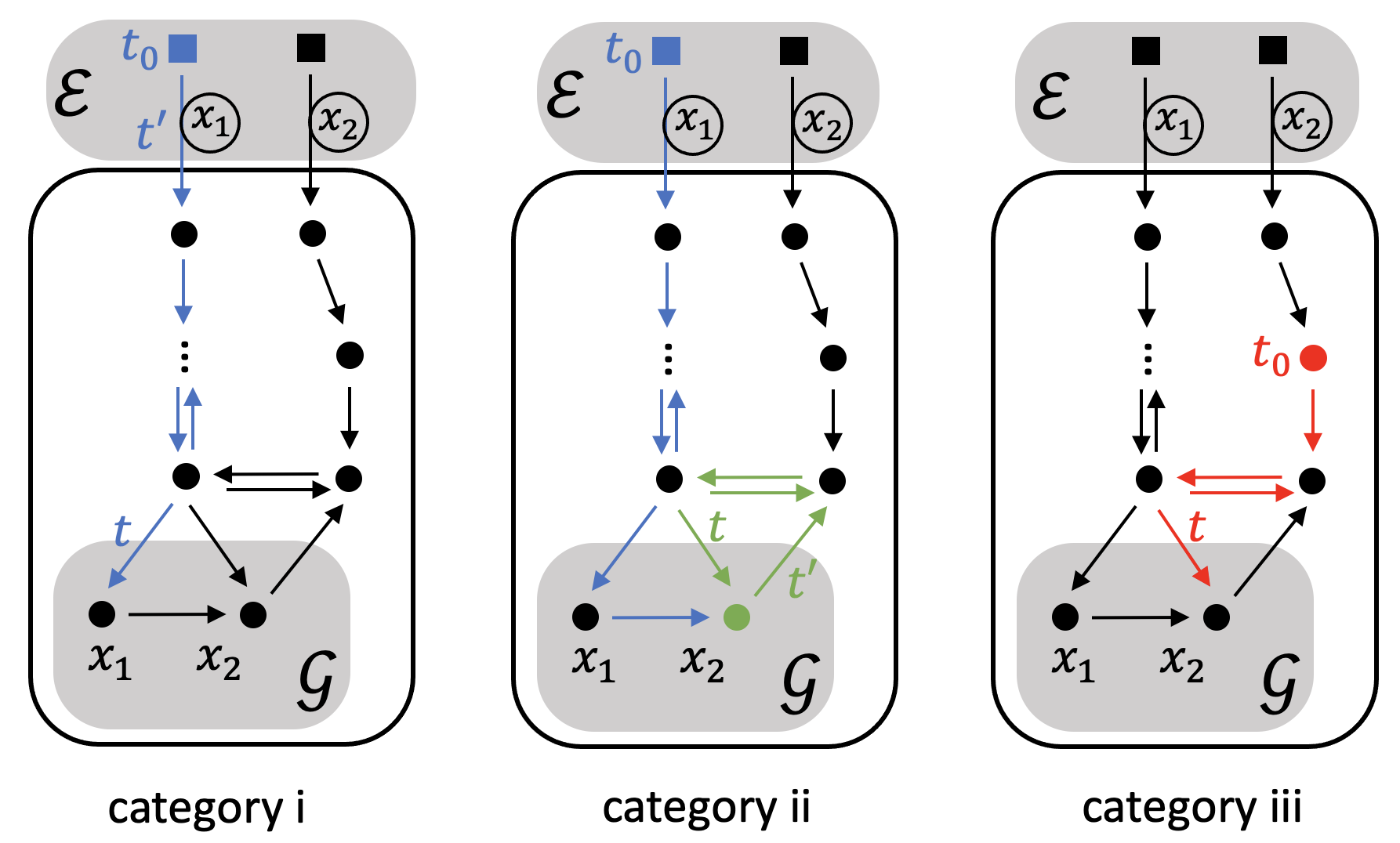} \\
\end{center}

\begin{enumerate}[label=(\roman*)]
	\item The biomass unit was located in the environment at time $t_0$, and never reaches $\mathcal{G}$ in $[t_0,t]$ (blue path in the figure above). In this case, let $t'<t$ denote the time biomass unit entered the system. \\
	\item The biomass unit was located in the environment at time $t_0$, and has reached $\mathcal{G}$ in $[t_0,t]$ (green path in the figure above). In this case, let $t'<t$ denote latest time for the biomass unit to leave $\mathcal{G}$. \\
	\item The biomass unit was located in the system at time $t_0$ (red path in the figure above). \\
\end{enumerate}
	
Since three categories are mutually exclusive, one biomass unit can only belongs to one of them. Hence $m_{\mathcal{G}}^+(t)$ (see Definition 2.4) is the summation of contributions from each category. Notice that for category (i) each history corresponds to one reaction pathway $\pi \in \mathcal{F}(\mathcal{E},\mathcal{G})$, and each history has one $t' \in [0,t]$. Therefore, the total contribution of category (i) can be expressed by summing all relevant pathways and integrating all $t'$, that is,  
\begin{equation}
	\sum_{\pi\in \mathcal{F}(\mathcal{E},\mathcal{G})} 
	\int_{0}^t m_{\pi}(t',t)dt'.
\end{equation}\

Now, consider the contribution from category (ii), it is similar that each history corresponds to reaction pathway $\pi \in \mathcal{F}(\mathcal{G},\mathcal{G})$, and each history has one $t' \in [0,t]$ as the \textit{latest exiting time} from $\mathcal{G}$. Similarly, the total contribution of category (i) can be expressed by summing all relevant pathways and integrating all $t'$, that is,  
\begin{equation}
	\sum_{\pi\in \mathcal{F}(\mathcal{G},\mathcal{G})} 
	\int_{0}^t m_{\pi}(t',t)\,dt'.
\end{equation}


Finally, we consider the contribution from category (iii) and characterize $C_{ini}(t)$. In Lagrangian view, this is the first hitting time distribution of biomass unit from non-gatekeeper nodes to gatekeeper nodes. The total amount of biomass at time $t=0$ is $N(0)$, and these biomass could eventually exit the system or hit gatekeeper nodes at time $t \in [0,\infty)$. Since biomass is conserved, the integral satisfies $\int_0^\infty C_{ini}(t)\,dt \leq N(0)$. \\

We also need to find the expression for $m_{\mathcal{G}}^-(t)$. This is the sum of effluxes from gatekeeper nodes to non-gatekeeper nodes at time $t$. Note that one reaction $\phi_c$ can have multiple downstream nodes, while some of them are in $\mathcal{G}$ while the others are not. The fraction of biomass via $\phi_c$ being transferred into \textit{non-gatekeeper nodes} is $1 - \xi[\mathcal{G}, \phi_c]$ (see Definition 1.5). This allows us to express $m_{\mathcal{G}}^-(t)$ by 
\begin{equation}
	m_{\mathcal{G}}^-(t)
	= (-1)
	\sum_{x_g \in \mathcal{G}} \,
	\sum_{\phi_c \in out(x_g)} 
	(1-\xi[\mathcal{G}, \phi_c])\,
	S[x_g,\phi_c] \, J[\phi_c](t) \geq 0.
\end{equation}
Note that there is a $(-1)$ term in our definition since $S[x_g, \phi_c]<0$ in this cases. $\blacksquare$ \\

\vspace{10pt}

Lemma 2.5 decomposes $\frac{dZ}{dt}$ into expression of $m_{\mathcal{G}}^+$ and $m_{\mathcal{G}}^-$. Our next step is to express $m_{\mathcal{G}}^+$ and $m_{\mathcal{G}}^-$ in terms of $Z(t)$ and $Z(t-\tau)$ for large $t$. This can be shown explicitly from analyzing each reaction pathway. In below, we denote $S[u_k, \omega_a]$ as the stoichiometry coefficient that correspond to node $u_k$ and reaction $\omega_a$. \\




\textbf{Definition 2.6 (Amplification rate).} Given a reaction pathway $\pi: u_0 \omega_0 u_1 \dots \omega_L u_{L+1}$, $\pi \in \mathcal{F}(\mathcal{E},\mathcal{G}) \cup \mathcal{F}(\mathcal{G},\mathcal{G}) $, we define $\kappa_\pi(Y(t'))$,  the \textit{amplification rate} at time $t'$ by

\begin{equation}
\kappa_{\pi}(t') := 
\frac{ \text{mass flux arrived $u_1$}}
{ \text{gatekeeper mass } }
=  \frac{S[u_1, \omega_0]\,J[\omega_0](t')}{Z(t')}.
\end{equation} \

Note that the unit of $\kappa_{\pi}$ is [1/time]; conceptually, it  represent the "mass flux facilitated by per unit of gatekeeper biomass". \\


\begin{center}
	\includegraphics[scale=0.4]{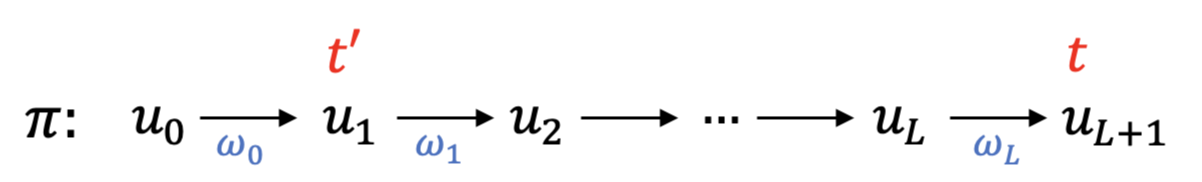} \\
\end{center}



\textbf{Definition 2.7 (Arrival function).} Consider a group of biomass units transferred along a reaction pathway $\pi$ (see the figure above), where the biomass unit arrived $u_1$ and $u_{L+1}$ at time $t'$ and $t$, respectively. Assume the number of biomass unit are large such that we can use a continuous function to describe the statistics. We define the \textit{arrival function} $H_{\pi}$ as 
\begin{equation}
\begin{aligned}
&H_{\pi}(t',t) :=
\frac{
\text{mass flux arrived $u_1$ at time $t'$, transferred via $\pi$, and arrived $u_{L+1}$ at time $t$} }
{\text{mass flux arrived $u_1$ at time $t'$ } }.
\end{aligned}
\end{equation}
Note that the unit of $H(t',t)$ follows $\frac{\text{[mass][time]}^{-2}} {\text{[mass][time]}^{-1}}=\frac{1}{\text{[time]}}$. \\

\vspace{10 pt}



\textbf{Proposition 2.8.} Given an LRN, consider a reaction pathway $\pi: u_0\,\omega_0\, u_1\cdots u_{L+1}$ with $\pi \in \mathcal{F}(\mathcal{E},\mathcal{G}) \;\cup\; \mathcal{F}(\mathcal{G},\mathcal{G})$. We have
\begin{equation} \label{E2.9}
m_{\pi}(t',t) 
= \kappa_{\pi}(t') \,H_{\pi}(t',t) \,Z(t').
\end{equation}


\textbf{Proof.} From Definition 2.4, we have 
\begin{equation}
\begin{aligned}
m_{\pi}(t',t) = \{
\text{mass flux arrived $u_1$ at time $t'$, transferred via $\pi$, and arrived $u_{L+1}$ at time $t$}\}.
\end{aligned}
\end{equation}

The formula in (\ref{E2.9}) is a direct consequence of Definition 2.6 and 2.7. $\;\blacksquare$ \\ 


In the definition below, we summarized the results of arrival functions and first-hitting pathways and construct DDEs for linear reaction networks. \\


\textbf{Definition 2.9.} Consider a reaction network with satisfying assumptions in Definition 1.4, and assume all boundary reactions has a gatekeeper. Denote $\tau = t-t'$, we define the \textit{catalytic kernel} $\alpha^{\bigstar}$ by
\begin{equation} 
\alpha^{\bigstar}(\tau; t) :=
\sum_{\pi \in \mathcal{F}(\mathcal{E}, \mathcal{G}) 
\cup  \mathcal{F}(\mathcal{G}, \mathcal{G})}
\kappa_{\pi}(t-\tau)\,
H_{\pi}(t-\tau,t),  
\quad t \geq \tau,
\end{equation} 
and define $\alpha^{\bigstar}(\tau; t) :=0$ otherwise. We define  the \textit{gatekeeper degradation rate} $\beta^{\bigstar}$ by
\begin{equation}
\beta^{\bigstar}(t) := (-1)
\sum_{x_g \in \mathcal{G}} \;
\sum_{\phi_c \in out(x_g)}
(1-\xi[\mathcal{G}, \phi_c])\;
S[x_g,\phi_c]\;
\frac{J_c(X(t))}{Z(t)}.
\end{equation}

The units of the two quantities follows $\alpha^{\bigstar}(\tau,t) \sim [time]^{-2}$ and $\beta^{\bigstar} \sim [time]^{-1}$. In summary, by assuming the existence of gatekeeper nodes for every boundary reactions, the system with positive growth rate must also have positive growth rate of gatekeeper biomass. Since gatekeeper biomass can be backtracked from the environment or from the recycling pathways (which correspond to $\mathcal{F}(\mathcal{E}, \mathcal{G})$ and $\mathcal{F}(\mathcal{G}, \mathcal{G})$, respectively), where $\tau$ represent the delay of biomass transfer. In this way, we can express $\frac{dZ}{dt}$ as a DDE. \\

\begin{center}
\includegraphics[scale=0.5]{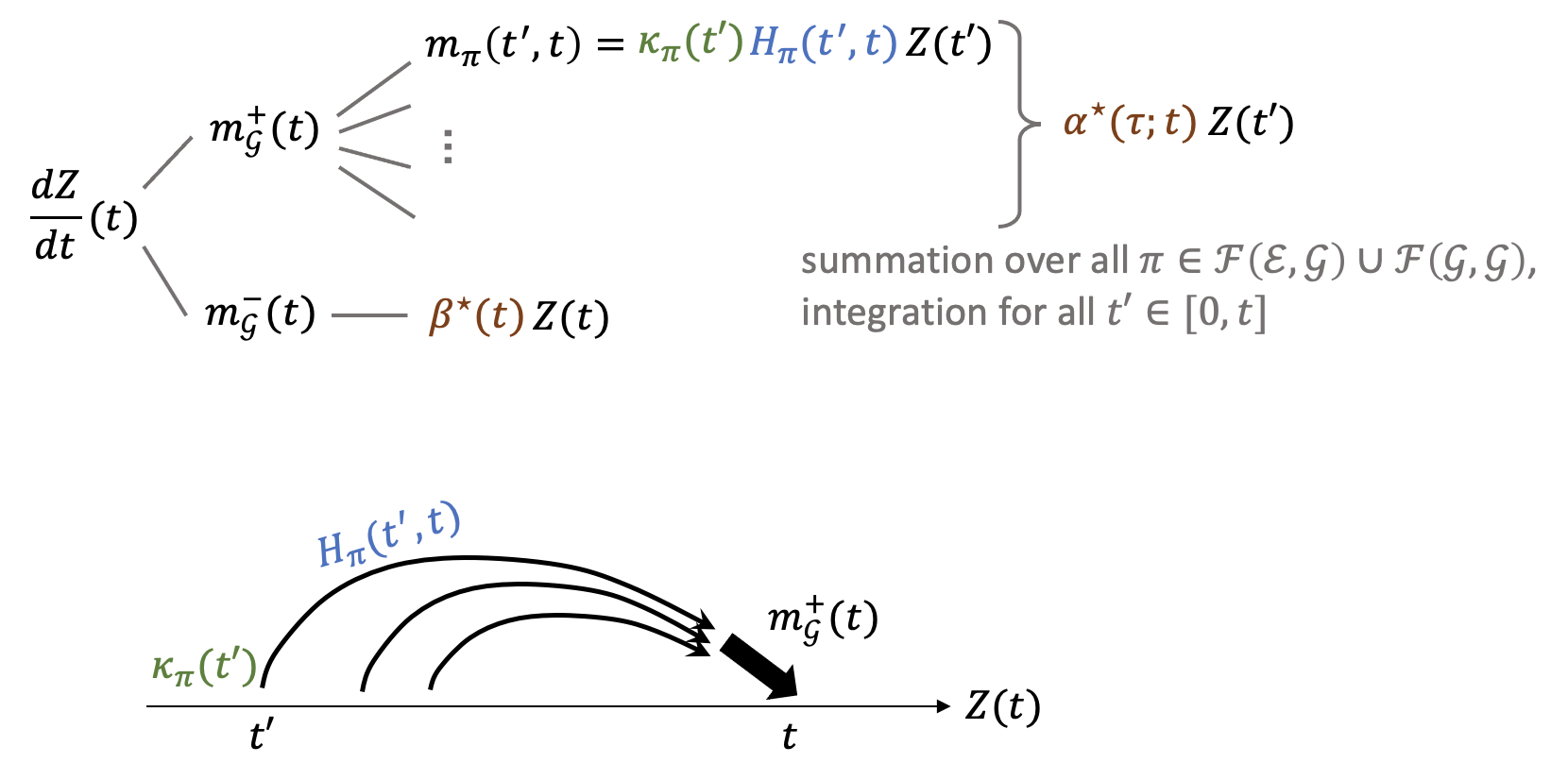} \\
\end{center}

\vspace{10 pt}

%
%
%


\textbf{Lemma 2.10.} Using the assumptions in Definition 2.9,  the gatekeeper biomass $Z(t)$ follows the equation   
\begin{equation} \label{E4.8a}
\frac{dZ}{dt}(t) = -\beta^{\bigstar}(t)\,Z(t) + 
\int_0^t \alpha^{\bigstar}(\tau; t)
\,Z(t-\tau)\,d\tau + C_{ini}(t),
\end{equation}
with $\int_0^\infty C_{ini}(t) \,dt \leq N(0)$ bounded by initial biomass. \\ 

\textbf{Proof.} We combine Lemma 2.5 and Proposition 2.8, and   replaces $t'$ with $t-\tau$. $\;\blacksquare$ \\ 


The equation (\ref{E4.8a}) is still not a typical DDE in the sense that $\alpha^{\bigstar}(\tau; t)$ depends not only on $\tau$ but also a second variable $t$. Also, we have not fully characterize the contribution of $C_{ini}(t)$. These uncertainties will be pinned down after imposing stronger conditions for the flux function $J(X)$. As we will shown later, for linear or scalable reaction networks, by their long-term property and phase average, equation (\ref{E4.8a}) can lead to an effective DDE. \\

In the following, we analyze the arrival function $H(t',t)$ and catalytic kernel $\alpha^{\bigstar}(\tau; t)$ in detail, which will be used in latter sections. The following results are derived from the  transition probability in Proposition 1.6. \\


\textbf{Proposition 2.11.} Consider a node $x_j$ in a reaction network with a unique efflux reaction $\phi_a$. Define the first hitting transition probability as
\begin{equation}
\begin{aligned}
	P(t_0,t_1) := \{
	&\text {A biomass unit located on $x_j$ at time $t_0$}, \\ 
	&\text {and arrived $x_k$ at time $t_1>t_0$ for the first time} \}.
\end{aligned}
\end{equation}
Define $r(t) := |S_{ja}|J[\phi_a]/X_j$, and assume $r(t)$ is bounded for all $t \geq 0$, we have 
\begin{equation}
	P(t_0,t_1) = r(t_1) \exp\bigg( 
	-\int_{t_0}^{t_1} r(t')\,dt' 
	\bigg ).
\end{equation}	

\textbf{Proof.} For a biomass unit on $x_j$ at time $t_1$, the probability for the biomass to react via $\phi_a$ at time $[t_1, t_1+\delta t]$ is $r(t_1)\delta t$. Also, the survival probability for the biomass \textit{not} react with $\phi_a$ until time $t_1$ can be approximated by spitting $[t_0,t_1]$ into small time intervals with length $h$ (where $(t_1-t_0)/h := N_h$ as integers), which gives
\begin{equation}
\lim_{h\rightarrow 0} \prod_{k=0}^{N_h}
\bigg( 1-  r(kh) \cdot h\bigg)
=\lim_{h\rightarrow 0} \prod_{k=0}^{N_h}
e^{ - r(kh) \cdot h}
=\lim_{h\rightarrow0} \exp\bigg(
\sum_{k=0}^{N_h} r(kh) \cdot h
\bigg)
= \exp\bigg(-\int_{t_0}^{t_1} r(t')dt'\bigg).
\end{equation}	
$\;\blacksquare$ \\


\textbf{Proposition 2.12.} Consider a node $x_j$ in a reaction network with several efflux reactions, and one of the efflux $\phi_a$ has downstream $x_k$. Define the first-hitting transition probability 
\begin{equation}
\begin{aligned}
	P(t_0,t_1|\,x_j,\phi_a,x_k) := \{
	&\text {A biomass unit located on $x_j$ at time $t_0$}, \\ 
	&\text {transferred via reaction $\phi_a$, } \\
	&\text {and arrived $x_k$ at time $t_1$ for the first time}. \}
\end{aligned}
\end{equation}
Assume the total transition rate of $out(x_j)$ is bounded for all $t \geq 0$, denoted by 
\begin{equation}
	r_{tot}(t) := 
	\frac{1}{X_j}\sum_{\phi_c \in out{(x_j)} } |S_{jc}|\, J_c(t),
\end{equation}
and also define
\begin{equation}
	r_{ja}(t) := 
	\frac{1}{X_j} |S_{ja}|\, J_a(t),
\end{equation}
then we have
\begin{equation}
	P(t_0,t_1|\,x_j,\phi_a,x_k) 
    = 
	r_{ja}(t_1)\,\exp\bigg( 
	-\int_{t_0}^{t_1} r_{tot}(t') \,dt' 
	\bigg )
     \times\, \xi[x_j,\phi_a],
\end{equation}
where $\xi[x_j,\phi_a] $ is the downstream fraction defined in Definition 1.5. \\

\textbf{Proof.} When there are multiple downstream, the transition probability that biomass enters a specific downstream node $x_k$ is a conditional probability. The conditional waiting time is the calculated by total transition rate $r_{tot}(t)$ (similar to the derivation in the Gillespie algorithm), and the transition rate from $x_j$ to $\phi_a$ at time $t$ is $r_{ja}(t)$. By Definition 1.4 (iv), the biomass unit from upstream to downstream is equally distributed, and hence the downstream fraction follows $\xi[x_k, \phi_a]$ as in Proposition 1.5.  $\;\blacksquare$ \\


\textbf{Definition 2.13 (Bounded transition rates).} For a reaction network with flux functions $J_1(X),\cdots, J_m(X)$, we say the reaction network has \textit{bounded transition rates} if exist constants $M_1, M_2$ such that 
\begin{enumerate}[label=(\roman*)]
	\item If $x_k \in dw(\phi_a)$, then $M_1<\frac{J_a(X)}{X_k} < M_2$ for all $t$.  
	\item If $x_g$ is a gatekeeper for a boundary reaction $\phi_a$, then $M_1<\frac{J_a(X)}{X_g} < M_2$ for all $t$. 
\end{enumerate}\



The arrival function $H(t',t)$ can be regarded as conditional waiting time distribution for a random variable $T_{\pi}$, indicated in the figure below. Note that by definition, $t'$ is the time for biomass to \textit{arrive} $u_1$, and hence the sum $T_{\pi} = T_1+\cdots T_L$ does not include $T_0$. \\

\begin{center}
	\includegraphics[scale = 0.45]{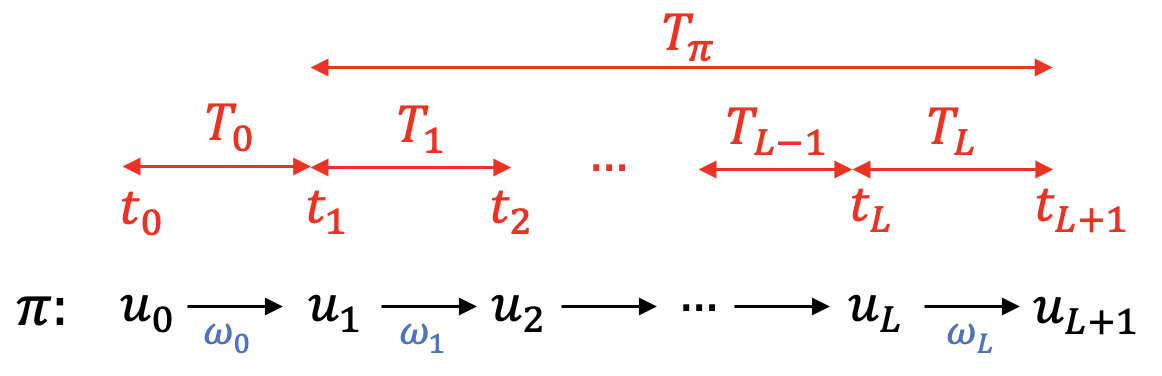}
\end{center}



\textbf{Proposition 2.14.} Consider a reaction network with bounded transition rates. For a reaction pathway 
\begin{equation}
\pi: u_0\, \omega_0\, u_1 \cdots u_L \, \omega_L \,u_{L+1} 
\end{equation}
Let $t_j \;(j=1,\cdots,L+1)$ be the time points where the biomass arrive $u_j$ and set $t':=t_1$ and $t:=t_{L+1}$. \\

If $\pi$ has only one reaction ($L=0$), i.e. $\pi: u_0 \omega_0 u_1$, then the arrival function is $H(t',t) = \delta (t-t')$ as a Dirac delta measure. \\

If $\pi$ has multiple reactions ($L \geq 1$), then the arrival function is given by
\begin{equation}
\begin{aligned}\
H_{\pi}(t',t) 
= \int_{t_1\leq t_2 \leq\cdots\leq t_L \leq t_{L+1}}
& P(t_1,t_2|\;u_1,\omega_1,u_2)\;P(t_2,t_3|\; u_2,\omega_2,u_3) \\
&\cdots P(t_L,t_{L+1}|\;u_L,\omega_L,u_{L+1})\; 
	dt_2\cdots dt_L.
\end{aligned}
\end{equation}

\textbf{Proof.} The existence of $H(t',t)$ is guaranteed by the assumption of bounded transition rate. The equation for $P_{\pi}(t',t)$ is a consequence of Chapman-Kolmogorov equation (\cite{klenke_probability_2013}). $\;\blacksquare$ \\


%
%
%

\newpage
\section{Biomass transfer on simple autocatalytic pathways (SAPs)}

We discussed simple autocatalytic pathways in the main text. Here, we state some of the results again for completeness. A simple autocatalytic pathways is defined as a reaction network with a linear topology and one influx reaction ($\phi_n$). We assume $\mathfrak{m}(x_j)=1$ for all nodes, and all flux functions are in form of $J[\phi_k](X) = c_kX_k, c_k>0$, as depicted in the figure below: 


\begin{center}
	\includegraphics[scale = 0.5]{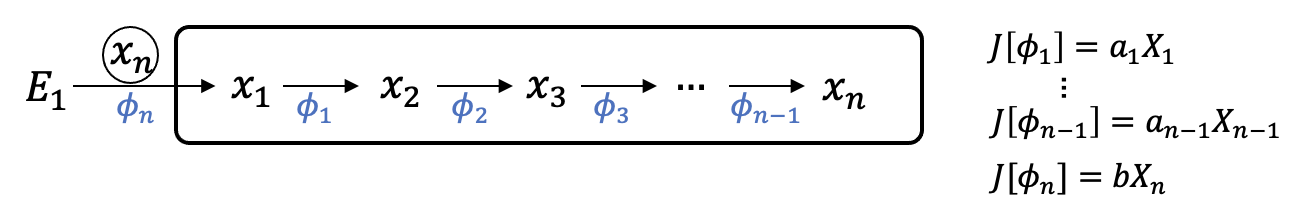}
\end{center}


The SAP contains $n$ nodes $\{x_1,...,x_n\}$ are called SAP of length $n$. By definition, the last node $x_n$ serves as the gatekeeper. The dynamics of an SAP follows linear ODEs $\frac{dX}{dt} = MX$, with
\begin{equation}
M=
\begin{pmatrix}
-a_1 & 0 & \cdots &\cdots & b  \\
a_1 & -a_2 & 0 &\cdots & 0 \\
0 & a_2 & \ddots &\ddots & \vdots \\
\vdots & \ddots & \ddots & -a_{n-1} & 0\\
0 & \cdots & 0 & a_{n-1} & 0\\
\end{pmatrix}.
\end{equation}\

Define $a_{\max} := \max{\{a_1,\cdots,a_{n-1}}\}$. Then the matrix $M+a_{\max}I$ is nonnegative and irreducible (note that the topology of simple reaction pathway makes the directed graph of $M$ strongly connected). We say an eigenvalue is \textit{dominant} if it has the largest real part among all eigenvalues. By the Perron--Frobenius theorem (see Theorem 8.4.4 in \cite{horn_matrix_2012}) the dominant eigenvalue of $M+a_{\max}I$ (and hence $M$) is real with algebraic multiplicity one, and the dominant eigenvector of $M+a_{\max}I$ (and hence $M$) is unique (up to a multiplication scalar) and positive. \\

\textbf{Proposition 3.1.} For an SAP, the dominant eigenvalue of $M$ is equal to the long-term growth rate $\lambda$, for all nonnegative initial conditions. The rescaled system $Y(t)$ converges to a unique eigenvector $Y^*$ of $M$ as $t\rightarrow\infty$. $\blacksquare$ \\

\textbf{Proof.} The properties of simple autocatalytic pathway can be shown from a more general results of LRNs (see Corollary 4.9). $\blacksquare$ \\


An SAP can be regarded as a reaction pathway on its own. The following proposition characterize the amplification rate and arrival function (as in Definition 2.6 and 2.7) of SAPs. In the following, we denote $f\ast g$ the convolution between $f$ and $g$, and $\mathcal{L}[f]$ as the Laplace transformation of $f$. \\

\textbf{Proposition 3.2.} For an SAP with $\frac{dX}{dt} = MX$, we have the amplification rate $\kappa = b$. For $n=1$, we have $h(\tau)=\delta(\tau)$, the delta measure. For $n>1$, we have
\begin{equation}
\begin{aligned}
	h(\tau) &:= h_1(\tau)\ast \dots \ast h_{n-1}(\tau), \\
	h_k(\tau) &:= a_k e^{-a_k\tau}, \qquad k=1,\cdots,n-1. \\
\end{aligned}
\end{equation}
The \textit{catalytic spectrum} is defined by $\tilde{\alpha}(s) := \mathcal{L}[\kappa h(\tau)]$. For $n=1$, $\tilde{\alpha}(s)=b$. For $n>1$, 
\begin{equation}
	\tilde{\alpha}(s) = b \; \prod_{j=1}^{n-1} \frac{a_j}{a_j+s}.
\end{equation}
Furthermore, the long-term growth rate $\lambda$ satisfies $\lambda = \tilde{\alpha}(\lambda)$. \\

\textbf{Proof.} The result is derived in the main text. The formula $\tilde{\alpha}(\lambda)=\lambda$ is a special case of Theorem B (see Section 5), here we derive this by simple calculation. Note that the characteristic polynomial of $M$ is 
\begin{equation}
    p_M(t)=(-1)^{n}[t(t-a_1)\dots(t-a_{n-1})-ba_1\dots a_{n-1}].
\end{equation}
Since $\lambda$ is the principal eigenvalue of $M$, $p_M(\lambda)=0$ and this implies $\tilde{\alpha}(\lambda)=\lambda$. While all eigenvalues of $M$ satisfies $\tilde{\alpha}(s)=s$, $\lambda$ is the only positive real solution for this relation since $\tilde{\alpha}(s)$ is non-increasing for $s>0$ and can only intersect with $y=s$ at a single point. $\;\;\blacksquare$ \\

\textbf{Note.} For SAPs, the system has one influx and no efflux, and hence we always have positive growth rate, i.e. $\mu(t)=\frac{1}{N}\frac{dN}{dt} > 0$. The long-term growth rate is time average of $\mu(t)$, which implies $\lambda > 0$. For SAP with length $L=2$, the long-term growth rate can be solved by $\frac{ba_1}{a_1+\lambda} = \lambda$ and we have 
\begin{equation}
	\lambda = \frac{a_1}{2} 
	\bigg(
	\sqrt{1+\frac{4b}{a_1}} - 1
	\bigg).
\end{equation} 

For SAP with length $L\geq 3$, the algebraic solution is either complicated or non-existed, and we may need to analyze $\lambda$ with other methods. One way to estimate $\lambda$ is to compare their catalytic spectra $\tilde{\alpha}(s)$. To simplify the notation, we denote $A(s):=\tilde{\alpha}(s)$ and use two notations interchangeably. We have the following proposition: \\


\textbf{Proposition 3.3.} Consider two SAPs with possibly different lengths and coefficients. Let $A_1(s),A_2(s)$ be their catalytic spectra and $\lambda_1, \lambda_2$ be their growth rates. Then, 

\begin{itemize}
	\item $A_1(\lambda_2) \geq A_2(\lambda_2)$ implies $\lambda_1 \geq \lambda_2$.
	\item $A_1(\lambda_1) \leq A_2(\lambda_1)$ implies $\lambda_1 \leq \lambda_2$.
\end{itemize}
\begin{center}
	\includegraphics[scale = 0.34]{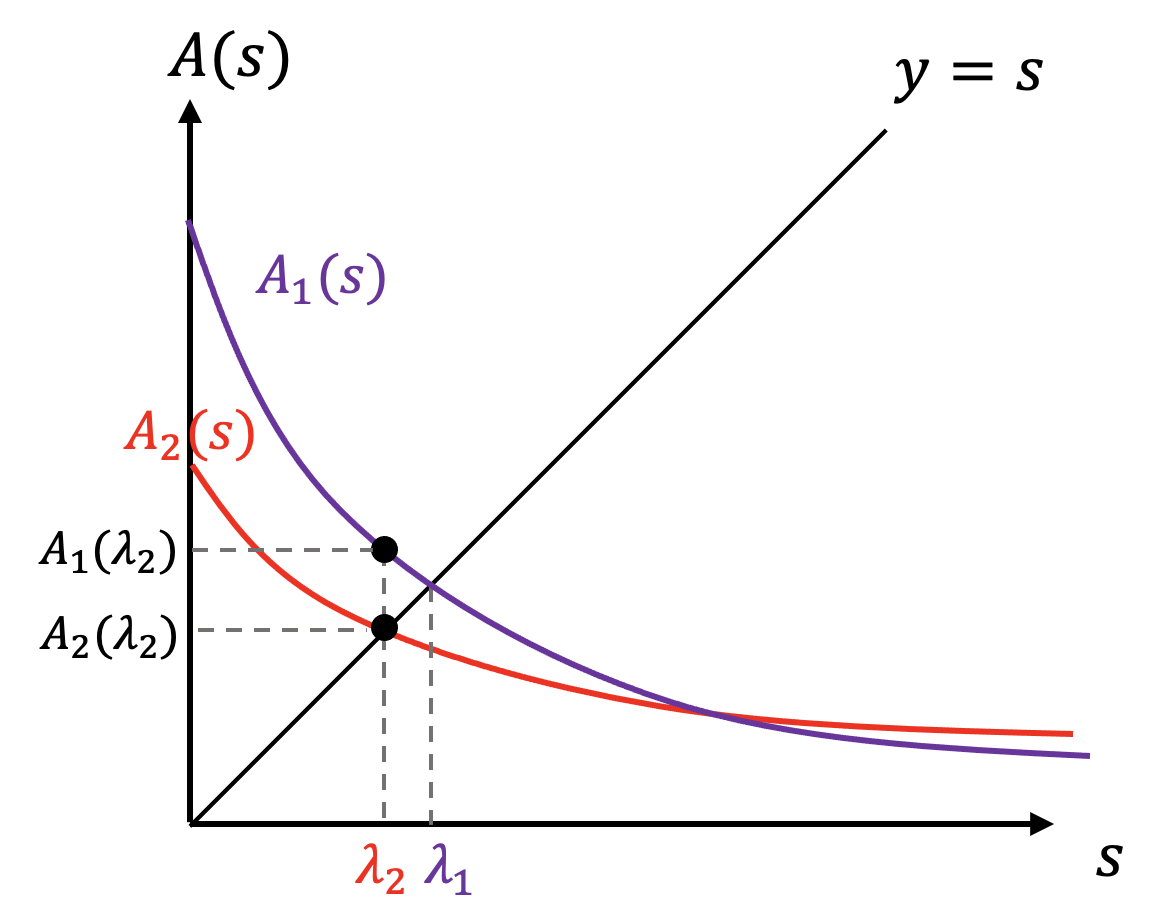}
\end{center}
\textbf{Proof.} Note that the inequalities in Proposition 3.3 has a geometric interpretation (see the figure in above). For a rigorous proof, first we assume $A_1(\lambda_2) \geq A_2(\lambda_2)$ and show that $\lambda_1 \geq \lambda_2$. Suppose the contrary that $\lambda_1 < \lambda_2$. Since $A_1(s)$ is deceasing function for $s\geq 0$, by assuming $\lambda_1 < \lambda_2$ we have
\begin{equation} \label{3-3B}
	A_1(\lambda_1) > A_1(\lambda_2). 
\end{equation}
From the Proposition 3.2 we have 
\begin{equation} \label{3-3C}
	A_1(\lambda_1) = \lambda_1, \qquad
	A_2(\lambda_2) = \lambda_2. 
\end{equation}
By combining (\ref{3-3B}), (\ref{3-3C}) with the assumption $A_1(\lambda_2) \geq A_2(\lambda_2)$, we have 
\begin{equation}
	\lambda_1 = A_1(\lambda_1) \geq A_1(\lambda_2) 
	\geq A_2(\lambda_2) = \lambda_2,
\end{equation}
which contradicts to $\lambda_1 < \lambda_1$. Hence we must have  $\lambda_1 \geq \lambda_2$. This proves the first statement. Now, interchange $A_1,A_2$ and $\lambda_1,\lambda_2$ proves the second statement. $\;\blacksquare$ \\ 

 In general, we could use simpler SAPs with equivalent coefficients to approximate general SAPs. This is illustrated by the next Proposition. \\


\textbf{Proposition 3.4.} Consider an SAP of length $n$ with catalytic spectrum 
\begin{equation}
A(s) = b \; \prod_{j=1}^{n-1} \frac{a_j}{a_j+s},
\end{equation}
the growth rate $\lambda$ of this SAP can be bounded by the growth rates of two SAPs with length 2. Specifically, consider two SAPs having growth rate $\lambda_U, \lambda_L$ with catalytic spectra
\begin{equation}
A_U(s):= \frac{ba_U}{a_U+s}, \qquad 
A_L(s):= \frac{ba_L}{a_L+s}.
\end{equation}  
Choose 
\begin{equation}
\begin{aligned}
	& a_U^{-1} := a_1^{-1}+\cdots+a_{n-1}^{-1}, \\
	& a_L^{-1} := (A(b)-1)/b.
\end{aligned}
\end{equation}
Then we have $\lambda_L \leq \lambda \leq \lambda_U$. \\ 

\textbf{Proof}. First we show $\lambda_U \geq \lambda$. Our approach is to establish inequalities about $A(s)$. First, we have
\begin{equation} \label{2-3A}
\begin{aligned}
\frac{b}{A(s)} 
& = \bigg(1+\frac{s}{a_1}\bigg)\cdots\bigg(1+\frac{s}{a_{n-1}}\bigg) \\
& \geq 1 + s\; \bigg(\frac{1}{a_1}+\cdots+\frac{1}{a_{n-1}}\bigg) \\
& = 1+\frac{s}{a_U} \\
& = \frac{b}{A_U(s)}.
\end{aligned}
\end{equation}
Hence, we have $A_U(s) \geq A(s)$ for all $s\geq 0$ and this implies $A_U(\lambda) \geq A(\lambda)$ (since for SAP we have $\lambda>0$). By Proposition 3.3, we have $\lambda_U > \lambda$. \\ 

Next we show $\lambda \geq \lambda_L$ with $a_L$ defined as $a_L^{-1} :=(A(b)-1)/b$. First we express $A(\lambda)$ as the following form:
\begin{equation}
\begin{aligned}
\frac{b}{A(\lambda)} 
& = \bigg(1+\frac{\lambda}{a_1}\bigg)\cdots\bigg(1+\frac{\lambda}{a_{n-1}}\bigg) 
  =: 1+\lambda P(\lambda),
\end{aligned}
\end{equation}
where $P(\lambda) := \frac{1}{\lambda} \,
\bigg( (1+\frac{\lambda}{a_1})\cdots(1+\frac{\lambda}{a_{n-1}}) - 1 
\bigg)$ is a polynomial with nonnegative coefficients. Using the inequality $A(s) \leq b$ for all $s \geq 0$, we have  $\lambda =A(\lambda) \leq b$. Since the coefficients in $P(\lambda)$ are nonnegative, we have $P(\lambda) \leq P(b)$. Hence,
\begin{equation}
    \frac{b}{A(\lambda)} \leq 1+\lambda P(b).
\end{equation}
Now, we define $a_L^{-1} := P(b) = (A(b)-1)/b$. This gives 
\begin{equation}
\frac{b}{A(\lambda)} \leq 1+\frac{\lambda}{a_L}.
\end{equation}
Note that we have $A_L(\lambda)=\frac{ba_L}{a_L+\lambda}$ for an SAP with length 2. Rearranging the terms yields the inequality 
\begin{equation}
\frac{b}{A(\lambda)} \leq 1+\frac{\lambda}{a_L} \leq \frac{b}{A_L(\lambda)}.
\end{equation}
This implies $A(\lambda) \geq A(\lambda_L) $. By Proposition 3.3, we have $\lambda \geq \lambda_L$.  $\blacksquare$ \\ 

The definition of upper bound coefficient $a_U$ is the harmonic mean of $a_j$'s. Equivalently, this is using the sum of mean waiting time of each step ($\mathbb{E}[T_j] = 1/a_j$) as a coarse-grained waiting time, and define $a_U^{-1} := \sum_j \mathbb{E}[T_j]$. This estimation is exact if $\lambda/a_j \rightarrow 0$, and there is a mathematical connection with equivalent resistance of resisters in series. However, the current flux is at steady state and hence $\lambda=0$. For $\lambda > 0$, this formula gives an over-estimation. \\


\textbf{Example 3.5.} Consider an SAP with $(a_1,a_2,a_3,a_4,a_5) =(0.8, 0.4, 0.5, 0.6, 0.9)$ and $b=0.2$. The coarse-grained  coefficients are $a_U \approx 0.117, \, a_L \approx 0.061$. The exact growth rate, its upper bounds and its lower bound are $\lambda = 0.0949, \,\lambda_U = 0.1054$ and $\lambda_L = 0.0841$. \\
\begin{center}
	\includegraphics[scale = 0.55]{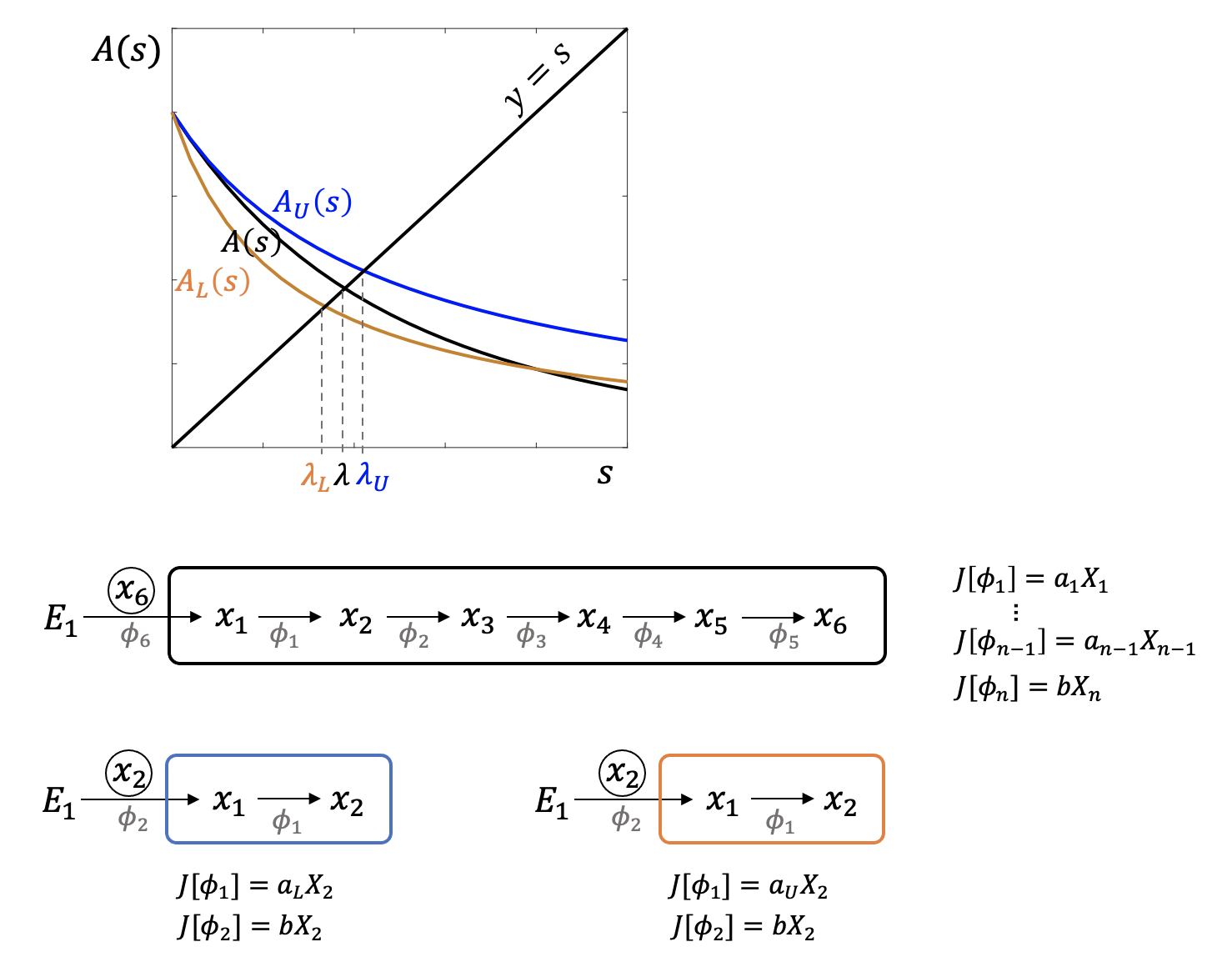}
\end{center}


%
%
%


\newpage
\section{Long-term property of linear reaction networks (LRNs)}

In this section, we briefly review our definition of linear reaction networks (LRNs) and prove Theorem A. These proofs are rather technical and only use basic matrix analysis. Readers may skip this section at the first time when reading this manuscript. \\ 


\textbf{Definition 4.1}: An LRN is a reaction network satisfying the following \textit{LRN conditions}: 

\begin{enumerate}[label=(\roman*)]
	\item Each reaction $\phi_a$ has exactly one upstream node, denoted as $up(\phi_a)$. 
	\item If $up(\phi_a)=x_k$ is a system node, then its flux function is $J_a = R_a X_k$. 
	\item If $up(\phi_a)=E_k$ is an environmental node, then the flux function follows $J_a = R_a X_g$ for some node $x_g$ in the system. In this case, we say $x_g$ \textit{gatekeeps} $\phi_a$.
\end{enumerate}

The dynamics of an LRN can be represented in matrix form. Let $\mathcal{M}^{n,m}$ denote the collection of $n$-by-$m$ matrices. Assume the LRN has $n$ nodes and $m$ reactions. Then, for the stoichiometry matrix $S$ belongs to $\mathcal{M}^{n,m}$. Define $P\in \mathcal{M}^{m,n}$ by
\begin{equation}
	P_{ak} = 
	\begin{cases}
		R_a,  & \text{ if } J_a(X) = R_a X_k, \\
		0,    & \text{ otherwise. }
	\end{cases}
\end{equation}
Then we have $\frac{dX}{dt} = MX$ as linear ODEs with $M=SP\in \mathcal{M}^{n,n}$. \\

\textbf{Note.} One of the important properties of an LRN is that all of its off-diagonal entries are nonnegative. This is due to LRN condition (ii) which requires all effluxes from $x_k$ must be function in form of $R_a X_k$, which implies all negative coefficients can only appear in diagonal entries of $M$. Matrices with this property are called \textit{essentially nonnegative}. For linear ODEs with essentially nonnegative matrices, the nonnegative orthant $\mathbb{R}^n_{\geq0}$ is forward-invariant in time, that is, if $X(0) \in \mathbb{R}^n_{\geq0}$, then $X(t) \in \mathbb{R}^n_{\geq0}$ for all $t>0$. This is a direct consequence of
\begin{equation}
	\lim_{X_k \rightarrow 0}
	\frac{dX_k}{dt}
	= \lim_{X_k \rightarrow 0} \sum_{p\neq k} M_{kp}X_p
	+ \lim_{X_k \rightarrow 0} M_{kk} X_k
	\geq 0,
\end{equation}
for all $X_k$, and hence the trajectory $X(t) \in \mathbb{R}^n_{\geq0}$ cannot leave $\mathbb{R}^n_{\geq0}$ in finite time. Our first main result is about the long-term behavior of the rescaled trajectory $Y(t) = X(t)/N(t)$ (see Definition 1.7). \\


\textbf{Theorem A.} Consider an LRN $\frac{dX}{dt} = MX$ with an  initial condition $X(0) \in \mathbb{R}^n_{\geq0}$. The long-term growth rate $\lambda$ equals a real eigenvalue of $M$. Furthermore, the rescaled trajectory $Y(t)$ converges to an eigenvector $Y^\#$ affiliated with $\lambda$. $\blacksquare$ \\


We will need several propositions for proving Theorem A. Note that the long-term growth rate may depend on initial conditions, as shown in the example below. \\

\textbf{Example.} Consider the following LRN with rate constants $b_k$ labeled: \\
\begin{center}
	\includegraphics[scale = 0.45]{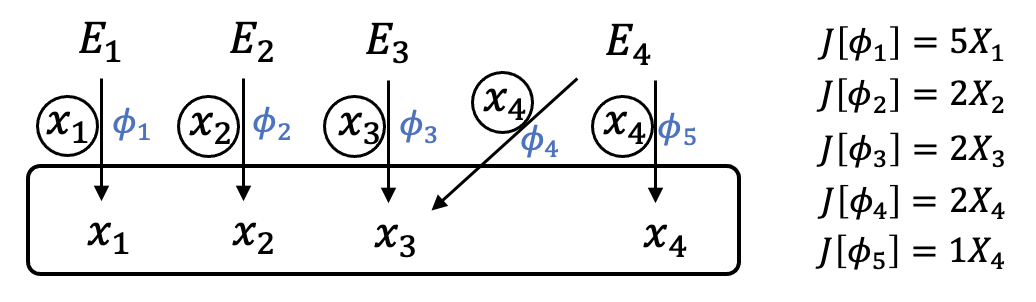}
\end{center}

Suppose $X(0) = (c_1, c_2, c_3, c_4)^T$. The matrix $M$ and solution $X(t)=e^{Mt}X(0)$ are 
\begin{equation}
M=
\begin{pmatrix}
5 & 0 & 0 & 0  \\
0 & 2 & 0 & 0 \\
0 & 0 & 2 & 1 \\
0 & 0 & 0 & 2\\
\end{pmatrix}, \;\;
e^{Mt}=
\begin{pmatrix}
e^{5t} & 0 & 0 & 0  \\
0 & e^{2t} & 0 & 0 \\
0 & 0 & e^{2t} & te^{2t} \\
0 & 0 & 0 & e^{2t}\\
\end{pmatrix}, \;\;
X(t) = 
\begin{pmatrix}
c_1e^{5t}\\
c_2e^{2t}\\
(c_3+tc_4)e^{2t}\\
c_4 e^{2t}
\end{pmatrix}.
\end{equation}

If $c_1>0$, then $\lambda = 5$ and $Y^\# = (1,0,0,0)^T$. However, if $X(0) = (0,c_2,c_3,c_4)$, then the system grows in the eigenspace corresponding to the eigenvalue 2. If $c_1=0$ and $c_4>0$, then $Y^\# = (0,0,1,0)^T$. If $c_1=c_4=0$, then $Y^\# = \frac{1}{c_2+c_3}(0, c_2, c_3, 0)^T$. In all cases, $Y^\#$ are eigenvector of $M$, but could belong to different eigenspaces. $\blacksquare$ \\

In general, we could express $M$ in Jordan form $M=PJP^{-1}$ with the solution $X(t)=Pe^{Jt}P^{-1} X(0)$. Assume $M$ has total $Q$ distinct eigenvalues $b_1,\cdots,b_Q$ ($Q \leq n$), the matrix $J$ is composed by $Q$ Jordan blocks. For a Jordan block with size $L$, it is represented by
\begin{equation}
B_q =
\begin{pmatrix}
b_q & 1 & 0 & \cdots & 0  \\
0 & b_q & 1 & &  \vdots \\
\vdots & 0 & \ddots & \ddots & 0 \\
& & \ddots & b_q & 1\\
0 & \cdots & & 0 & b_q\\
\end{pmatrix}
,\;\;\;\;
e^{B_qt}= e^{b_qt}
\begin{pmatrix}
1 & t & \frac{t^2}{2} & \cdots & \frac{t^{L-1}}{(L-1)!}  \\
0 & 1 & t & \ddots & \vdots \\
0 &  & \ddots & \ddots & \frac{t^2}{2}  \\
\vdots & & \ddots & 1 &t\\
0 & \cdots & & 0 & 1\\
\end{pmatrix}.
\end{equation}\

Denote these Jordan blocks as $B_1,\cdots,B_Q$ and the block diagonal form of $J$ as $\mathrm{diag}(B_1,\cdots,B_Q)$, the matrix $e^{Jt}$ has corresponded block diagonal form $\mathrm{diag}(e^{B_1t},\cdots,e^{B_Q t})$. Our next proposition is a sufficient condition for the long-term convergence in LRN.  \\

  

\textbf{Proposition 4.2.} Consider an LRN with $\frac{dX}{dt} = MX$ where $M \in \mathcal{M}^{n,n}$. Assume $f(Y)$ is a continuous function. Then, for every solution trajectory $X(t)$, the time average $\langle f\rangle_t$ exists. \\

\textbf{Proof.} For the LRN, the $k$-th component of the solution $X_k(t)$ can be expressed as 
\begin{equation}
	X_k(t) = \sum_{j=1}^n  p_{kj}(t)\,e^{\alpha_j t}\,
	\big(\cos(\theta_jt)+i \sin(\theta_j t)\,\big) \geq 0.
\end{equation}
where $\alpha_j + i \theta_j$ is the $j$-th eigenvalue of $M$, with $\alpha_j, \theta_j \in \mathbb{R}$, and $p_{kj}(t)$ are polynomials with degree less than $n$ (if $M$ is diagonalizable, then $p_{kj}(t)$ are constants). Consider the rescaled system $Y$, we have 
\begin{equation}
	Y_k(t) = \frac{
	\sum\limits_{j=1}^n \;
	p_{kj}(t)\,
	e^{\alpha_j t}\,
	\big(\cos(\theta_j t)+i \sin(\theta_j t)\big)}
	{\sum\limits_{k=1}^n \sum\limits_{j=1}^n \;
	p_{kj}(t)\, e^{\alpha_j t}\, 
	\big( \cos(\theta_j t)+i \sin(\theta_j t) \big)}.
\end{equation}

It can be seen that for large $t$, the attractor of $Y(t)$ is either a fixed point or a limit torus in $\Delta^{n-1}$. For a polynomial $p(t)$, we define $\mathrm{deg}(p(t))$ as the degree of $p(t)$. We can see that in the long-term the numerator terms in $Y_k$ that do not decay to zero are the terms with $\alpha_j = \max_{j'}\{\alpha_{j'}\}$ and $\mathrm{deg}(p_{kj}) = \max_{j',k'}\{\mathrm{deg}(p_{k'j'})\}$. In this way, $Y_k(t)$ converges to a quasi-periodic function $f_k(\theta_1,\cdots,\theta_n)$. If all eigenvalues are real, we have $f_k$ as a constant for all $k$ and the attractor of $Y$ to be a fixed point. \\

Since $\mathbb{R}^n_{\geq0}$ is invariant for LRNs, the solution  $X(t)$ is confined in the $\mathbb{R}^n_{\geq0}$ and the $Y(t)$ is confined in simplex $\Delta^{n-1}$. Since the attractor is a fixed point or a limit torus in $\Delta^{n-1}$, by ergodic theory (Section 3.1, Theorem 2 in \cite{cornfeld_ergodic_2012}) the long-term average $f(Y(t))$ converges. $\blacksquare$ \\

\textbf{Note.} The quantities $Y_k = X_k/N$ and $\mu(X) = \frac{1}{N}\frac{dN}{dt}$ are scale-invariant and hence can be expressed as $f(Y)$. From Proposition 3.2, these quantities in LRNs have converged time-averages. In particular, $\lambda = \langle \mu \rangle_t$ is the long-term growth rate. We say the component $Y_k$ is \textit{persistent in fraction} if $\langle Y_k \rangle_t > 0$. Finally, we should keep in mind that these averages may depend on the initial condition, which will be assume to be in the nonnegative orthant. \\



The matrix $M$ of an LRN is essentially nonnegative. The next two propositions utilize this property of LRNs.  \\ 

\textbf{Proposition 4.3.} Consider an LRN with $\frac{dX}{dt}=MX$. For every $k=1,\cdots,n$, if $\langle Y_k\rangle_t > 0$, then there exist a time point $t_k>0$ such that $Y_k(t) >0$ for all $t>t_k$. \\

\textbf{Proof.} Since $\langle Y_k\rangle_t > 0$, there must be a time point $t_k$ where $X_k(t_k)>0$. Consider a time point $t_h>t_k$ and assume the contrary that $X_k(t_h) = 0$. Divide $dX/dt$ by $X_k(t)$ and integrate from $t_k$ to $t_h$, we have
\begin{equation}
	\int_{t_k}^{t_h} 
\frac{1}{X_k(t)}\frac{dX_k}{dt}(t)
\;dt
=\log(X_k(t_h)) - \log(X_k(t_k)).
\end{equation}

Since $X_k({t_h})=0$, the above integral diverges to $-\infty$. However, we can also express the differential equation of $X_k$ as $ \frac{dX_k}{dt} = \sum_{j \neq k} M_{kj} X_j + M_{kk} X_k$. By the LRN propery, $M_{kj} \geq 0 $ for all $ k\neq j$. This gives 
\begin{equation}
\int_{t_k}^{t_h} 
\frac{1}{X_k(t)}\frac{dX_k}{dt}(t)
\;dt
= \int_{t_k}^{t_h} 
\frac{1}{X_k(t)}\sum_{j \neq k} M_{kj} X_j(t) 
\;dt + M_{kk} (t_h-t_k). 
\end{equation}
Note that the integral $I := \int_{t_k}^{t_h} \frac{1}{X_k(t)}\sum_{j \neq k} M_{kj} X_j(t)$ has a nonnegative integrand, and the term $|M_{kk}(t_h-t_k)|$ is bounded, the above equation cannot diverge to $-\infty$ for a finite $t_h$. This leads to a contradiction, and we conclude that no such $t_h \in \mathbb{R}$ can exist. Therefore, we have $X_k(t) > 0$ for all $t>t_k$. $\blacksquare$ \\ 

In the following, we denote $\rho(M) \in \mathbb{R}$ as the spectral radius of matrix $M$. We say $b \in \mathbb{C}$ is the \textit{dominant eigenvalue} of $M$ if the real part of $b$ is greater then the real parts of other eigenvalues of $M$. \\



\textbf{Proposition 4.4.} For a real matrix $M$ which is essentially nonnegative, exists an eigenvalue $b \in \mathbb{R}$ which is strictly larger than the real part of all other eigenvalues. Let $\delta := \min_k\{M_{kk}\}$, we have $b = \rho(M-\delta I)+\delta \in \mathbb{R}$ as the dominant eigenvalue of $M$. Furthermore, there exist a nonnegative eigenvector $v$ such that $Mv = bv$. \\

\textbf{Proof.} By definition, $M-\delta I$ is a nonnegative matrix. From the Perron--Frobenius theorem (Theorem 8.3.1 in \cite{horn_matrix_2012}), the spectral radius $\rho(M-\delta I)$ is an eigenvalue of $M-\delta I$ and all other eigenvalues of $M-\delta I$ are located in the disc with radius $\rho(M-\delta I)$. Therefore, all other eigenvalues of $M-\delta I$ has real part strictly smaller than $\rho(M-\delta I)$. Since the eigenvalues between $M$ and $M-\delta I$ have one-to-one correspondence (differed by $\delta$), $\rho(M-\delta I) + \delta \in \mathbb{R}$ is an eigenvalue of $M$ and is strictly larger than the real parts of all other eigenvalues of $M$. The existence of nonnegative eigenvector $v$ is also a consequence of Perron--Frobenius theorem. $\blacksquare$ \\

The Perron--Frobenius theorem on nonnegative matrix guarantees that there is at least one nonnegative eigenvector in the eigenspace of $b$. For an essentially nonnegative matrix, $b$ is the dominant eigenvalue. Still, the eigenvector corresponding to $b$ may not be unique. In the following, we obtain a basic result on the long-term growth rate for a special class of LRNs with $\langle Y_k \rangle_t > 0$ for all $k$. \\



\textbf{Lemma 4.5.} Consider an LRN with $\frac{dX}{dt} = MX$. If $\langle Y_k \rangle_t > 0$ for all $k = 1,\dots,n$, then the long-term growth rate $\lambda$ is equal to the dominant eigenvalue of $M$. Furthermore, $Y(t)$ converges to an eigenvector $Y^*$ in the eigenspace of $\lambda$. \\

\textbf{Proof.} From Proposition 4.3, there is a time point $t_p \geq 0$ such that $Y_k(t_p)>0$ for all $k=1,\dots,n$. From Proposition 4.4, $M$ has one dominant eigenvalue (denoted as $b$) with a nonnegative eigenvector (denotes as $Y_{top}$). We can express $Y(t_p)$ as 
$$ Y(t_p) = c_1 Y_{top} + Y_{res} $$
with $c_1 > 0$ and $Y_{res}$ nonnegative. Since the system is linear and the component along $Y_{top}$ grows exponentially with the rate $b$, the long-term growth rate $\lambda$ is at lease $b$. Since $b$ is the dominant eigenvalue, $\lambda$ cannot exceed $b$. This concludes that $\lambda = b$. \\

Now, we recall the expression in Proposition 4.2
\begin{equation}
	X_k(t) = \sum_{j=1}  p_{kj}(t)\,e^{\alpha_j t}\,
	\big(\cos(\theta_jt)+isin(\theta_j t)\,\big) \geq 0,
\end{equation}
where $\alpha_j + i \theta_j$ is the $j^{th}$ eigenvalue of $M$, with $\alpha_j, \theta_j \in \mathbb{R}$, and $p_{kj}(t)$ are polynomials with degree less than $n$. Since the dominant eigenvalue $b$ is unique and real, for large $t$ we have 
\begin{equation}
	X_k(t) \rightarrow 
	\sum_{\{j:\;\alpha_j = b\}}  p_{kj}(t)\,e^{\alpha_j t}\,
	+ o(e^{bt}),
\end{equation}
with $o(.)$ the Landau little-o notation. Consider the rescaled system $Y$, we have 
\begin{equation}
	Y_k(t) \rightarrow
	\frac{
	\sum\limits_{\{j:\alpha_j=b\}} \;
	p_{kj}(t) \, e^{\alpha_j t } + o(e^{bt}) }
	{ \sum\limits_{k=1}^n \sum\limits_{\{j:\alpha_j=b\}} \;
	p_{kj}(t)\, e^{\alpha_j t } + o(e^{bt}) }.
\end{equation}

Since $p_{kj}(t)$ are polynomials, the limit converges to the ratio of the leading coefficients. The exponential factors $e^{bt}$ in numerator and denominator cancel out and $Y_k(t)$ approaches to a constant in $\Delta^{n-1}$ for large $t$. Let $Y^* := \lim_{t\rightarrow\infty}Y(t)$, it remains to show that  $Y^*$ is an eigenvector. Recall that $\frac{dY}{dt} = MY-\mu(Y)Y$. Since $Y(t) \rightarrow Y^*$ as $t \rightarrow \infty$, we have $\lim_{t\rightarrow\infty} \frac{dY}{dt}(t) = 0$. Also, we have  $\lim_{t\rightarrow\infty}\mu(Y(t)) = \mu(Y^*) = \lambda$. Therefore, taking $t\rightarrow\infty$ we have  $MY^* = \lambda Y^*$ and $Y^*$ is in the eigenspace of $\lambda$. $\;\blacksquare$ \\

Lemma 4.5 applies for special cases where $\langle Y_k \rangle_t > 0$ for all $k$. In the following, we proceed for the general results. We first locate the region for the attractor of $Y(t)$. \\



\textbf{Proposition 4.6.} Consider an LRN and a component $Y_k$. We have $\langle Y_k \rangle_t = 0$ if and only if $\lim_{t\rightarrow\infty} Y_k(t) = 0$. \\

\textbf{Proof.} By Proposition 4.2, $\langle Y_k \rangle_t $ exists for every $k$. Note that for LRN, we have $0 \leq Y_k \leq 1$ and hence $0 \leq \langle Y_k \rangle_t \leq 1$. Suppose $Y_k(t) \rightarrow 0$ as $t \rightarrow \infty$. Then, for every $\varepsilon > 0$ there is a constant $t_0$ such that $Y_k(t) < \varepsilon$ for all $t > t_0$. This implies 
\begin{equation}
	\frac{1}{T} \int_0^T Y_k(t)\,dt \leq 
\frac{1}{T} \int_0^{t_0} Y_k(t)\,dt  
+ \frac{1}{T}(T-t_0)\varepsilon.
\end{equation} 
Taking $T\rightarrow \infty$, we conclude $\langle Y_k \rangle_t \leq \varepsilon$. Since $\varepsilon>0$ can be arbitrarily small, we have $\langle Y_k \rangle_t =0$. \\

Now, assume $\langle Y_k \rangle_t = 0$. If the attractor of $Y(t)$ is a fixed point $Y^*$, we have $Y_k^*(t) = 0$ and this implies $\lim_{t\rightarrow 0} Y_k(t) = 0$. If the attractor of $Y(t)$ is a limit torus $\Omega$, then the solution approaches to a quasi-periodic trajectory which is bounded between $Y_k^{(\min)} := \min_{Y \in \Omega} Y_k$ and $Y_k^{(\max)} := \max_{Y \in \Omega} Y_k$. For an LRNs, the vector field $MY-\mu(Y)Y$ is continuous and bounded on $\Delta^{n-1}$. Suppose $Y_k^{(\min)} < Y_k^{(\max)}$. The long-term time fraction for the quasi-periodic trajectory $Y(t)$ to have $Y_k(t) > (Y_k^{(\max)}+Y_k^{(\min)})/2 =: Y_k^{(\text{mid})}$ is positive, since $\frac{dY_k}{dt}$ is bounded and a positive time is required for $Y_k(t)$ to increase from $Y_k^{(\text{mid})}$ to $Y_k^{(\max)}$. This contradicts to the assumption $\langle Y_k \rangle_t = 0$. Therefore, we must have $Y_k^{(\min)} = Y_k^{(\max)}$. If $Y_k^{(\min)} > 0$, we have $\langle Y_k \rangle_t > 0$, contradicting to the assumption. Hence $Y_k^{(\min)} = Y_k^{(\max)} = 0$, and this implies $\lim_{t\rightarrow\infty} Y_k(t) = 0$. $\blacksquare$ \\


\textbf{Proposition 4.7.} Consider an LRN with $\frac{dX}{dt} = MX$ and an initial condition $X(0)$. Assume that $\langle Y_k\rangle_t > 0$ for $k=1,\dots, h$ with $h \leq n$, and $\langle Y_k\rangle_t = 0$ otherwise. Then, we have $M_{qp} = 0$ for all $p\leq h$ and $q > h$. That is, $M$ can be represented as 

\begin{equation}
M := 
\begin{pmatrix}
M^{+} & * \\
O & * 
\end{pmatrix},
\end{equation}  
where $M^{+} \in \mathcal{M}^{h,h}$ and $O \in \mathcal{M}^{n-h,h}$, and $\ast$ represent general matrix blocks.  \\

\textbf{Proof.} Fix $p,q$ such that $p \leq h$ and $q > h$. By assumption, $\langle Y_p \rangle_t >0$ and $\langle Y_q \rangle_t =0$. We show that $M_{qp}>0$ will lead to a contradiction. We first calculate $\frac{dY_q}{dt}$ as 
\begin{equation}\label{E3.7}
\begin{aligned}
\frac{dY_q}{dt} 
&= (MY)_q -\mu(Y)Y_q \\
&= \sum_{k\neq q} M_{qk}Y_k + M_{qq}Y_q-\mu(Y)Y_q.
\end{aligned}
\end{equation}

Since $\langle Y_q \rangle_t = 0$, by Proposition 4.6, we have $\lim_{t\rightarrow\infty} Y_q(t) = 0$. Therefore, $M_{qq}Y_q(t)$ and $\mu(Y(t))Y_q(t)$ approach to zero. Now, assume $M_{qp}> 0$. Since $M_{qk}$ are nonnegative for $q\neq k$ and we assume $\langle Y_p \rangle_t > 0$, the first term implies the inequalities 
\begin{equation}
\bigg\langle \sum_{k\neq q} M_{qk}Y_k \bigg \rangle_t \geq 
\langle M_{qp} Y_p \rangle_t \geq
M_{qp} \langle Y_p \rangle_t > 0.
\end{equation}

Hence we have $\langle \frac{dY_q}{dt} \rangle_t > 0$. However, this implies $\lim_{t\rightarrow \infty} Y_q(t) > 0$ by Proposition 4.6 and contradicts to the assumption $\langle Y_q \rangle_t = 0$. $\blacksquare$ \\


\textbf{Proposition 4.8.} Consider an LRN with $\frac{dX}{dt} = MX$ and fix an initial condition $X(0)$. Assume that $\langle Y_k\rangle_t > 0$ for $k=1,\cdots,h \leq k$ and $\langle Y_k\rangle_t = 0$ otherwise. Partition the matrix $M$ as

\begin{equation}
M := 
\begin{pmatrix}
M^{+} & * \\
O & * 
\end{pmatrix},
\end{equation}  
with $M^{+} \in \mathcal{M}^{h,h}$. Define a subsystem as

\begin{equation}
	\begin{aligned}
		& X^+ := (X_1,\cdots,X_h)^T, \\
		& N^+ := X_1+\cdots+X_h, \\
		& Y^+ := \frac{X^+}{N^+}, \\
		& \nu(Y^+) := \frac{1}{N^+}\frac{dN^+}{dt}. 
	\end{aligned}
\end{equation}
Then in the long-term, the dynamics of $Y(t)$ follows
\begin{equation}
	\begin{aligned}
		& \lim_{t \rightarrow \infty} 
		\bigg(
		\frac{dY_k}{dt} (t) -
		\frac{d(Y^+)_k}{dt} (t)
		\bigg) = 0, \\
		&\frac{d(Y^+)_k}{dt} (t) = (M^+ Y^+(t))_k - \nu(Y^+(t)) Y_k^+(t), 
		\qquad k=1,\dots,h, \\
	\end{aligned}
\end{equation}
and
\begin{equation}
	\begin{aligned}
		& \lim_{t \rightarrow \infty}
		\frac{dY_k}{dt} (t) = 0, 
		\qquad k > h.
	\end{aligned}
\end{equation}
In particular, the long-term growth rate of system $X(t)$ and subsystem $X^+(t)$ are the same. \\

\textbf{Proof.} First, since $\langle Y_k \rangle_t = 0$ for $k>h$, we have $\lim_{t\rightarrow \infty}Y_k(t) = 0$ for $k>h$ by Proposition 4.6. This implies $\lim_{t\rightarrow\infty} \frac{N^+(t)}{N(t)} = 1$ and $\lim_{t\rightarrow\infty} \frac{(Y^+)_k(t)}{Y_k(t)} = 1$ for all $k \leq h$. This also implies $\lim_{t\rightarrow\infty} \big( \frac{dY_k}{dt} (t) -  \frac{d(Y^+)_k}{dt}(t) \big) = 0$ for all $k$.  \\

To proceed, we show that $(MY)_k = (M^+Y^+)_k$ and $\nu(Y^+) = \mu(Y)$. The first equality is immediate, since 
\begin{equation}
	\lim_{t\rightarrow\infty}
	\bigg(
	\sum_{j=1}^n M_{kj}(Y^+)_k - 
	\sum_{j=1}^h M_{kj}Y_k
	\bigg) = 0,
\end{equation}
and the second equality can be seen from 
\begin{equation}
\mu(Y(t))-\nu(Y^+(t))
=\frac{d}{dt} 
\bigg(
\log \frac{N(t)}{N^+(t)} 
\bigg)
\rightarrow 0 
\qquad \text{as } t\rightarrow\infty, 
\end{equation}
since we have $\lim_{t\rightarrow\infty} \frac{N^+(t)}{N(t)} = 1$. Therefore, we have
\begin{equation}
	\begin{aligned}
		\lim_{t\rightarrow\infty}
		\frac{dY_k}{dt} &= 
		\lim_{t\rightarrow\infty} (MY)_k -
		\lim_{t\rightarrow\infty} \mu(Y)Y_k \\
 		&= (M^+Y^+)_k - \nu(Y^+) (Y^+)_k,
	\end{aligned}
\end{equation} 
for all $k\leq h$. For $k>h$, since $Y_k(t)$ is continuously differentiable with respect to $t$ and $\lim_{t\rightarrow\infty} Y_k(t) = 0$, by definition, we have $\lim_{t\rightarrow\infty} \frac{dY_k}{dt} = 0$. \\

To show the long-term growth rates of $X(t)$ (denoted as $\lambda$) and $X^+(t)$ (denoted as $\lambda^+$) are the same, we notice that since $\lim_{t\rightarrow\infty}(N^+(t)/N(t)) = 1$, we have
\begin{equation}
	\lambda^+ - \lambda
= \lim_{t\rightarrow\infty} \frac{1}{t}\, 
\log \bigg(
\frac{N^+}{N}
\bigg) = 0.
\end{equation}
$\blacksquare$ \\


Proposition 4.8 states that the long-term dynamics of  $\frac{dY}{dt} = MY - \mu(Y)Y$ approaches to the long-term dynamics of the subsystem $\frac{dY^+}{dt} = M^+Y^+ - \nu(Y^+)Y^+$. Notably, the subsystem dynamics can be regard as a sub-reaction network, and $M^+$ is also essentially nonnegative. In this way, we can apply Lemma 4.5 on the subsystem and obtain similar results. This leads to the proof of Theorem A. \\


\textbf{Proof of Theorem A.} Without loss of generality, we can re-index the node of LRN such that $\langle Y_k \rangle_t$ follows the description in Proposition 4.8. By Lemma 4.5, the solution converges to an eigenvector $Y^* \in \mathbb{R}^{h}$ in the subsystem $\frac{dY^+}{dt} = M^+Y^+ - \nu(Y^+)Y^+$ and $\lambda$ is equal to a real eigenvalue of $M^+$. It remains to show that, first, $\lambda$ is an eigenvalue of $M$, and second, the limit vector $\lim_{t\rightarrow\infty} Y(t)$ is an eigenvector of $M$ corresponding to $\lambda$. \\

Define $Y^\# := (Y^*_1,\dots,Y^*_h, 0,\dots,0)^T \in \mathbb{R}^n$. From Proposition 4.8, we have $\lim_{t\rightarrow\infty} Y(t) = Y^\#$ and the long-term growth rate of the subsystem $\lambda^+$ equals $\lambda$. Therefore,

\begin{equation}
MY^\# = 
\begin{pmatrix}
M^{+} & * \\
O & * 
\end{pmatrix}
\begin{pmatrix}
Y^* \\ 0
\end{pmatrix}
= 
\begin{pmatrix}
M^+Y^* \\ 0
\end{pmatrix}
=
\begin{pmatrix}
\lambda^+ Y^* \\ 0
\end{pmatrix}
= \lambda Y^\#.
\end{equation}  
$\blacksquare$ \\



\vspace{10 pt}


\textbf{Corollary 4.9.} Consider an LRN with $\frac{dX}{dt}=MX$. If $M$ is irreducible, then for all nonnegative initial conditions, we have $\langle Y_k\rangle_t > 0$ for all $k$. Furthermore, for all nonnegative initial conditions, $\lambda$ is the dominant eigenvalue of $M$, with the solution $Y(t)$ approaching to a unique (up to scaling constant) eigenvector $Y^*$ corresponding to $\lambda$. \\

\textbf{Proof.} First, we show $\langle Y_k\rangle_t > 0$ for all $k$. Assume the contrary. Then, there exist some $Y_k$ with $\langle Y_k \rangle_t = 0$. With suitable re-index of nodes and by Proposition 4.7, $M$ has a matrix partition with off-diagonal zero matrix block, and this contradicts to the assumption that $M$ is irreducible (see 6.2.21 in \cite{horn_matrix_2012}). Therefore, we have $\langle Y_k\rangle_t > 0$ for all $k$. \\

By Lemma 4.5, $\lambda$ is the dominant eigenvalue of $M$, and by Perron--Frobenius theorem the dominant eigenvalue of irreducible matrix is unique with algebraic multiplicity 1 (see Section 8.4 in \cite{horn_matrix_2012}). Lemma 4.5 guarantees that $ \lim_{t\rightarrow\infty} Y(t) = Y^*$ and $Y^*$ belongs to the eigenspace of $\lambda$. In this case, the eigenspace of $\lambda$ has dimension 1, corresponding to the unique dominant eigenvector of $M$. $\;\blacksquare$ \\



In the following, we establish some estimations about LRN dynamics, which will be used in the later sections. \\

 
\textbf{Proposition 4.10.} Consider an LRN of $n$ nodes with long-term growth rate $\lambda$. For initial condition $X(0)\in \mathbb{R}_{\geq0}^n$, $X_k(t)$ can be expressed as 
\begin{equation}
	X_k(t) = c_k(t) \, e^{\lambda t}\,
	\big( 1 + E_k(t) \big),
\end{equation}
where $c_k(t)$ a polynomial with degree less than $n$, and $E(t)$ is an error term. We can choose the signs such that both $c_k(t)$ and $1+E_k(t)$ are positive for all $t \geq 0$. In addition, the error term is bounded by 
\begin{equation}
\begin{aligned}
	|E_k(t)| &\leq A_k, \qquad &t \in [0,T_k],  \\
	|E_k(t)| &\leq A_k \,t^n\,e^{-\delta t} , 
	\qquad &t \in (T_k,\infty), 
\end{aligned}
\end{equation}
where $A_k, T_k, \delta$ are positive constants. \\

\textbf{Proof.} Suppose the LRN follows the ODE $\frac{dX}{dt}=MX$. By the property of LRN, the matrix $M$ is essentially nonnegative and hence the nonnegative orthant is invariant for $X(t)$ with $X(0)\in \mathbb{R}_{\geq0}^n$. By the proof in Proposition 4.3, $X_k(0)>0$ implies $X_k(t) > 0$ for every finite $t$. Therefore, we can choose $c_k(t)$ and $1+E_k(t)$ as both positive quantities. \\

Note that we do not require $M$ to be irreducible. Therefore, depending on the initial condition, the long-term growth rate $\lambda$ may not be the eigenvalue with largest real part of $M$ and may have multiplicity. Using the formula in Proposition 4.2, we express $X_k$ as
\begin{equation} \label{E4.10A}
\begin{aligned}
	X_k(t) 
	& = \sum_{\alpha_j > \lambda} 
	p_{kj}(t)\, e^{\alpha_j t} 
	\big(
	\cos(\theta_j t) + i \sin( \theta_j t)
	\big) \\	
	& + \sum_{\alpha_j = \lambda} 
	p_{kj}(t)\, e^{\alpha_j t} 
	\big(
	\cos(\theta_j t) + i \sin( \theta_j t)
	\big) \\
	& + \sum_{\alpha_j < \lambda} 
	p_{kj}(t)\, e^{\alpha_j t} 
	\big(
	\cos(\theta_j t) + i \sin( \theta_j t)
	\big). \\
\end{aligned}
\end{equation}
Here, $p_{kj}(t)$ is a polynomial with degree less than $n$, and $b_j := \alpha_j + i\theta_j$ is the $j$th eigenvalue of matrix $M$, with $\alpha_j, \theta_j \in \mathbb{R}$. Note that the coefficients in $p_{kj}(t)$ depends on initial condition, and may be complex-valued. Still, after the summation $X_k(t)$ must be real-valued. \\

Since $\lambda$ is the long-term growth rate, the first summation in (\ref{E4.10A}) must be zero (otherwise the system grows faster than $\lambda$). For the second summation, note that an eigenvalue $b_j$ with $\alpha_j = \lambda$, it must have $\theta_j = 0$ (i.e. $b_j \in \mathbb{R}$). Otherwise, $X_k(t)$ has oscillatory trajectory which contradicts to Theorem A. Therefore, by defining $c_k(t):= \sum_{\alpha_j = \lambda} p_{kj}(t)$, equation (\ref{E4.10A}) can be further simplified as  
\begin{equation} \label{E4.10}
\begin{aligned}
	X_k(t) &= c_k(t)\,e^{\lambda t} \;
	\bigg(1 + 
	\sum_{\alpha_j < \lambda}
	\frac{p_{kj}(t)}{c_k(t)} \, e^{(\alpha_j-\lambda) t} \,
	\big(
	\cos(\theta_j t) + i \sin( \theta_j t)
	\big  )
	\bigg ) \\ 
	&:= c_k(t)\,e^{\lambda t} 
	\big ( 
	1+E_k(t)
	\big ).
\end{aligned}
\end{equation}
In the term $E_k(t)$, every value of $\alpha_j - \lambda$ is negative, and we define 
\begin{equation}
	\delta := \min_{\alpha_j <\lambda} 
	\{ \lambda - \alpha_j \} > 0,
\end{equation}
which is the slowest exponential decay rate. Also, exist a constant $A'_{kj}$ such that
\begin{equation}
\begin{aligned}
	\lim_{t\rightarrow\infty} 
	\bigg |
    \frac{p_{kj}(t)}{c_k(t)}
    \big(
    \cos(\theta_j t) + i \sin( \theta_j t)
    \big)
    \bigg |
 	\leq A'_{kj}\,t^{\,|\deg p_{kj} - \deg c_{k}| } 
	\leq |A'_{kj}|\, t^n.
\end{aligned}
\end{equation}
Define $A'_k:= \sum_{\alpha_j < \lambda} |A'_{kj}|$. There exist $T_k>0$ such that for all $t>T_k$,
\begin{equation} \label{E4.10C}
	|E_k(t)| \leq A'_k\, t^n\, e^{-\delta t}. 
\end{equation} 


Now, we consider $E_k(t)$ in the time range of $[0,T_k]$. Since $X_k(t)$ is a solution of ODE, it is bounded for any finite time. Hence, $E_k(t)$ is a continuous, bounded function on $[0,T_k]$. This implies exist a positive constant $A''_k$ such that $|E(t)| \leq A''_k,\; t\in[0,T_k]$. Taking $A_k := \max \{A_k',A_k'' \}$ completes the proof. $\; \blacksquare$ \\

\vspace{15 pt}


\textbf{Proposition 4.11.} Consider an LRN of $n$ nodes with long-term growth rate $\lambda$. Let $x_g$ be a gatekeeper nodes and consider the gatekeeper sub-fraction $W_g(t) = \frac{X_g(t)}{Z(t)}$ (see Definition 2.1). Then, $W_g(t)$ converge to a constant $W_g^*$ as $t \rightarrow \infty$, and there exists a constant $T>0$ such that 
\begin{equation}
\begin{aligned}
	|W_g(t)-W_g^*| & \leq B_0, 
	\qquad  &t \leq T, \\ 
	|W_g(t)-W_g^*| & 
	\leq \frac{B_1}{t} + B_2\, t^n\, e^{-\delta t},
	\qquad  &t > T.
\end{aligned}
\end{equation}
where $\delta > 0$ and $B_0, B_1, B_2$ are nonnegative constants. \\

\textbf{Proof.} By Theorem A, the normalized vector $Y(t)$ of the LRN converges to a constant vector $Y^*$. The gatekeeper sub-fraction follows
\begin{equation}
	\lim_{t\rightarrow\infty} 	W_g(t) 
	=	\lim_{t\rightarrow\infty} 
	\frac{Y_g} 
	{\sum_{x_g \in \mathcal{G}} Y_g} 
	= \frac{Y^*_g} 
	{\sum_{x_g \in \mathcal{G}} Y^*_g} := W_g^*.
\end{equation}
By Proposition 4.10, we express $W_g(t)$ as
\begin{equation} \label{E4.11A}
	W_g(t) 
	= \frac{c_g(t) \, e^{\lambda t} \,\big( 1 + E_g(t) \big)}
	{c_Z(t) \, e^{\lambda t} \,\big( 1 + E_Z(t) \big)},
\end{equation}
where $c_Z = \sum_{x_g \in \mathcal{G}} c_g$ and $E_Z = \sum_{x_g \in \mathcal{G}} E_g$ in Proposition 4.10. By definition, we have $\deg c_Z(t) \geq \deg c_g(t)$ since $c_Z(t)$ is a summation over $c_g$ for all gatekeeper nodes. Therefore, we can express the ratio as 
\begin{equation} \label{E4.11B}
	\frac{c_g(t)}{c_Z(t)} = K_1 + \frac{K_2}{t p_g(t)},
\end{equation}
where $p_g(t)$ is a polynomial and $K_2$ is a constant (possibly zero). Noticing that $\lim_{t\rightarrow\infty} W_g(t) \ = W_g^*$, and comparing (\ref{E4.11A}) and (\ref{E4.11B}), we found that the constant $K_1 = W_g^*$. Substitute (\ref{E4.11B}) back to (\ref{E4.11A}), and using Proposition 4.10, we have a constant $T>0$ such that for all $t>T$, 
\begin{equation}
\begin{aligned}
	W_g(t) &= 
	\bigg( W_g^* + \frac{K_2}{t p_g(t)} \bigg )\, 
	\big( 1+E_g(t)\big )\,
	\big( 1-E_Z(t)+E_Z(t)^2-... \big ) \\
	&\leq 
	\bigg( W_g^* + \frac{K_2}{t} \bigg )\,
	\big( 1+ K_3 \,t^n e^{-\delta t} \big ),
\end{aligned}
\end{equation}
where $K_2, K_3$ are constants. By rearrange the term, there exist nonnegative constant $B_1,B_2$ such that
\begin{equation}
	|W_g(t) -  W_g^*| \leq 
	\frac{B_1}{t} + 
	B_2 \,t^n e^{-\delta t},
\end{equation}
for $t>T$. Finally, we can take $B_0 := \max_{t \in [0,T]} |W_g(t)-W_g^*|$. It is clear that $B_0$ exists since $W_g(t)$ is continuous and $W_g(t) \in [0,1]$ for all $t$. $\;\blacksquare$ \\

\vspace{15 pt}


\textbf{Proposition 4.12.} Consider an LRN of $n$ nodes with long-term growth rate $\lambda$. Let $Z(t)$ denote the gatekeeper biomass and consider $\tau$ as a parameter range between $\in [0,t]$. Consider the ratio 
\begin{equation}
	f(t;\tau) := \frac{Z(t-\tau)}{Z(t)}. 
\end{equation}
Then, we have
\begin{equation}
	\lim_{t\rightarrow\infty}  f(t;\tau)
	= e^{-\lambda\tau},
\end{equation}
and there are constants $T>0$, $M>0$ such that when $t>T$, 
\begin{equation}
	f(t;\tau) \leq Me^{-\lambda \tau}  
	\;\;\text{ for all } \tau \in [0,t].
\end{equation}\


\textbf{Proof.} We express the ratio as in Proposition 4.10:

\begin{equation}
\begin{aligned}
	f(t;\tau) &= 
	\frac{ 
	\sum_{x_g \in \mathcal{G}} \;\;
	c_g(t-\tau) \, e^{\lambda (t-\tau)}\,
	\big( 1 + E_g(t-\tau) \big)
	}
	{
	\sum_{x_g \in \mathcal{G}} \;\;
	c_g(t) \, e^{\lambda (t)}\,
	\big( 1 + E_g(t) \big) 
	}.
\end{aligned}
\end{equation}
Fixed $\tau$ and taking the $t\rightarrow\infty$ limit, we see that $\lim_{t\rightarrow\infty}  f(t;\tau) = e^{-\lambda\tau}$. Now, define 
\begin{equation}
	c_Z(t):=\sum_{x_g \in \mathcal{G}} c_g(t), 
	\qquad
	Q_Z(t):= c_Z(t)^{-1} \sum_{x_g \in \mathcal{G}} c_g(t) \, E_g(t), 
\end{equation}
the ratio can be expressed as 
\begin{equation}
	f(t;\tau)= 
	\frac{ c_Z(t-\tau) }{ c_Z(t)}\;\,
	\frac{ 1+ Q_Z(t-\tau) } 
	{ 1 + Q_Z(t)  } \,
	e^{-\lambda \tau},
\end{equation}
where $c_Z(t)$ is a polynomial with degree $n_0 \leq n$. By Proposition 4.10, both $c_g(t)$ and $1+E_g(t)$ are positive for all $t>0$. This implies $c_Z(t)>0$ and $E_g(t)> (-1)$ for all $t>0$. Direct calculation shows that $1+Q_Z(t)>0$ for all $t>0$ and hence $f(t;\tau)$ is nonnegative. By Proposition 4.10, $Q_Z(t)$ is bounded as
\begin{equation}
\begin{aligned}
	|Q_Z(t)| &\leq A^*, 
	\qquad  &t \in [0,T] \\
	|Q_Z(t)| &\leq A^* t^n e^{-\delta t},
	\qquad  &t \in (T,\infty] \\
\end{aligned}
\end{equation}
where $A^*, T, \delta$ are positive constants. Therefore, taking $A^{**} := \max_{t\in (T,\infty)} \{ A^*t^n \,e^{-\delta t}\}$ and $A := \max \{A^*,A^{**} \}$, we have
\begin{equation}
	1+Q_Z(t-\tau) \leq 1+A
\end{equation}
for every $\tau \in [0,t]$. It is clear that exist $T_1>0$ such that $At^n e^{-\delta t} < \frac{1}{2}$. Therefore, \begin{equation}
	\frac{1}{2}
	\leq 
	1+Q_Z(t) 
	\leq 
	1+ \frac{1}{2},
\end{equation}
when $t>T_1$. Combining the above two equations, we have
\begin{equation}
	\frac{ 1+ Q_Z(t-\tau) } { 1 + Q_Z(t)  }
	\leq 2(1+A),
\end{equation}
for all $\tau \in [0,t]$ when $t>T$. Finally, calculating the limit yields that
\begin{equation}
	\lim_{t\rightarrow\infty} 
	\frac{c_Z(t-\tau)}{c_Z(t)} 
	= \lim_{t\rightarrow\infty} 
	\bigg( 1-\frac{\tau}{t} \bigg)^{n_0}
	\leq 1, \;\;\text{ for all } \tau \in [0,t],
\end{equation}
where $n_0$ is the degree of polynomial $c_Z(t)$. Therefore, exist $T_2$ such that $\frac{c_Z(t-\tau)}{c_z(t)} $ when $t>T_2$. The proposition is proved by setting $M := 2(1+A)$ and $T=\max \{T_1,T_2 \}$. $\; \blacksquare$ \\


%
%


%
%


\newpage
\section{Biomass transfer on LRNs}

In this section, we describe how to formulate and calculate the biomass transfer of LRNs using the Lagrangian perspective. The main idea is to employ Theorem A, i.e. for LRNs the rescaled system $Y(t)$ converge to a fixed point $Y^*$ in the long-term. \\

\textbf{Theorem B.} Consider an LRN and assume given an initial condition with $\lambda>0$. Then, the gatekeeper biomass follows the long-term dynamics
		\begin{equation}
			\frac{dZ}{dt}(t)= -\beta Z(t)
			+ \int_0^t 
			\alpha(\tau)\,Z(t-\tau)\,d\tau
			+ \gamma(t), 
		\end{equation}
with $\lim_{t\rightarrow\infty} \frac{\gamma(t)}{Z(t)}=0$. Furthermore, define $\tilde{\alpha}(s):=\mathcal{L}[\alpha(\tau)]$ as the catalytic spectrum, we have 
\begin{equation}
	\lambda + \beta = \tilde{\alpha}(\lambda).
\end{equation} 
$\;\blacksquare$ \\

We will prove Theorem B in the end of this section. Our goal is to give explicit descriptions for $\alpha(\tau)$ and $\beta$ in terms of rate constants in reaction pathways (i.e. the entries in $M$ in Definition 4.1). First, we explicitly specify the arrival functions of LRNs in below. \\



\textbf{Definition 5.1.} Consider a reaction pathway in an LRN, denoted by $\pi: u_0 \omega_1 u_1 \dots \omega_L u_{L+1}$, with $L\geq1$. For a reaction $\phi_c$ with upstream node $X_k$, we denote the constant pre-factor of flux function by $R[\phi_c]$, i.e. $J[\phi_c]=R[\phi_c]X_k$. Several quantities of $\pi$ are considered in below: \\

(i) The \textit{pathway probability} for biomass being transferred via $\pi$, defined as 
\begin{equation}
\begin{aligned}
&q_{\pi} :=
q_1 q_2\cdots q_{L}, \\
&q_k := 
\frac{\text{mass efflux of $u_k$ via $\omega_k$} }
{\text{total mass efflux of } u_k}
= 
\frac{S[u_k,\omega_k]R[\omega_k]}
{\sum_{\phi_c \in out(u_k)} S[u_k,\phi_c] R[\phi_c]}
\in [0,1].
\end{aligned}	
\end{equation}

(ii) The \textit{transmission efficiency} for biomass being transferred via $\pi$, defined as (via downstream fraction, see Definition 1.5)
\begin{equation}
\begin{aligned}
&\theta_{\pi} =
\theta_0 \theta_1\cdots\theta_L, \\
&\theta_k := \xi[u_{k+1}, \omega_k] \in [0,1].
\end{aligned}	
\end{equation}

(iii) The \textit{conditional waiting time} (a random variable between $[0,\infty)$) for biomass being transferred via $\pi$, define as   
\begin{equation}
T_{\pi} = T_1 + T_2 +\cdots+T_L, 
\end{equation}
where $T_k$ is a random variables of exponential waiting times, namely,
\begin{equation}
T_k \sim c_k\,e^{-c_k t}, \;\; 
c_k := (-1)
\sum_{\phi_{c}\in out(u_k)} S[u_k,\phi_c]\, R[\phi_c] 
\geq 0. 
\end{equation}\

\textbf{Note.} For LRNs, $q_{\pi}, \theta_{\pi}$ and $c_k$ are all constants in time. The formulae for (i), (ii) are straightforward for their definitions. By definition, the stoichiometry coefficient of efflux $S[u_k,\omega_k]$ is negative, and hence for formula of $c_k$ carries a $(-1)$ sign. \\


\begin{center}
	\includegraphics[scale = 0.45]{FigS5.1.png}
\end{center}




\textbf{Proposition 5.2 (Arrival function).} Consider a reaction pathway $\pi$ with $L\geq 1$ in an LRN, the arrival function $H(t',t)$ (see Definition 2.7) can be expressed as a function of $\tau := t-t'$. Specifically, with the notation in Definition 5.1, 
\begin{equation}
	H_{\pi}(t',t) :=
	h_{\pi}(\tau) = q_{\pi}\, \theta_{\pi}\, f_{T_\pi}(\tau),
\end{equation}
where $f_T(.)$ is the probability density function of random variable $T$. \\


\textbf{Proof.} First we notice that by definition, $h_{\pi}$ is a conditional waiting time distribution for biomass to transferr along reaction pathway $\pi$. Qualitatively speaking, a reaction pathway can have two reasons to "loss" or "dissipate" the biomass during the transfer process. The first reason is that for each node, there may be multiple efflux reactions, and only a fraction of biomass being transferred along the reaction of $\pi$. This is quantified by $q_{\pi}$, the pathway probability. The second reason is that even within a single reaction, there may be multiple downstream nodes and the biomass is partitioned among these downstream nodes. Therefore, only a fraction of biomass being transferred for the specific node of the pathway $\pi$. This is quantified by $\theta_{\pi}$, the transmission efficiency. \\

Therefore, the arrival function must include the pre-factor $q_{\pi} \theta_{\pi}$ as defined in Definition 5.2. For calculating the conditional waiting time, we notice that under the stochastic framework a discrete biomass unit is transferred by the \textit{earliest} reaction event. Suppose a node $x_k$ has downstream reactions $\phi_1,\cdots,\phi_M$, by the LRN property (ii) all these flux functions have the form $\phi_c = R_cX_k$. Therefore, for each reaction the unconditional waiting time $T_{k,c}$ follows probability density function $r_c e^{-r_c t}$ with rate constants
\begin{equation}
	r_c := |S[x_k,\phi_c]|R_c \geq 0.
\end{equation} 

For the earliest waiting time to leave $x_k$, we define the random variable $T_{k,\min} = \min\{T_{k,1},\cdots,T_{k,M}\}$. Without loss of generality, we can assume the first reaction is the earliest, Define $f_k(t_1)$ as the probability density function for the conditional waiting time $\{ T_{k,1}:\;T_{k,1}=T_{k,\min} \}$. We have
\begin{equation}
\begin{aligned}
	f_k(t_1) 
	&= C'\int_{t_2=t_1}^{\infty} \cdots
	\int_{t_M=t_1}^{\infty} r_1 r_2\cdots r_M\,
	e^{-(r_1t_1+\cdots+r_Mt_M)} \,dt_2\cdots  dt_M \\
	&= C'' e^{-(r_1+\cdots+r_M)t_1},
\end{aligned}
\end{equation}
where $C', C''$ are normalization constants. A similar argument applies for each reactions in $\{\phi_1,\cdots,\phi_M\}$, and hence the conditional waiting time for \textit{every} reactions in $\{\phi_1,\cdots,\phi_M\}$ to be the earliest are the same, and all follow the probability density function $f_k$. Hence, $f_k(\tau)$ is the probability density function of the conditional waiting time $T_k$. Now, apply the formula on each node $u_k$ in the pathway $\pi$, we obtained the formula of $f_{T_\pi}$. \\

Finally, note that all the coefficients in $f_{T_\pi}$ are time-independent, and therefore the distribution only depends on $\tau := t-t'$, which is the conditional waiting time of mass transfer between the first reaction $\omega_0(t_1)$ and the last reaction $\omega_L(t_{L+1})$ (see Figure in Definition 5.1). This is consistent with the Definition 2.7. Together, we have $h_{\pi}(\tau) = q_{\pi}\theta_{\pi}f_{T_\pi}(\tau)$. $\;\blacksquare$ \\


 
\textbf{Proposition 5.3 (Amplification rate).} Consider an LRN with a  reaction pathway $\pi\in \mathcal{F}(\mathcal{E},\mathcal{G})\cup \mathcal{F}(\mathcal{G},\mathcal{G})$, $\pi: u_0 \omega_0 u_1 \dots \omega_L u_{L+1}$. The amplification rate $\kappa_{\pi}(t')$ converge to a constants $\kappa^*$ as $t' \rightarrow \infty$, namely 
\begin{equation}
	\kappa^* = S[u_1,\omega_0]\,R[\omega_0] \; W_g^*, 
    \qquad  W_g^* :=\frac{Y^*_g}{\sum_{x_k \in \mathcal{G}}Y_k^*}.
\end{equation}

\textbf{Proof.} By definition, for $\pi\in \mathcal{F}(\mathcal{E},\mathcal{G})\cup \mathcal{F}(\mathcal{G},\mathcal{G})$ the flux function of first reaction (denoted by $J[\omega_0]$) must be proportional to a gatekeeper node biomass, i.e., $J[\omega_0] = R[\omega_0]X_g$ with $R[\omega_0]$ a constants and $x_g \in \mathcal{G}$. With Definition 2.6, the amplification rate can be expressed as 

\begin{equation}
\kappa_{\pi}(t') 
= S[u_1,\omega_0] \, \frac{J[\omega_0](t')}{Z(t')}
= S[u_1,\omega_0] \, R[\omega_0] \,\frac{X_g(t')}{Z(t')}.
\end{equation} 

For LRNs, we have $\lim_{t'\rightarrow\infty} Y(t')=Y^*$ by Theorem A. Therefore, the gatekeeper sub-fraction $W_g = X_g/Z $ (see Definition 2.1) converges to a constant in the long-term, i.e.

\begin{equation}
	W_g^* = 
	\lim_{t'\rightarrow\infty} W_g(t') 
	= \lim_{t'\rightarrow\infty} \frac{X_g(t')}{Z(t')} 
	= \lim_{t'\rightarrow\infty} 
	\frac{Y_g(t')}{\sum_{x_k \in \mathcal{G}}Y_k(t')}
	= \frac{Y_g^*}{\sum_{x_k \in \mathcal{G}}Y_k^*}.
\end{equation}
Therefore, we have 
\begin{equation}
	\lim_{t'\rightarrow\infty} \kappa_{\pi}(t')
	= S[u_1,\omega_0] \, R[\omega_0] \,W^*_g =:\kappa^*.
\end{equation}
$\;\blacksquare$ \\


 
\textbf{Proposition 5.4 (Catalytic kernel and gatekeeper degradation rate).} Consider an LRN, define 
\begin{equation}\begin{aligned}
	\alpha(\tau)&:=
	\lim_{t\rightarrow\infty} 
	\alpha^{\bigstar}(\tau; t), \\
	\beta &:=
	\lim_{t\rightarrow\infty} 
	\beta^{\bigstar}(t), \\
\end{aligned}
\end{equation}
as the (long-term) catalytic kernel and gatekeeper degradation rate. Then, we have
\begin{equation}
	\alpha(\tau) = 
	\sum_{\pi \in \mathcal{F}(\mathcal{E}, \mathcal{G}) 
	\cup  \mathcal{F}(\mathcal{G}, \mathcal{G})}
	\kappa^*_{\pi}\, h_{\pi}(\tau)
\end{equation}

\begin{equation}
\beta = (-1)
\sum_{x_g \in \mathcal{G}} \;
\sum_{\phi_c \in out(x_g)}
(1-\xi[\mathcal{G}, \phi_c])\;
S[x_g,\phi_c]\;
R_c W_g^*,
\end{equation}
where $R_c$ are constant pre-factors of flux function $J_c(X)$, and $W^*_g$ is the long-term gatekeeper sub-fraction. \\

\textbf{Proof.} For LRNs the arrival function $H(t',t)$ can be expressed by $h(\tau)$, a function of $\tau$ (Proposition 5.2).  Other time-dependent $\alpha^{\bigstar}, \beta^{\bigstar}$ are linear combinations of $W_g(t)$ and hence converge to constants as $W_g(t) \rightarrow W_g^*$. $\;\blacksquare$ \\

\textbf{Note.} The catalytic kernel $\alpha$ is defined by fixed $\tau$ and taking $t\rightarrow\infty$; there are some details for taking the limit. If there are more than one system nodes in a reaction pathway, the arrival function $H(t',t)$ is a continuous function and we interpret the limit $\alpha(\tau) = \lim_{t\rightarrow\infty} \alpha^{\bigstar}(\tau; t)$ as a point-wise limit for each $\tau$. If there is only one system node in the pathway, the waiting time follows a Dirac's delta measure $\delta(\tau)$ and $h(\tau)=\delta(\tau)$. In this case, the limit is a sequence of measures in form of $\kappa(t-\tau)\,\delta(\tau)$, which point-wisely converges to $\kappa^*\,\delta(\tau)$ as $t \rightarrow \infty$. In either case, the point-wise limit of $t \rightarrow \infty$ is well-defined. \\

 
\textbf{Proposition 5.5.} Consider an LRN and let $\alpha(\tau)$ be its catalytic kernel in Proposition 5.4. We have the followings results:

\begin{enumerate} [label=(\roman*)]
	\item $\alpha(\tau)$ is a nonnegative measure on $\mathbb{R}_{\geq 0}$. 
	\item $\int_0^\infty \alpha(\tau)d\tau \leq M_{\alpha}$ for a constant $M_{\alpha}>0$.
	\item The Laplace transform $\mathcal{L}[\alpha(\tau)] := \tilde{\alpha}(s)$ is nonnegative and non-increasing on $s \geq 0$.
\end{enumerate}

\textbf{Proof.} From Proposition 5.2, the function $h_{\pi}(\tau)$ of a reaction pathway $\pi$ is either proportional to a delta measure (if $\pi$ has only one system node) or a convolution of exponential decay functions in form of $ce^{-c\tau}$ (if $\pi$ has more than one system nodes). Since $\alpha$ is a linear combination of $h_{\pi}(\tau)$, it is a nonnegative measure on $\mathbb{R}_{\geq 0}$. \\ 

Next, we show $\int_0^\infty \alpha(\tau)d\tau$ is bounded. For each reaction pathway $\pi$ we have
\begin{equation}
	\int_0^\infty h_{\pi}(\tau)d\tau
	= q_{\pi}\theta_{\pi}
	\int_0^\infty f_{T_{\pi}}(\tau)d\tau
	= q_{\pi}\theta_{\pi}.
\end{equation}
since $f_{T_{\pi}}$ is a conditional probability density function. Considering all reaction pathways in $\mathcal{F}(\mathcal{E},\mathcal{G}) \;\cup\; \mathcal{F}(\mathcal{G},\mathcal{G})$. The number of reaction pathways may be infinite, but we could group them according to their first reaction. Let $\mathcal{P}[\phi_c]$ denote the pathway collection of all reaction pathways that have $\phi_c$ as their first reaction. We claim that  
\begin{equation} \label{Eq5.5}
	\sum_{\pi\in\mathcal{P}[\phi_c]} 
	\kappa^*_{\pi} q_{\pi} \theta_{\pi}
	\leq \kappa^*_{\pi}.
\end{equation} 
The above inequality is a consequence of mass conservation along reaction pathways. By definition, all reaction pathways can only have first or last node as environmental node. Also, LRN can only have one upstream node of each reaction, hence the branches only happen for effluxes, not influxes. Therefore, the only source for biomass in the reaction pathway is the first reaction. This implies when tracking biomass along each pathways the biomass is non-increasing, where the ratio of biomass remained to the last reaction is $q_{\pi}\theta_{\pi}$. Consider all biomass transferred via $\phi_c$. Notice that all these biomass can undergoes different reaction pathways (sen example illustrated in below) but for each pathway the biomass is non-increasing as stated in above. Therefore, summation up for all reaction pathways we obtain the equation (\ref{Eq5.5}).  \\ 

\includegraphics[scale=0.45]{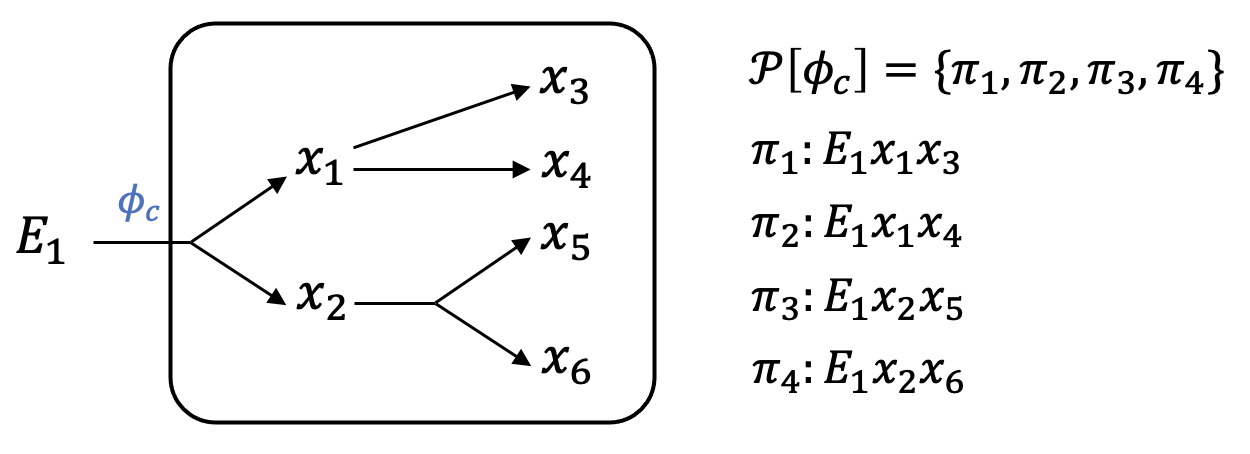} \\

By Definition 5.3, we have 
\begin{equation}
	\kappa^*_{\pi} = S[x_g,\phi_c] R[\phi_c] \, W_g^*
	\leq S[x_g, \phi_c] R[\phi_c],
\end{equation}
and therefore, consider all reactions $\phi_c$ as the first reactions in all $\pi \in \mathcal{F}(\mathcal{E},\mathcal{G}) \;\cup\; \mathcal{F}(\mathcal{G},\mathcal{G})$, and denote these reactions as the collection $K_{(1)}$, then  

\begin{equation}
	\int_0^\infty \alpha^*(\tau)d\tau
	= \sum_{\pi \in \mathcal{F}(\mathcal{E},\mathcal{G}) \;\cup\; \mathcal{F}(\mathcal{G},\mathcal{G})}
	\kappa_{\pi}^* q_{\pi}\theta_{\pi}
	\leq \sum_{\phi_c \in K_{(1)}} 
	S_{max}R[\phi_c]:= M_{\alpha}, 
\end{equation}
where $S_{max}:=\max{|S_{ij}|}$ for all $i,j$. Finally, we show the property about the Laplace transform of $\alpha$. For reaction pathways $\pi$ the arrival function $h_{\pi}$ is either proportional to a delta measure or a convolutions of exponential factors (denoted as $c_j e^{-c_j\tau}$). For the former case, we have $\mathcal{L}[h_{\pi}(\tau)](s)$ as a constant, while for the latter case,
\begin{equation}
	\mathcal{L}[h_{\pi}(\tau)](s)
	= \kappa^* \prod_j \mathcal{L}[c_j e^{-c_j\tau}]
	= \kappa^* \prod_j \frac{c_j}{s+c_j},
\end{equation}
which is a nonnegative and decreasing function for $s \geq 0$. The Laplace transform of $\alpha$ is a linear combination of $\mathcal{L}[h_{\pi}(\tau)]$. By (ii), $\mathcal{L}[\alpha(\tau)](s)|_{s=0}$ is finite, and hence $\mathcal{L}[\alpha(\tau)](s)$ is well-defined, nonnegative and non-increasing for $s \geq 0$. $\;\blacksquare$ \\
 
In the following, we state some propositions which will be used for the proof of Theorem B. \\



\textbf{Proposition 5.6} Consider an LRN and its reaction pathways. There is a universal bounds $\kappa^{\max}$ for all amplification rates $\kappa_{\pi}$. Namely, 
\begin{equation}
	\kappa_{\pi}(t-\tau) \leq \kappa^{\max}
\end{equation}
for all $\pi \in \mathcal{F}(\mathcal{E}, \mathcal{G}) \cup \mathcal{F}(\mathcal{G}, \mathcal{G})$ and for all $t \geq \tau \geq 0$. \\

\textbf{Proof.} From Proposition 4.11, exist nonnegative constants $B_1,B_2,B_3$ and positive constants $\delta, T$ such that 
\begin{equation} 
\begin{aligned}
		W_g(t-\tau) &\leq W_g^* + B_0, 
		\qquad &t-\tau \leq T, \\
		W_g(t-\tau) &\leq W_g^* + B_1/(t-\tau)+B_2 (t-\tau)^n e^{-\delta (t-\tau)} \qquad &t-\tau > T \\
\end{aligned}
\end{equation}
Choose $W_g^{\max}:= W_g^* + \max\{B_0, \;B_1/T+B_2T^n e^{-\delta T}\}$, then $W_g(t-\tau) \leq W_g^{\max}$ for all $t\geq \tau \geq 0$. \\

In Proposition 5.3, we have  
\begin{equation}
	\kappa_{\pi}(t-\tau) = 
	S[u_1, \omega_0]\, R[\omega_0] \, W_g(t-\tau). 
\end{equation}

For an LRN, the number of reaction is finite and there is a universal bound $R^{\max} > R_a$ for all linear flux functions in form of $J_a(X) = R_a X_k$. Also, there is a universal bounds $S^{\max} > |S_{ij}|$ for all entries in stoichiometry matrix. For all gatekeepers, we define $W^{\max} := \max \{ W_g^{\max}, x_g \in \mathcal{G} \}$. In this way, for any reaction pathway $\pi$, 
\begin{equation}
	\kappa_{\pi}(t-\tau) 
	\leq S^{\max}\,R^{\max}\,W^{\max} := \kappa^{\max}
\end{equation}
for all $t\geq \tau \geq 0$. $\;\blacksquare$ \\



\textbf{Proposition 5.7.} For an LRN with $\lambda>0$, the catalytic kernel $\alpha^{\bigstar}(\tau; t)$ can be expressed by two terms, i.e.
\begin{equation}
	\alpha^{\bigstar}(\tau; t)
	= K_1(t-\tau)\, \delta(\tau)
	 + K_2(\tau, t),
\end{equation} 
where the first term is contributed by reaction pathways with only one reaction, and the second term is contributed by reaction pathways with multiple reactions. Furthermore, exist a universal constant $C>0$ such that $K_1(t-\tau) <C$, $K_2(\tau,t)<C$ for all $\tau \geq 0, t \geq \tau$. \\

\textbf{Proof.} First, we define $K_1(t-\tau)$ as 

\begin{equation}
	K_1(t-\tau) := \sum_{\pi' \in \mathcal{F}(\mathcal{E}, \mathcal{G}) \cup \mathcal{F}(\mathcal{G}, \mathcal{G})}
    \kappa_{\pi'}(t-\tau),
\end{equation}
with $\pi'$ are reaction pathways that has only one system nodes. Also, define $K_2(\tau,t)$ as
\begin{equation}
\begin{aligned}
    K_2(\tau,t) := \sum_{\pi'' \in \mathcal{F}(\mathcal{E}, \mathcal{G}) \cup \mathcal{F}(\mathcal{G}, \mathcal{G})}
    \kappa_{\pi''}(t-\tau)\, h_{\pi''}(\tau),
\end{aligned}
\end{equation}
with $\pi''$ are reaction pathways that has two or more system nodes. It is clear that $\alpha^{\bigstar}(\tau; t)
	= K_1(t-\tau)\, \delta(\tau) + K_2(\tau, t)$. To proceed, we need to show both functions has universal bounds. \\
	
First, we consider $K_1(t-\tau)$. Since the number of system influx is finite, the collection of reaction pathways with single system nodes is also finite (no more than the number of system fluxes), $K_1(t-\tau)$ is a finite sum and can be bounded by a universal constant $m \kappa^{\max}:=C_{(1)}$ where $m$ is number of fluxes in the system. \\ 

Next, we consider $K_2(\tau,t)$. We first show that for $\pi: u_0 u_1 \cdots u_{L+1}$ with $L\geq 1$, the arrival function $h_{\pi}(\tau)$ is bounded. Let $x_k:=u_1, x_g:=u_{L+1}$ denote the two nodes in the system. For LRNs, the waiting time distribution from $x_k$ to $x_g$ can be calculate by exact formula. For an LRN $\frac{dX}{dt} = SJ(X)$, let us consider a \textit{reduced system} by deleting all system influxes and effluxes from $x_g \in \mathcal{G}$. Let $S_{(r)}$ and $J_{(r)}$ denote the stoichiometry matrix and flux function vector of the reduced system, then we have a reduces LRN with 
\begin{equation}
\frac{dX}{dt} = S_{(r)}J_{(r)}(X) := M_{(r)}X	
\end{equation}
The matrix $M_{(r)}$ denote a transition matrix with $x_g$ as absorbing states, which is proper for calculating the first-hitting time of biomass. By construction of $M_{(r)}$, there is no system influxes and hence all eigenvalues of the reduced system is equal or less to zero. \\

Let $P(x_k, x_g)(t_0,t_1)$ denote the waiting time distribution for a biomass unit start at $x_k$ at time $t_0$ and arrive $x_g$ at time $t_1 \geq t_0$. Then we have 
\begin{equation}
	P(x_k, x_g)(t_0, t_1) = \bold{e}_g^\dagger \, e^{M_{(r)}(t_1-t_0)}\, \bold{e}_k.
\end{equation}
where $\bold{e}_k$ denote a unit column vector with the $k^{th}$ entry equal to 1 and other entries equal to 0, and  $\bold{e}_g^\dagger$ denotes a row vector. Since $P(x_k, x_g)(t_0,t_1)$ is a finite linear combinations of functions in form of $(t_1-t_0)^p\,e^{\mu_k (t_1-t_0)}$ with $\mu_k \leq 0$ and $0\leq p\leq n$, it is bounded by a universal constant (denoted as $C_{kg}$) for all $t_1\geq t_0 \geq 0$. \\

Note that the sum of all arrival functions of reaction pathways from node $x_k$ to node $x_g$ is $P(x_k, x_g)$. We have  
\begin{equation}
\begin{aligned}
    K_2(\tau,t) &= \sum_{\pi' \in \mathcal{F}(\mathcal{E}, \mathcal{G}) \cup \mathcal{F}(\mathcal{G}, \mathcal{G})}
    \kappa_{\pi''}(t-\tau)\, h_{\pi''}(\tau) \\
    &\leq \kappa^{\max} \times
    \sum_{\pi'' \in \mathcal{F}(\mathcal{E}, \mathcal{G}) \cup \mathcal{F}(\mathcal{G}, \mathcal{G})} h_{\pi''}(\tau) \\
    &= \kappa^{\max} \times
    \sum_{x_g \in \mathcal{G}}
    \sum_{k=1}^n
	P(x_k,x_g)(0,\tau) \\
	& \leq \kappa^{\max} \times
    \sum_{x_g \in \mathcal{G}}
    \sum_{k=1}^n
    C_{kg} =: C_{(2)}
\end{aligned}
\end{equation}
We can choose $C := \max \{C_{(1)}, C_{(2)} \}$ and this completes the proof. $\;\blacksquare$ \\



\textbf{Proposition 5.8.} Consider an LRN with $\lambda >0$. We have 
\begin{equation}
    I:= \lim_{t \rightarrow\infty} \int_0^t
    \alpha^{\bigstar}(\tau; t)
	\frac{Z(t-\tau)}{Z(t)}\,d\tau
    =\tilde{\alpha}(\lambda) 
    \in \mathbb{R}
\end{equation}

\textbf{Proof.} To calculate $I$, we first express the catalytic kernel in two terms, i.e. \begin{equation}
	\alpha^{\bigstar}(\tau; t)
	= K_1(t-\tau)\, \delta(\tau)
	  + K_2(\tau, t).
\end{equation} 
where the term $K_1$ is contributed by reaction pathways with only one reaction, and the term $K_2$ is contributed by reaction pathways with multiple reactions. For convenience, we extend the definition of $K_2(\tau,t)$ by
\begin{equation} \label{eq.K2}
\begin{aligned}
    K_2(\tau,t) 
    &= \sum_{\pi'' \in \mathcal{F}(\mathcal{E}, \mathcal{G}) \cup \mathcal{F}(\mathcal{G}, \mathcal{G})}
    \kappa_{\pi''}(t-\tau)\, h_{\pi''}(\tau), \qquad t\geq \tau,\\
    &= 0, \qquad \text{otherwise}.
\end{aligned}
\end{equation}
where $\pi''$ are reaction pathways that has two or more system nodes. By direct calculation, 
\begin{equation}
\begin{aligned}
    I&:=\lim_{t\rightarrow\infty} 
	\int_0^t K_1(t-\tau) \delta(\tau)
	\frac{Z(t-\tau)}{Z(t)}\,d\tau 
    + \lim_{t\rightarrow\infty} 
	\int_0^t K_2(\tau,t)
    \frac{Z(t-\tau)}{Z(t)}\,d\tau \\
    &= \lim_{t\rightarrow\infty} 
    K_1(t) +
    \lim_{t\rightarrow\infty} 
	\int_0^\infty K_2(\tau,t)
    \frac{Z(t-\tau)}{Z(t)}\,d\tau \\
    &=: K_1^*+ I_2
\end{aligned}
\end{equation}
Here, $K_1^* := \lim_{t\rightarrow\infty} K_1(t)$ is the summation of $\kappa_{\pi}^*$ where $\pi$ has only one system nodes and belongs to $\mathcal{F}(\mathcal{E}, \mathcal{G}) \cup \mathcal{F}(\mathcal{G}, \mathcal{G})$. Now, if there is an integrable function $U(\tau)$ on $\tau \in [0, \infty)$ such that 
\begin{equation}
    K_2(\tau,t)
	\frac{Z(t-\tau)}{Z(t)} 
    \leq U(\tau), 
\end{equation}
for all $t \geq \tau$, then by Lebesgue's Convergence Theorem we can exchange the limit and integration in $I_2$, i.e.
\begin{equation}
\begin{aligned}
	I_2 &= 
	\lim_{t\rightarrow\infty} 
	\int_0^\infty 
	K_2(\tau,t)
	\frac{Z(t-\tau)}{Z(t)}\,d\tau \\
	& = \int_0^\infty
	\lim_{t\rightarrow\infty} 
	\bigg(
	K_2(\tau,t)
	\frac{Z(t-\tau)}{Z(t)}
	\bigg ) \,d\tau.
\end{aligned}
\end{equation}
To perform the above operation, we notice that from Proposition 5.7, $K_2(\tau,t)$ is bounded by a universal constant $C_1$ for all $t\geq0, \tau \geq 0$. Also, from Proposition 4.12, $\frac{Z(t-\tau)}{Z(t)}$ is bounded by $Me^{-\lambda\tau}$ with a constant $M>0$ for large $t$. Together, the integrand $K_2(\tau,t) \frac{Z(t-\tau)}{Z(t)}$ is bounded by a integrable function $U(\tau)$ on $[0,\infty)$, and this allows us to invoke Lebesgue's Convergence Theorem to interchange limit and integration. By Proposition 4.12, $\lim_{t\rightarrow\infty}\frac{Z(t-\tau)}{Z(t)} = e^{-\lambda\tau} $. Therefore,
\begin{equation}
    \lim_{t\rightarrow\infty} K_2(\tau,t)
	\frac{Z(t-\tau)}{Z(t)}
    =e^{-\lambda\tau} \times
    \sum_{\pi'' \in \mathcal{F}(\mathcal{E}, \mathcal{G}) \cup \mathcal{F}(\mathcal{G}, \mathcal{G})}
    \kappa_{\pi''}^*\, h_{\pi''}(\tau).
\end{equation}

On the other hand, we can express the Laplace transform $\tilde{\alpha}(s)$ as 
\begin{equation}
\begin{aligned}
    \tilde{\alpha}(s) 
    &= \int_0^\infty \alpha(\tau) e^{-s\tau}\,d\tau \\
    &=\int_0^\infty 
    \bigg(
    K_1^*\delta(\tau) + \lim_{t\rightarrow\infty}K_2(\tau,t)
    \bigg) e^{-s\tau}\,d\tau \\
    &=K_1^* + 
    \int_0^\infty  
    \bigg(
    \sum_{\pi'' \in \mathcal{F}(\mathcal{E}, \mathcal{G}) \cup \mathcal{F}(\mathcal{G}, \mathcal{G})}
    \kappa_{\pi''}^*\, h_{\pi''}(\tau) 
    \bigg)
    e^{-s\tau}\, d\tau 
\end{aligned}
\end{equation}
where $\pi''$ are reaction pathways that has two or more system nodes. Therefore, $\tilde{\alpha}(\lambda) = K_1^* + I_2$. By Proposition 5.5 (iii), $I=\tilde{\alpha}(\lambda)$ is a real number. $\;\blacksquare$ \\

\newpage


\textbf{Proposition 5.9.} Consider an LRN with $\lambda >0$ and express the equation as in the equation in Lemma 2.10,
\begin{equation} \label{E5.9A0}
	\frac{dZ}{dt}(t)= -\beta^{\bigstar}(t)Z(t)
	+ \int_0^t \alpha^{\bigstar}(\tau; t)\,Z(t-\tau)\,d\tau
	+ C_{ini}(t).
\end{equation}
Then, we have $\lim_{t\rightarrow\infty} \frac{C_{ini}(t)}{Z(t)} = 0$. \\

\textbf{Proof.}  Use Proposition 2.2(iii), Proposition 5.4, and Proposition 5.8 we can express the limit of $C_{ini}/Z$ as 
\begin{equation} \label{E5.9A}
\begin{aligned}
\lim_{t\rightarrow\infty} 
	\frac{C_{ini}(t)}{Z(t)}
	&= \lim_{t\rightarrow\infty} 
	\frac{1}{Z(t)}\frac{dZ}{dt} 
	+ \lim_{t\rightarrow\infty} \beta^{\bigstar}(t)
	+ \lim_{t\rightarrow\infty} 
	\int_0^t \alpha^{\bigstar}(\tau; t)
	\frac{Z(t-\tau)}{Z(t)}\,d\tau \\
	&= \lambda + \beta
	+ \tilde{\alpha}(\lambda),
\end{aligned}
\end{equation}
where $\lambda$, $\beta$, $\tilde{\alpha}(\lambda)$ are constants. Therefore, $\lim_{t\rightarrow\infty} \frac{C_{ini}(t)}{Z(t)}$ must also converge to a constant. By definition, $\frac{C_{ini}(t)}{Z(t)} \geq 0$. Suppose $\lim_{t\rightarrow\infty} \frac{C_{ini}(t)}{Z(t)} := c_0 > 0$, then $C_{ini}(t)$ is unbounded (since we assume $\lambda>0$, $Z(t)$ is unbounded by Proposition 2.2(iii)). This contradicts to the fact that $\int_0^\infty C_{ini}(t) \leq N(0)$ in Lemma 2.10. Therefore, $\lim_{t\rightarrow\infty} \frac{C_{ini}(t)}{Z(t)} = 0$. $\;\blacksquare$ \\


\vspace{20 pt}


\textbf{Proof of Theorem B.}  To prove Theorem B, we have to show that for LRNs replacing $\alpha^{\bigstar}(\tau; t), \beta^{\bigstar}(t)$ by $\alpha(\tau), \beta$ only introduces an error which is negligible compared with $Z(t)$. Specifically, we have to show that 
\begin{equation} \label{E.ThmB-0}
\frac{dZ}{dt}(t) = -\beta Z(t) 
+ \int_{0}^{t} \alpha(\tau)\,Z(t-\tau)\,d\tau
+ \gamma(t),
\end{equation}
with the error term $\lim_{t\rightarrow\infty} \frac{\gamma(t)}{Z(t)} = 0$. That is, by comparing (\ref{E5.9A0}) and (\ref{E.ThmB-0}), we have 
\begin{equation} \label{E.ThmB-2}
\begin{aligned} 
	\lim_{t\rightarrow\infty} \frac{\gamma(t)}{Z(t)} 
	= \lim_{t\rightarrow\infty}
	(\beta - \beta^{\bigstar}(t)) 
	+ \lim_{t\rightarrow\infty}
	\frac{1}{Z(t)} \int_0^t 
	(\alpha^{\bigstar}(\tau; t) - \alpha(\tau)) \,
	Z(t-\tau) \,d\tau 
	+ \lim_{t\rightarrow\infty}
	\frac{C_{ini}(t)} {Z(t)}.
\end{aligned}
\end{equation} 
There are three terms at the right hand side. For the first term, by Definition 5.4, $\lim_{t\rightarrow\infty}(\beta - \beta^{\bigstar}(t)) =0$. For the second term, by Proposition 5.8 we also have 
\begin{equation}
\begin{aligned}
	&\lim_{t\rightarrow\infty}
	\frac{1}{Z(t)} 
	\int_0^t 
	(\alpha^{\bigstar}(\tau; t) - \alpha(\tau)) \,
	Z(t-\tau) \,d\tau =0.
\end{aligned}
\end{equation} 
For the third term, by Proposition 5.9, $\lim_{t\rightarrow\infty}\frac{C_{ini}(t)} {Z(t)} = 0$. This completes the proof of Theorem B. The formula $\lambda+\beta=\tilde{\alpha}(\lambda)$ was proved in Proposition 5.9, equation (\ref{E5.9A}). $\blacksquare$ \\


\newpage
\section{Examples of LRNs}

In this section, we study various examples of LRNs and use both Euler and Lagrangian views to analyze their long-term growth rate $\lambda$. Throughout this section, we assume all nodes have the same biomass content, i.e. $\mathfrak{m}(x_k)=1$ for all $k$. We consider LRNs with at least one boundary influxes and no boundary effluxes -- this implies the system influx is larger than the system efflux, which implies $\lambda > 0$ and allow us to apply Theorem B in the analysis. We follow the notation in Theorem B, where $\alpha(\tau),\beta$ denote catalytic kernel, and gatekeeper degradation rate, respectively. The catalytic spectrum is denoted by $\tilde{\alpha}(s)$ or $A(s)$. \\

The following proposition is useful for estimating $\lambda$. \\


\textbf{Proposition 6.1}. Consider an LRN with $\lambda > 0$ and denote $f(s) := \tilde{\alpha}(s) - \beta$. Suppose we have two functions $f^+(s), f^-(s)$ such that 

\begin{enumerate}[label=(\roman*)]
	\item $f^+(s), f^-(s)$ are continuous and decreasing,
	\item $f^+(0) >0$ and  $f^-(0)>0$,
	\item $f^-(s) \leq f(s) \leq f^+(s)$ for $s \in [0,\infty)$.
\end{enumerate}
Then there exist unique $\lambda^-, \in \mathbb{R}^+$, $\lambda^+ \in \mathbb{R}^+$ such that $\lambda \in [\lambda^-, \lambda^+]$, where $\lambda^- = f^-(\lambda^-)$, $\;\lambda^+ = f^+(\lambda^+)$.  \\ 

\begin{center}
	\includegraphics[scale=0.35]{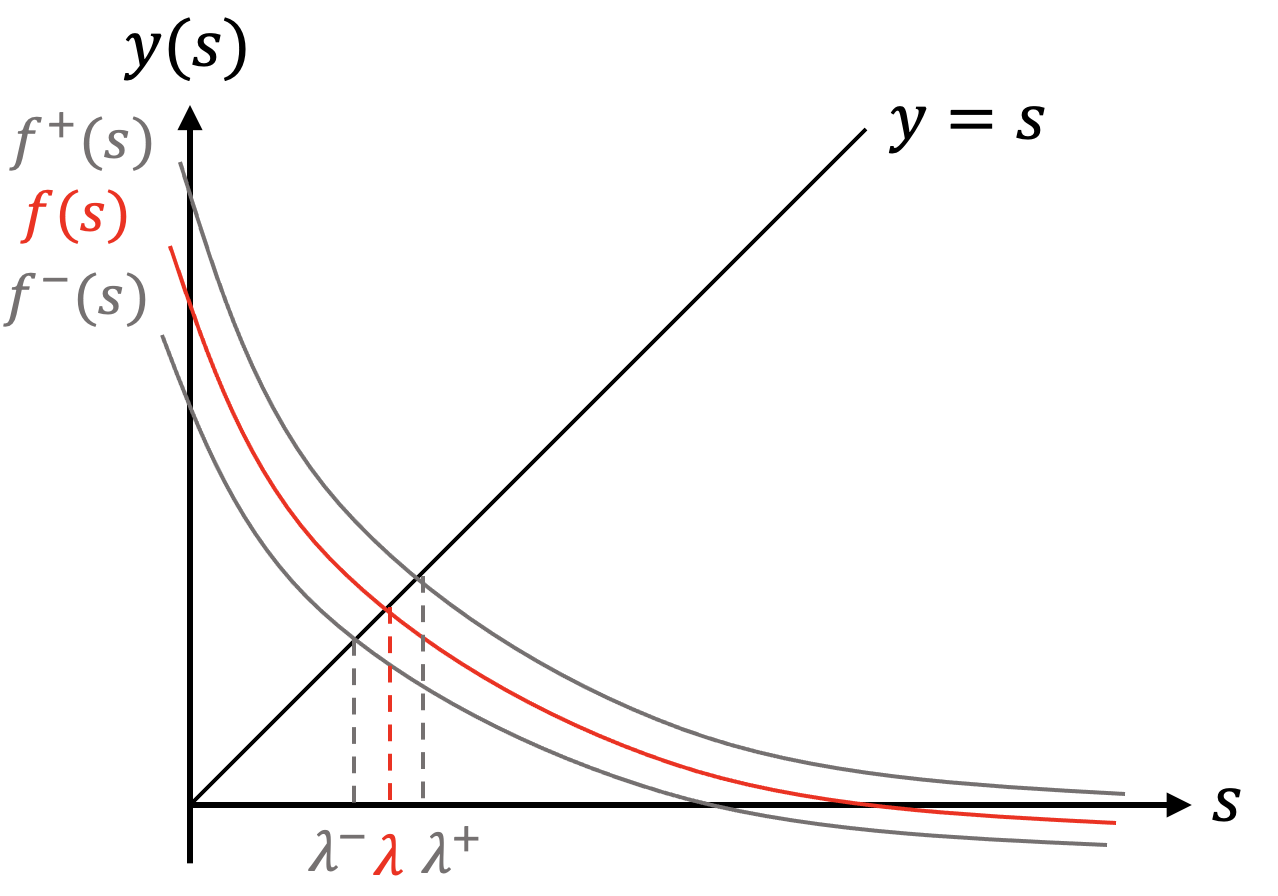} \\
\end{center}

\textbf{Proof}. Given an LRN, the catalytic spectrum $\tilde{\alpha}(s)$ is  decreasing by Proposition 5.5(iii). Hence $f(s)$ is also decreasing. Condition (i), (ii) guarantee that exist unique $\lambda^+ \in \mathbb{R}^+$ such that $\lambda^+=f^+(\lambda^+)$. Geometrically, it is the unique intersection point of $y=s$ and $y=f^+(s)$. To complete the proof, we need to show the intersection point $\lambda^+$ is no less than $\lambda$. Assume the contrary, where 
\begin{equation} \label{E6-1}
	f^+(\lambda^+) = \lambda^+ < \lambda,
\end{equation}
we show this leads to a contradiction. By condition (iii) we have $f^+(s) \geq f(s)$ and by condition (i) we have $f(s) \geq f(\lambda)$ for all $s \in [0,\lambda]$. This implies $f^+(s) \geq f(\lambda)$ for all $s \in [0,\lambda]$. Also, by Theorem B, $f(\lambda) = \lambda$. Together, we have 
\begin{equation}
	f^+(s) \geq \lambda, \qquad s \in [0,\lambda].
\end{equation}
So if $\lambda^+ < \lambda$, the formula above implies $f^+(\lambda^+) \geq \lambda $, contradicting to (\ref{E6-1}). Therefore we must have $\lambda^+ \geq \lambda$. The proof for $f^-(s)$ is similar. $\;\blacksquare$ \\ 

In the following examples, we assume all $\mathfrak{m}(x_j)=1$ for all nodes in all LRNs. \\


\newpage


\textbf{Example 6.2. LRNs with biomass dissipation in autocatalytic pathways}. In this example, we compare three LRNs (see the figure below). All LRNs have $\mathcal{G}=\{x_2\}$ as the gatekeeper node and have same reaction pathway $\pi \in \mathcal{F}(\mathcal{E},\mathcal{G})$, $\pi:E_1 x_1 x_2$. However, they have different dissipation along $\pi$: 
\begin{center}
	\includegraphics[scale=0.43]{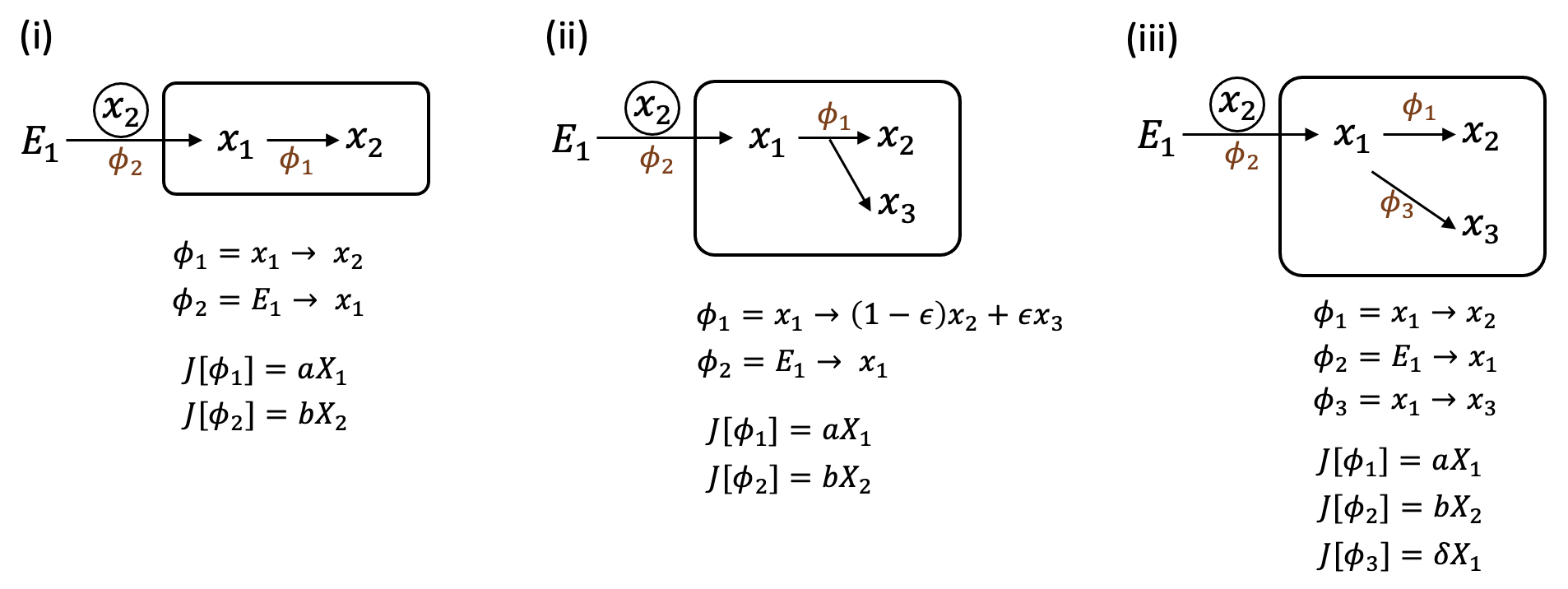} \\
\end{center}

In case (i), mass are transferred along $\pi$ and arrives $x_2$ without any dissipation. \\

In case (ii), the reaction $\phi_1$ has two downstream nodes. Based on the stoichiometric coefficients, only a fraction of biomass $(1-\varepsilon)$ arrive $x_2$, while the other fraction $\varepsilon$ arrives $x_3$. By Definition 5.1, the transmission efficiency for $\pi$ is $\theta_{\pi} = 1-\varepsilon$. \\

In case (iii), except $\phi_2$, there is another reaction $\phi_3$ competing with $\phi_2$. The relative flux magnitude is proportional to the reaction rates of reactions. Only a fraction of biomass ($\frac{a}{a+\delta}$) arrive $x_2$. By Definition 5.1, this is the pathway probability $q_{\pi} = \frac{a}{a+\delta}$. \\
\begin{center}
\begin{tabular}{l*{7}{|c}}
Case   & $\kappa_{\pi}$ & $f_{T_{\pi}}$& $q_{\pi}$ & $\theta_{\pi}$ & $h_{\pi}(\tau)=q_{\pi}\theta_{\pi}f_{T_{\pi}}$ & $\alpha(\tau)=\kappa_{\pi} h_{\pi}(\tau)$ & $\tilde{\alpha}(s)$ \\
\hline
(i) 	& $b$ & $ae^{-a\tau}$ & $1$ & $1$ 
& $ae^{-a\tau}$ & $bae^{-a\tau}$ & $\frac{ba}{s+a}$ \\
(ii)    & $b$ & $ae^{-a\tau}$ & $1$ & $1-\varepsilon$ 
& $a(1-\varepsilon)e^{-a\tau}$ & $b(1-\varepsilon)ae^{-a\tau}$ 
& $(1-\varepsilon)\frac{ba}{s+a}$ \\
(iii)   & $b$ & $(a+\delta)e^{-(a+\delta)\tau}$  & $\frac{a}{a+\delta}$ & $1$ 
& $a e^{-(a+\delta)\tau}$ & $b a e^{-(a+\delta)\tau}$
& $ \frac{ba}{s+a+\delta}$ \\
\end{tabular} 
\end{center}

\vspace{10 pt}
Since there is no effluxes from $x_2$, we have $\beta = 0$ for all cases. The long-term growth rate is given by the equation $\lambda = \tilde{\alpha}(\lambda)$. It is clear that the curves of $\tilde{\alpha}(s)$ for case (ii) and case (iii) are lower than the curve in case (i). Therefore, for case (ii), (iii), the curve $\tilde{\alpha}(s)$ intersect with $y=s$ at smaller values and has smaller $\lambda$ than in case (i) (see the figure below). \\

\begin{center}
	\includegraphics[scale=0.45]{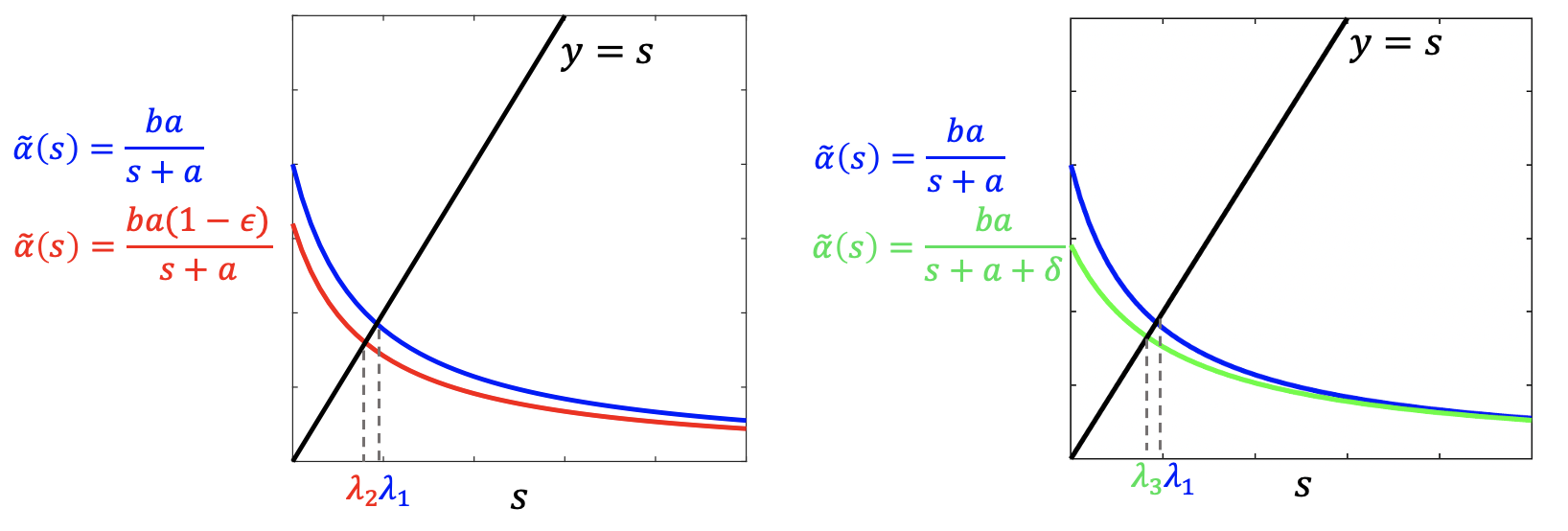} \\
\end{center}

For comparison, we also analyze the above three cases with Eulerian view. The matrices of the three cases are 
\begin{equation}
	M_1 = 
	\begin{pmatrix}
	-a & b \\
	a & 0 \\
	\end{pmatrix}, \;\;
	M_2 = 
	\begin{pmatrix}
	-a & b & 0 \\
	a(1-\varepsilon) & 0 & 0 \\
	a\varepsilon & 0 & 0
	\end{pmatrix}, \;\;
	M_3 = 
	\begin{pmatrix}
	-(a+\delta) & b & 0 \\
	a & 0 & 0 \\
	\delta & 0 & 0
	\end{pmatrix}.
\end{equation}

It can be verified that the characteristic equations $\det(M-\lambda I)=0$ gives the same formula as the Lagrangian view. We notice that the Lagrangian view is very convenience for studying the growth rate under network perturbations (for example, when $\varepsilon \rightarrow 0$ or $\delta \rightarrow 0$), while from the Eulerian view it is less clear, since matrix sizes are different in different cases. \\


\textbf{Example 6.3. LRNs with multiple gatekeepers}. In this example, we consider LRNs with two gatekeepers. \\
\begin{center}
	\includegraphics[scale=0.4]{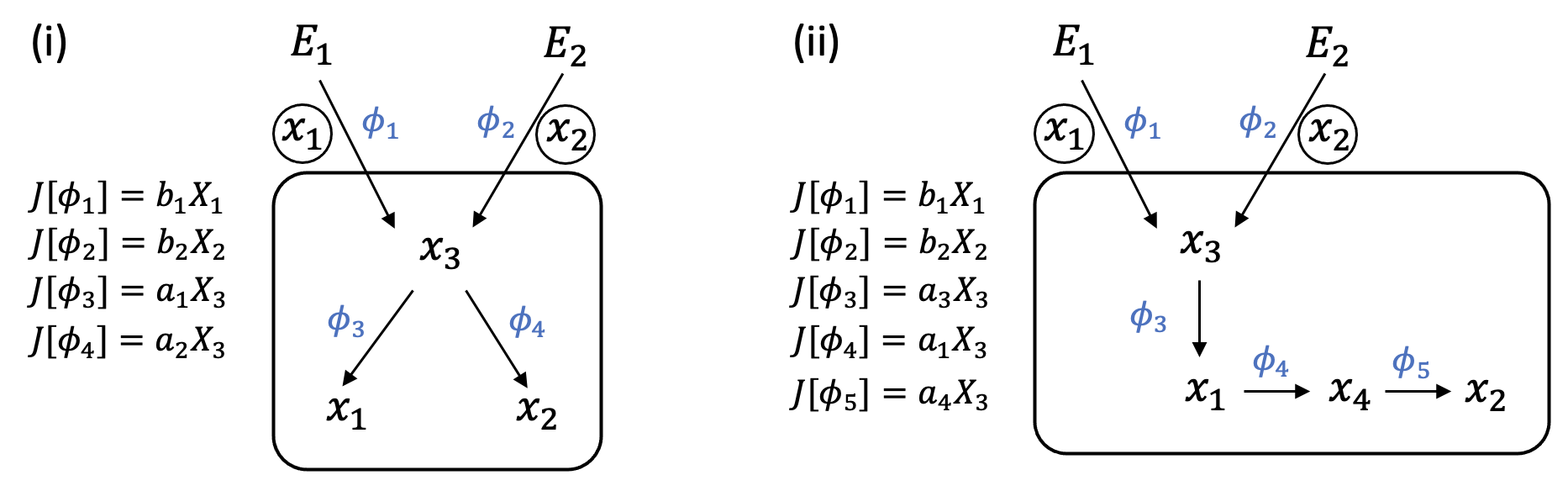} \\
\end{center}

\textbf{Case (i)}. The gatekeeper nodes are $\mathcal{G}=\{x_1, x_2\}$. There are four reaction pathways in $\mathcal{F}(\mathcal{E},\mathcal{G})$, from combinations of $\{E_1, E_2\}$ and $\{x_1,x_2\}$. The property of these pathways are summarized in the table below. Note that $W_1 := \frac{X_1}{X_1+X_2}=\frac{a_1}{a_1+a_2}$ in this case, since $x_1,x_2$ are synthesized proportionally from $x_3$ with rate $a_1, a_2$. Similarly, $W_2 := \frac{X_2}{X_1+X_2}=\frac{a_2}{a_1+a_2}$. \\
\begin{center}
\begin{tabular}{l*{7}{|c}}
         $\pi$  & $\kappa_{\pi}$& $q_{\pi}$ & $\theta_{\pi}$ & $\alpha_{\pi}(\tau)$ \\
\hline
$\pi_1:E_1\cdots x_1$ 
& $b_1W_1$ & $W_1$ & $1$ 
& $b_1W_1^2(a_1+a_2)e^{-(a_1+a_2)\tau}$ \\
$\pi_2:E_1\cdots x_2$ 
& $b_1W_1$ & $W_2$ & $1$ 
& $b_1W_1W_2(a_1+a_2)e^{-(a_1+a_2)\tau}$ \\
$\pi_3:E_2\cdots x_1$ 
& $b_2W_2$ & $W_1$ & $1$ 
& $b_2W_1W_2(a_1+a_2)e^{-(a_1+a_2)\tau}$ \\
$\pi_4:E_2\cdots x_2$ 
& $b_2W_2$ & $W_2$ & $1$ 
& $b_2W_2^2(a_1+a_2)e^{-(a_1+a_2)\tau}$ \\
\end{tabular} \\
\end{center}

Summing up all pathways and using $W_1+W_2=1$, we obtain 

\begin{equation}
	\alpha(\tau)=(a_1b_1+a_2b_2)\,e^{-(a_1+a_2)\tau}.
\end{equation}
Using the formula $\tilde{\alpha}(\lambda)=\lambda$, we obtain the equation $\lambda(\lambda+a_1+a_2) = a_1b_1+a_2b_2$. On the other hand, from the Eulerian view, we have 
\begin{equation}
	M = 
	\begin{pmatrix}
	0 & 0 & a_1 \\
	0 & 0 & a_2 \\
	b_1 & b_2 & -(a_1+a_2)  
	\end{pmatrix},
\end{equation}
and $\det(\lambda I-M)=0$ gives the same formula. \\

\textbf{Case (ii)}. The gatekeeper nodes are $\mathcal{G}=\{x_1, x_2\}$. There are two reaction pathways in $\mathcal{F}(\mathcal{E},\mathcal{G})$, e.g. $\pi_1:E_1\cdots x_1$ and $\pi_2:E_2\cdots  x_1$, and one reaction pathway in $\mathcal{F}(\mathcal{G},\mathcal{G})$, e.g. $\pi_3: x_1\cdots x_2$. Note that $\pi': E_2\,x_3\,x_1\,x_4\,x_2$ is also a reaction pathway but it does not belong to $\mathcal{F}(\mathcal{E},\mathcal{G})$ or $\mathcal{F}(\mathcal{G},\mathcal{G})$ by definition of first-hitting pathways. The property of these pathways are in the table below:

\begin{center}
\begin{tabular}{l*{7}{|c}r}
         $\pi$  & $\kappa_{\pi}$& $q_{\pi}$ & $\theta_{\pi}$ & $\alpha_{\pi}(\tau)$ \\
\hline
$\pi_1:E_1\cdots x_1$ 
& $b_1W_1$ & $1$ & $1$ 
& $b_1W_1\,a_3e^{-a_3\tau}$ \\
$\pi_2:E_2\cdots x_1$ 
& $b_2W_2$ & $1$ & $1$ 
& $b_2W_2\,a_3e^{-a_3\tau}$ \\
$\pi_3:x_1\cdots x_2$ 
& $a_1W_1$ & $1$ & $1$ 
& $a_1W_1\,a_4e^{-a_4\tau}$ \\
\end{tabular} \\
\end{center}

Summing up for all pathways, we obtain 

\begin{equation}
	\tilde{\alpha}(s)-\beta
=(b_1W_1^*+b_2W_2^*)\frac{a_3}{s+a_3} - 
a_1 W_1^*
\bigg( 1-  \frac{a_4}{s+a_4} \bigg).
\end{equation}

Up to this point, there is no obvious way to simplify the formula. It can be verified that the algebraic expression $W_1^*, W_2^*$ are quite complicated (by solving the linear ODE). However, we could acquire upper and lower bounds for $\lambda$. Define $b^+ := \max\{b_1,b_2\}, \; b^- := \min\{b_1,b_2\} $. We have a crude estimation 
\begin{equation}
	\lambda = \mu(Y^*) = b_1Y_1^* + b_2Y_2^* \in [b^-,b^+].
\end{equation}

Using the Lagrangian view, we get improved estimation. With the fact $W_1+W_2=1$, we have the inequality $\tilde{\alpha}(s)-\beta \leq  \frac{b^+a_3}{s+a_3} $. By Proposition 6.1, we have the upper bound 
\begin{equation} \label{E5-2}
	\lambda \leq \frac{a_3}{2} \bigg( 
\sqrt{1+\frac{4b^+}{a_3}} - 1
\bigg),
\end{equation}
which can be a better estimation than the crude upper bound ($b^+$). For example, when $b^+=3, a_3=4$, we have $\lambda \leq 3$ from the crude estimate, while we have $\lambda \leq 2$ from the inequality (\ref{E5-2}) in above. \\

In comparison, from the Eulerian view the matrix is

\begin{equation}
	M = 
	\begin{pmatrix}
	-a_1 & 0 & a_3 & 0 \\
	0 & 0 & 0 & a_4 \\
	b_1 & b_2 & -a_3 & 0 \\
	a_1 & 0 & 0 & -a_4   
	\end{pmatrix}.
\end{equation}
The characteristic equation $\det(M-\lambda I)$ is a nontrivial quartic equation and the algebraic solutions are complicated. There is no obvious way to estimate $\lambda$ from the Eulerian view. \\ 



\vspace{20 pt}

\textbf{Example 6.4. LRNs with multiple gatekeepers (continued)}. In this example, we consider LRNs with two gatekeepers of different configurations (see figure below). \\
\begin{center}
	\includegraphics[scale=0.4]{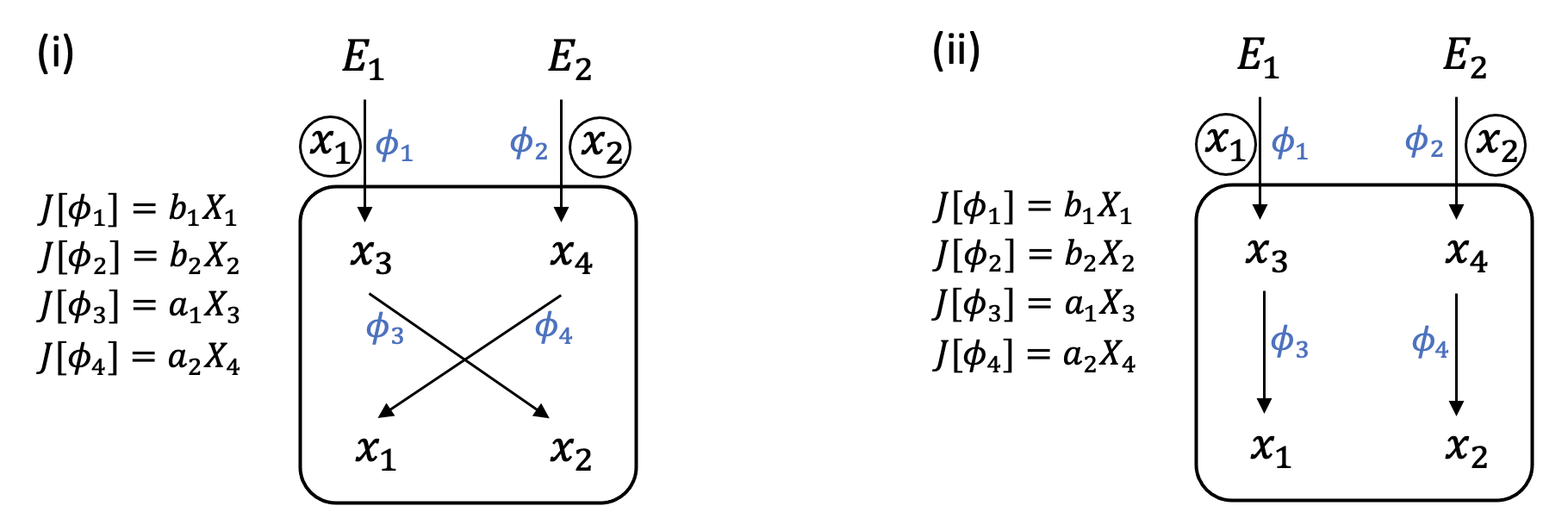} \\
\end{center}

\textbf{Case (i)}. The gatekeeper nodes are $\mathcal{G}=\{x_1, x_2\}$. There are two reaction pathways in $\mathcal{F}(\mathcal{E},\mathcal{G})$, and the gatekeepers $x_1,x_2$ promote the synthesis of each other. The quantities for reaction pathways are summarized in below (note that $W_1 := \frac{X_1}{X_1+X_2}$, $W_2 := \frac{X_2}{X_1+X_2}$). \\

\begin{center}
\begin{tabular}{l*{7}{|c}r}
         $\pi$  & $\kappa_{\pi}$& $q_{\pi}$ & $\theta_{\pi}$ & $\alpha_{\pi}(\tau)$ \\
\hline
$\pi_1:E_1\cdots x_2$ 
& $b_1W_1$ & $1$ & $1$ 
& $a_1b_1W_1e^{-a_1\tau}$ \\
$\pi_2:E_2\cdots x_1$ 
& $b_2W_2$ & $1$ & $1$ 
& $a_2b_2W_2e^{-a_2\tau}$ \\
\end{tabular} \\
\end{center}

\vspace{10 pt}

Unlike Example 6.3, case (i), we cannot solve $W_1, W_2$ easily, since they are promoted by different nodes. However, we still have $W_1+W_2 = 1$. Summing up $\pi_1$ and $\pi_2$ and calculate $\tilde{\alpha}(s)$, we have the following equation for the growth rate:

\begin{equation} \label{E6.4a}
	\tilde{\alpha}(\lambda)= W_1^* \frac{a_1b_1}{\lambda+a_1} 
+ W_2^* \frac{a_2b_2}{\lambda+a_2} = \lambda.
\end{equation}\

Since we do not solve $W_1^*,W_2^*$, we cannot solve $\lambda$ explicitly. However, we can qualitatively analyze (\ref{E6.4a}) geometrically with Proposition 6.1. Notice that the curve $\tilde{\alpha}(s)$ is bounded between two curves $\tilde{\alpha}_1(s) := \frac{a_1b_1}{s+a_1}$ and $\tilde{\alpha}_2(s) := \frac{a_2b_2}{s+a_2}$. Therefore, the growth rate is bounded between two growth rates $\lambda_1, \lambda_2$ satisfying $\lambda_1=\tilde{\alpha}_1(\lambda)$, $\lambda_2=\tilde{\alpha}_2(\lambda)$, respectively (see the figure below). This allows us to localize the range of $\lambda$:
\begin{equation}
\begin{aligned}
\lambda \in [\min\{\lambda_1, \lambda_2\}, \max\{\lambda_1, \lambda_2\}], \\	
\lambda_1 = \frac{a_1}{2}[\sqrt{1+(4b_1/a_1)}-1], \\
\lambda_2 = \frac{a_2}{2}[\sqrt{1+(4b_2/a_2)}-1]. \\
\end{aligned}
\end{equation}

In comparison, with the Eulerian view we calculate $\det(\lambda I - M)$ with 
\begin{equation}
	M = 
	\begin{pmatrix}
	0 & 0 & 0 & a_2 \\
	0 & 0 & a_1 & 0 \\
	b_1 & 0 & -a_1 & 0 \\
	0 & b_2 & 0 & -a_2 \\
	\end{pmatrix}.
\end{equation}
This gives explicit formula $\lambda^2(\lambda+a_1)(\lambda + a_2)=a_1a_2b_1b_2$. However, the analytical solution for this quartic equation is quite complicated and it is even unclear which root is the largest. Without numerical calculation, it is difficult to proceed for further analysis on $\lambda$. \\


\textbf{Case (ii)}. The gatekeeper nodes are $\mathcal{G}=\{x_1, x_2\}$. There are two reaction pathways in $\mathcal{F}(\mathcal{E},\mathcal{G})$. Unlike Case (i), the gatekeepers $x_1,x_2$ promote their own synthesis. The quantities for reaction pathways summarized in the table below: \\
\begin{center}
	\begin{tabular}{l*{7}{|c}r}
         $\pi$  & $\kappa_{\pi}$& $q_{\pi}$ & $\theta_{\pi}$ & $\alpha_{\pi}(\tau)$ \\
\hline
$\pi_1:E_1\cdots x_1$ 
& $b_1W_1$ & $1$ & $1$ 
& $a_1b_1W_1e^{-a_1\tau}$ \\
$\pi_2:E_2\cdots x_2$ 
& $b_2W_2$ & $1$ & $1$ 
& $a_2b_2W_2e^{-a_2\tau}$ \\
\end{tabular} \\
\end{center}

The catalytic spectrum $\tilde{\alpha}(s)$ has the same form as in Case (i), see equation (\ref{E6.4a}). However, the fixed point fraction $W_1^*, W_2^*$ (which depends on parameter $a_1,a_2,b_1,b_2$) are different from Case (i). \\

With the same idea in Case (i), we can locate the range of $\lambda$. Note that this LRN system can be  decoupled into two sub-systems, with nodes $\{x_1, x_3\}$ and $\{x_2,x_4\}$. Let $\lambda_1$ denote the growth rate of $X_1+X_3$ and $\lambda_2$ denote the growth rate of $X_2+X_4$. We could analyze each sub-system separately and have  
\begin{equation}
		\lambda_1 = \frac{b_1 a_1}{a_1 + \lambda_1},  \;\;
		\lambda_2 = \frac{b_2 a_2}{a_2 + \lambda_2}, \\
\end{equation}
and this gives $\lambda_1 = \frac{a_1}{2}(\sqrt{1+(4b_1/a_1)}-1),\;\;\lambda_2 = \frac{a_2}{2}(\sqrt{1+(4b_2/a_2)}-1)$. Since two sub-systems are not coupled, they grow separately with rates $\lambda_1, \lambda_2$, respectively. The long-term growth rate is $\lambda = \max\{\lambda_1, \lambda_2\}$. Note that in this case we have either $W_1^* = 1$ or $W_2^* =1$, and hence either $\kappa_{\pi_2} = 0$ or $\kappa_{\pi_1} = 0$. Therefore, the formula for the entire system is still valid. In comparison, using the Eulerian view, we calculate $\det(\lambda I - M)$ with 
\begin{equation}
	M = 
	\begin{pmatrix}
	0 & 0 & a_1 & 0 \\
	0 & 0 & 0 & a_2 \\
	b_1 & 0 & -a_1 & 0 \\
	0 & b_2 & 0 & -a_2 \\
	\end{pmatrix}.
\end{equation}\
This gives the formula $(\lambda(\lambda+a_1)-a_1b_1)(\lambda(\lambda+a_2)-a_2b_2) = 0$, which yields the identical result as above. \\


\vspace{20 pt}


\textbf{Example 6.5: LRNs with reversible pathways}. Consider the following LRNs (i) -- (iii). For all cases, we denote the ODEs by $\frac{dX}{dt} = MX$ and let $p_M(\lambda)$ be the characteristic polynomial of $M$. \\

\begin{center}
	\includegraphics[scale=0.5]{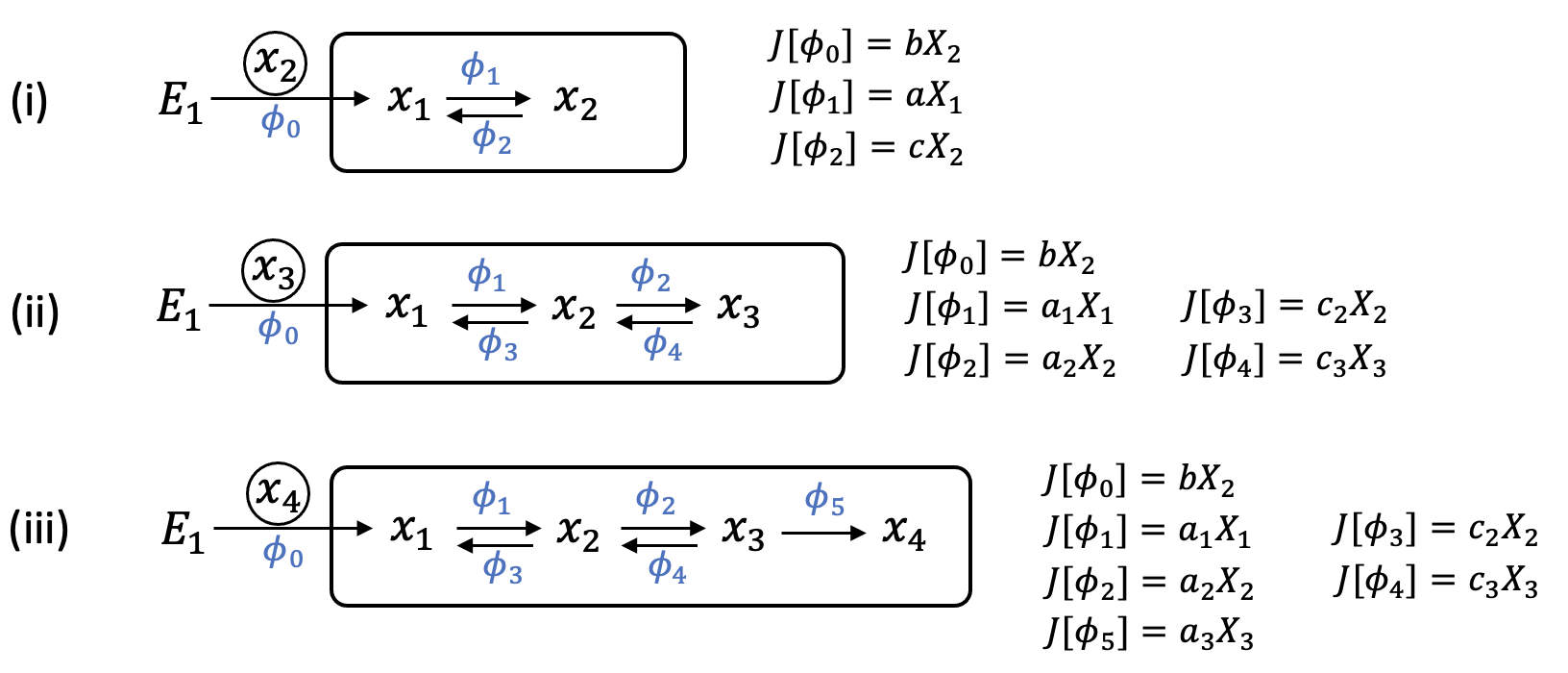} \\
\end{center} 

\textbf{Case (i)}. Using the Eulerian view, we have 
\begin{equation}
M = 
\begin{pmatrix}
-a & b+c \\
a &-c \\
\end{pmatrix},
\end{equation}
and hence $\lambda$ satisfies $p_M(\lambda) = (\lambda+a)(\lambda+c)-a(b+c)=0$. In the Lagrangian view, we consider the first-hitting reaction pathways of $\mathcal{G}=\{x_2\}$. There is only one pathway from the environment to the gatekeeper, that is, 
\begin{equation}
\mathcal{F(\mathcal{E},\mathcal{G})} = \{\pi_0\}, \;\;
\pi_0 = E_1 x_1 x_2.
\end{equation}
For recycling reaction pathways, there is one pathway starting from $x_2$ and back to $x_2$:
\begin{equation}
\mathcal{F(\mathcal{G},\mathcal{G})} = \{\pi_1\}, \;\;
\pi_1 = x_2 x_1 x_2.
\end{equation}
For $\pi_0$, we have $\kappa_{\pi_0}=b$, $h_{\pi_0}(\tau) = ae^{-a\tau}$ and therefore $A_{\pi_0}(s) = \frac{ba}{s+a}$. \\
For $\pi_1$, we have $\kappa_{\pi_1}=c$, $h_{\pi_1}(\tau) = ae^{-a\tau}$ and therefore $A_{\pi_1}(s) = \frac{ca}{s+a}$.  \\

Therefore, we have $\tilde{\alpha}(s) = \frac{(b+c)a}{s+a}$ and $\beta = c$. Using Theorem B, we have $\lambda + c = \tilde{\alpha}(\lambda)$, leading to the same formula as the Eulerian analysis. \\


\textbf{Case (ii)}. Using the Eulerian view, we have 
\begin{equation}
M = 
\begin{pmatrix}
-a_1 & c_2 & b \\
a_1 & -a_2-c_2 & c_3 \\
0 & a_2 & -c_3 \\
\end{pmatrix},
\end{equation}
and $\lambda$ satisfies $p_M = (\lambda+a_1)(\lambda+c_3)(\lambda+a_1+c_2) - a_1 a_2 b - a_2c_3(\lambda+a_1)-a_1c_2(\lambda+c_3)=0$. In general, it is not easy to analyze the three roots of $p_M$ algebraically. \\

In the Lagrangian view, we analyze the first-hitting reaction pathways. There is only one gatekeepers, i.e. $\mathcal{G}=\{x_3\}$. For autocatalytic pathways, they are in form of 
\begin{equation}
\mathcal{F(\mathcal{E},\mathcal{G})} = \{\pi_0,\pi_1,\cdots\};
\;\;
\pi_k = E_1 x_1 x_2 \, (x_1 x_2)_k\, x_3; 
\;\; k = 0,1,\cdots
\end{equation}

Note that there are infinite pathways, where the pathway $\pi_k$ made $k$ "extra loops" by reversible reactions $\phi_3$ and $\phi_2$. The pathway probability of $\pi_k$ decreases geometrically with $k$; specifically, $q[\pi_k] = (\frac{c_2}{c_2+a_2})^k(\frac{a_2}{c_2+a_2})$. Let $H_k(s) := \mathcal{L}[\,h_{\pi_k}(\tau)]$ and $A_{\pi}(s) := \mathcal{L}[\kappa_{\pi}h_{\pi}(\tau)]$. With the conditional waiting time formula (c.f. Proposition 5.5), we have 
\begin{equation}
\begin{aligned}
H_k(s) 
&= \frac{a_1}{s+a_1} 
\bigg( \frac{a_2+c_2}{s+a_2+c_2} \frac{a_1}{s+a_1}\bigg)^k
\frac{a_2+c_2}{s+a_2+c_2}
\bigg(\frac{c_2}{a_2+c_2}\bigg)^k
\frac{a_2}{a_2+c_2} \\
&= \frac{a_1}{s+a_1} 
\bigg( \frac{c_2}{s+a_2+c_2} \frac{a_1}{s+a_1}\bigg)^k
\frac{a_2}{s+a_2+c_2}.
\end{aligned}
\end{equation}

Summing up all $\pi_k$, and noticing that $\kappa[\pi_k] = b$ for all $k$, we have 

\begin{equation}
\begin{aligned}
\sum_{\pi \in \mathcal{F}(\mathcal{E},\mathcal{G})}
A_\pi(s)
&= b \sum_{k=0}^\infty H_k(s)
= b\;\frac{1}{1-r(s)}\;
\frac{a_1}{s+a_1} \;
\frac{a_2}{s+a_2+c_2}, \\
r(s) & := \frac{c_2}{s+a_2+c_2} \;\frac{a_1}{s+a_1}.
\end{aligned}
\end{equation}\

For the recycling pathways, they are all in the form of 
\begin{equation}
\mathcal{F(\mathcal{G},\mathcal{G})} = \{\pi'_0,\pi'_1,\dots\};
\;\;
\pi'_k = x_3 x_2 \,(x_1 x_2)_k\, x_3; 
\;\; k = 0,1,\dots
\end{equation}
Each pathway $\pi'_k$ made $k$ "extra loops" by reversible reaction $\phi_3$ and $\phi_2$. Using the similar analysis, and noticing that $\kappa[\pi'_k] = c_3$ for all $k$, we have

\begin{equation}
\begin{aligned}
\sum_{\pi' \in \mathcal{F}(\mathcal{G},\mathcal{G})}
A_{\pi'}(s)
&= \frac{c_3}{1-r(s)} \;
\frac{a_2}{s+a_2+c_2}.
\end{aligned}
\end{equation}\
Finally, we have 
$\tilde{\alpha}(s) = \sum_{\pi \in \mathcal{F}(\mathcal{E},\mathcal{G})}
A_{\pi}(s)+ \sum_{\pi' \in \mathcal{F}(\mathcal{G},\mathcal{G})}
A_{\pi'}(s)$ and $\beta = c_3$. Using the formula $\tilde{\alpha}(\lambda) = \lambda + \beta $, with some simplification, we obtain the same formula of $p_M(\lambda)=0$. \\


\textbf{Case (iii)}. Using the Eulerian view, we have 
\begin{equation}
M = 
\begin{pmatrix}
-a_1 & c_2 & 0 & b \\
a_1 & -a_2-c_2 & c_3 & 0\\
0 & a_2 & -c_3-a_3 & 0 \\
0 & 0 & a_3 & 0 
\end{pmatrix},
\end{equation}
and $\lambda$ satisfies 
\begin{equation}
	p_M(\lambda)=\lambda(\lambda+a_3+c_3)((\lambda+a_1)(\lambda+a_2+c_2)-a_1c_2) - \lambda(\lambda+a_1)a_2c_3 - a_1a_2a_3b = 0.
\end{equation}

Now we analyze the system with the Lagrangian view. In this case, there are two pairs of reversible reactions $\{\phi_1,\phi_3\}$ and $\{\phi_2,\phi_4\}$. The biomass can make "extra loops" with these reversible reactions, before first-hitting to the gatekeeper $x_4$. Let $\mathcal{V}_{k,\ell}$ denotes the pathway class, with reaction pathways in the form of 
\begin{equation}
E_1x_1x_2 \;
\{*\}_{k,\ell} 
\;x_3 x_4
\end{equation}
where the pathway $\{*\}_{k,\ell}$ contains total $k$ loops of $(x_1 x_2)$ and $\ell$ loops of $(x_3 x_2)$. By combinatorics, there are in total of $\binom{k+\ell}{k}$ ways to arrange the loops and the class $\mathcal{V}_{k,\ell}$ contains $\binom{k+\ell}{k}$ different reaction pathways, and all of them have exactly the same pathway probability $q$ and the same conditional waiting time. Therefore, different loops in sequential order can be viewed as "multiplicity" of reaction pathways. Let $\pi_{k,\ell}$ represent one reaction pathway in $\mathcal{V}_{k,\ell}$, and denote $H_{k,\ell}(s) := \mathcal{L}[h_{\pi_{k,\ell}}(\tau)]$. We have 
\begin{equation}
\begin{aligned}
	H_{k,l}(s) &= f(s) (r_1(s))^k (r_2(s))^{\ell}, \\
	f(s) & := \frac{a_1}{s+a_1} \frac{a_2}{s+a_2+c_2} \frac{a_3}{s+a_3+c_3}, \\
	r_1(s) & := \frac{a_1}{s+a_1} \,\frac{c_2}{s+a_2+c_2}, \\
	r_2(s) & := \frac{a_2}{s+a_2+c_2} \,\frac{c_3}{s+a_3+c_3}. \\ 
\end{aligned} 
\end{equation}

Using the formula 
\begin{equation}
\begin{aligned}
\sum_{k,\ell=0}^{\infty}
\binom{k+\ell}{k} r_1^k \;r_2^{\ell} =\frac{1}{1-r_1-r_2},
\end{aligned} 
\end{equation}
we have 
\begin{equation}
\begin{aligned}
\tilde{\alpha}(s)=
\sum_{\pi \in \mathcal{F}(\mathcal{E},\mathcal{G})}
A_\pi(s)
&= b \sum_{k,l=0}^\infty
\binom{k+\ell}{k} H_{k,\ell}(s)
= \frac{bf(s)}{1-r_1(s)-r_2(s)}. 
\end{aligned}
\end{equation}
In this case we have $\beta=0$ and hence $\tilde{\alpha}(\lambda) = \lambda$. After simplification, we obtained the same formula as $p_{M}(\lambda) = 0$. \\

\textbf{Summary.} In general, both Eulerian and Lagrangian views have advantages and disadvantages; researchers may benefit from both, by applying them in suitable scenarios. \\

(1) The Eulerian method is more straightforward, since the matrix determinant and characteristic polynomial can be calculated by an algebraic program. Typically, it is unclear which root of the characteristic polynomial has the largest real part, and hence numerical solutions for all roots are required. \\

(2) The Lagrangian method requires analysis on arrival functions and amplification rates of the relevant reaction pathways. The calculation is more complicated, but can also be programmed. The catalytic spectrum $\tilde{\alpha}(s)$ allows an comparison between LRNs of different dimensions, and provides geometric intuition. \\ 

(3) The Lagrangian method provides additional physical/biochemical meaning on biomass transfer dynamics; one obtains more information from the catalytic kernel $\alpha(\tau)$ then simply $\lambda$. It is more intuitive to study how $\alpha$ changes with kinetic constants in the system. In comparison, the Eulerian analysis requires knowledge from perturbation theory on the principle eigenvalue and matrix coefficients. \\


%
%
%


\newpage
\section{Interlude: time average and phase average}

In the next sections, we will generalize our ODE analysis from linear reaction networks (LRNs) to scalable reaction networks (SRNs). For LRNs, the solution $Y(t)$ converges to a constant vector $Y^* \in \Delta^{n-1}$ and the long-term growth rate converges to $\mu(Y^*)$, as shown in Theorem B. For SRNs, due to nonlinearity the long-term average $\lambda$ may not exist, and when it exists the attractors of $Y(t)$ can be much more complicated than a fixed point. Interestingly, with suitable conditions, the Lagrangian view still applies for many nonlinear systems. The concept of catalytic kernel $\alpha(\tau)$ and gatekeeper decay rate $\beta$ can be generalized by phase averaging on ergodic measure in some  SRNs. In this section, we summarize some known results with rigorous terminology. \\


\textbf{Definition 7.1.} Consider a systems ODE: $\frac{dY}{dt}=G(Y)$, where the solution $Y(t), \; t\in [0,\infty)$ is confined in bounded space $S \subseteq \mathbb{R}^n$. Given a function $f(Y(t)):\mathbb{R}^n \rightarrow \mathbb{R}$, we define its \textit{time average} by
\begin{equation}
	\langle f \rangle_t :=
	\lim_{T\rightarrow\infty} \frac{1}{T} \int_0^T f(Y(t))\,dt,
\end{equation}
if the limit exist. We define its \textit{phase average} with respect to a probability measure $\omega$ by
\begin{equation}
	\langle f \rangle_\omega :=
	\int_S f(Y) \,\omega(dY).
\end{equation}

Our central questions are: (1) Do these averages exist? (2) Can we find a probability measure $\omega$ to satisfy $\langle f \rangle_t = \langle f \rangle_\omega$? The candidate for such probability measure is obtained from the occurrence statistics of the trajectory. Let $\chi_B(Y)$ be the characteristic function of a set $B \subseteq \mathbb{R}^n$, with $\chi_B(Y)=1$ if $Y\in B$ and $\chi_B(Y)=0$ otherwise. We have the following definition: \\ 


\textbf{Definition 7.2.} Consider the same condition as in Definition 7.1, and let $B \subseteq S$  be a Borel set. Given a trajectory $Y(t), t \in[0,\infty)$, we define 
\begin{equation}
	\rho(B) :=
	\langle \chi_B \rangle_t
	= \lim_{T\rightarrow\infty} 
	\frac{1}{T}
	\int_{0}^T
	\chi_B(Y(t)) \,dt.	
\end{equation}
as its \textit{occurrence frequency measure}. If the above limit exists for every Borel set $B \subseteq S$, we say $Y(t)$ is $\rho$\textit{-regular}. \\

It is clear that $\rho$ is a probability measure if exists. Since we assume $Y(t)$ is confined in $S$, $\chi_S(Y(t)) = 1$ for all $t \geq 0$ and this implies 

\begin{equation}
\begin{aligned}
	\int_S \rho(dY) = \rho(S) 
	= \lim_{T\rightarrow\infty} 
	\frac{1}{T}
	\int_{0}^T
	\chi_S(Y(t)) \,dt	
	= \lim_{T\rightarrow\infty} 
	\frac{1}{T}
	\int_{0}^T
	1 \,dt
	= 1.
\end{aligned}
\end{equation}

Note that Definition 7.2 depends on the systems ODE as well as the initial condition $Y(0)$, which determines the trajectory $Y(t)$. For example, if the ODE has a global fixed-point attractor $Y^*$, then for all initial conditions the trajectory converge to $Y^*$ and we have $\rho=\delta(Y-Y^*)$ as the Dirac's $\delta$-measure on $Y^*$. In general, an occurrence probability measure $\rho$ can be defined on limit cycle, limit torus, or even chaotic attractors. \\


\textbf{Proposition 7.3.} Consider the same condition as in Definition 7.1. If $Y(t)$ is $\rho$-regular, and $f$ is continuous, then both $\langle f \rangle_\rho$ and $\langle f \rangle_t$ exist and are equal to each others. \\

\textbf{Proof.} Since $Y(t)$ is $\rho$-regular, the probability measure $\rho$ is well-defined. Also, $f$ is continuous on bounded space $S$, the integral $\langle f \rangle_\rho$ is well-defined. \\

Next we show the existence of $\langle f \rangle_t$, which involves three steps. First, fixed $\varepsilon_1 > 0$, we can partition $S$ into disjoint regions $\mathcal{B}(\varepsilon_1):= \{B_1,\cdots,B_{M(\varepsilon_1)} \}$ such that for each $B_k$,
\begin{equation}
	\sup_{Y\in B_k} f(Y) - 
	\inf_{Y\in B_k} f(Y) 
	\leq \varepsilon_1.
\end{equation}
By the continuity of $f$ and boundedness of $S$, the members of $\mathcal{B}(\varepsilon_1)$ can be finite. Let 
\begin{equation}
	f^{(k)} := \frac{1}{2} 
	\bigg( 
	\sup_{Y\in B_k} f(Y) + \inf_{Y\in B_k} f(Y) 
	\bigg),
\end{equation}
for all $k=1, \cdots, M(\varepsilon_1)$. This implies  
\begin{equation} \label{E7.3A}
	\bigg |
	\int_S f(Y)\,\rho(dY) 
	-
	\sum_{B_k \in \mathcal{B}} 
	f^{(k)}\, 
	\rho(B_k) \;
	\bigg | 
	\leq 
	\rho(S) \,\varepsilon_1 
	= \varepsilon_1, 
\end{equation}
since $\rho$ is a probability measure. Second, since each $\rho(B_k)$ is a convergence of long-term statistics, given $\varepsilon_2$ there exist $T^*>0$ such that 
\begin{equation}
	\bigg| \,
	\rho(B_k) 
	- 
	\frac{1}{T} \int_0^T \chi_{B_k}(Y(t))\,dt\;
	\bigg|
	\leq \varepsilon_2,
\end{equation}
for all $T>T^*$ and all $k=1, \cdots, M(\varepsilon_1)$. Summing up for all $B_k \in \mathcal{B}(\varepsilon_1)$, this implies 
\begin{equation} \label{E7.3B}
	\bigg |
	\sum_{B_k \in \mathcal{B}} 
	f^{(k)}\, 
	\rho(B_k) \; 
	-
	f^{(k)}
	\frac{1}{T} \int_0^T \chi_S(Y(t))\,dt\;
	\bigg | 
	\leq f_{max} M(\varepsilon_1)\, \varepsilon_2,
\end{equation}
for all $T>T^*$, where $f_{max}:= \max_{x \in S}\{|f(x)|\}$. Third, by the earlier choice of $B_k$, for each region $B_k$ we have $|f^{(k)}-f(Y(t))| \leq \varepsilon_1$ for each region $B_k$. Summing up all $B_k$ and weighted with the characteristic function $\chi_{B_k}$, we have 
\begin{equation} \label{E7.3C}
\begin{aligned}
    &\bigg | \bigg (
    \sum_{B_k \in \mathcal{B}(\varepsilon_1)} 
	f^{(k)} \frac{1}{T}\int_0^T \chi_{B_k}(Y(t))\,dt\; \bigg )
	-\frac{1}{T} \int_0^T f(Y(t))\,dt\;
	\bigg |  \\
    = &\bigg | 
    \sum_{B_k \in \mathcal{B}(\varepsilon_1)} 
    \bigg ( 
	f^{(k)} \frac{1}{T}\int_0^T \chi_{B_k}(Y(t))\,dt\; 
	-  \frac{1}{T} \int_0^T f(Y(t))\,\chi_{B_k}(Y(t))\,dt\; \bigg )
	\bigg |  \\
    \leq & 
    \sum_{B_k \in \mathcal{B}(\varepsilon_1)} 
    \bigg (
    |f^{(k)}-f(Y(t))| \times
    \bigg |  \frac{1}{T}\int_0^T \chi_{B_k}(Y(t))\,dt\; \bigg | \bigg )  \\
    \leq &\,\varepsilon_1\,
    \sum_{B_k \in \mathcal{B}(\varepsilon_1)} 
    \bigg |  \frac{1}{T}\int_0^T \chi_{B_k}(Y(t))\,dt\; \bigg |  \\
    = &\, \varepsilon_1 \, 
    \bigg |\frac{1}{T}\int_0^T \chi_{S}(Y(t))\,dt  \bigg | \\
    = &\, \varepsilon_1 \,
    \bigg |\frac{1}{T}\int_0^T 1\,dt  \bigg | 
    = \varepsilon_1.
\end{aligned}
\end{equation}
Now we can show $\langle f \rangle_\rho = \langle f \rangle_t$. Given every $\varepsilon>0$, from above derivation we can choose $T>T^*$ and a partition $\mathcal{B}(\varepsilon_1)$ fine enough such that 
\begin{equation}
\varepsilon_1 < \frac{\varepsilon}{3}, \qquad
\varepsilon_2 < \frac{\varepsilon}{3 |f_{max}| M(\varepsilon_1)}, \qquad
\end{equation}
Combining equations (\ref{E7.3A}),(\ref{E7.3B}),(\ref{E7.3C}), we have 
\begin{equation} 
	\bigg |
	\int_S f(Y)\,\rho(dY) 
	- 
	\frac{1}{T}
	\int_0^T f(Y(t))\,dt\;
	\bigg | 
	< \varepsilon.  
\end{equation}
$\;\;\blacksquare$ \\

Note that it is possible that $\langle f \rangle_t$ exists but $\langle f \rangle_\rho$ does not exist. A trivial example is for constant function $f(Y) = C$ for all $Y \in S$, where $\langle f \rangle_t = C$ but $\langle f \rangle_\rho$ may not exist. It is also possible that both averages do not exist, showing by the following example. \\


\textbf{Example 7.4. (May-Leonard) (\cite{may_nonlinear_1975})} Consider the systems ODE
\begin{equation}
	\begin{aligned}
		Y_1'(t) &= Y_1 (1-Y_1-\alpha Y_2 - \beta Y_3),  \\
		Y_2'(t) &= Y_2 (1-\beta Y_1- Y_2 - \alpha Y_3), \\
		Y_3'(t) &= Y_3 (1- \alpha Y_1- \beta Y_2 - Y_3).\\
	\end{aligned}
\end{equation}

The system is symmetric under cyclic permutation of ${Y_1,Y_2,Y_3}$. For parameter $\beta < 1 < \alpha$ and $\alpha + \beta > 2$, every trajectory $Y(t)$ with $Y(0) \in \mathbb{R}^3_{>0}$ converges to a heteroclinic cycle (HC). The HC consists of three saddle points $Y_1^*=(1,0,0)$, $Y_2^*=(0,1,0)$, $Y_3^*=(0,0,1)$ and their connections, while $Y(t)$ stay closed to  each saddle point for longer and longer times (see the figure in below). \\

The trajectory $Y(t)$ is non-periodic, and it has been show (\cite{gaunersdorfer_time_1992}) that the time average $\langle Y_j \rangle_t$ for $j=1,2,3$ do not converge for the HC trajectories. Actually, the time average does not converge for most of the functions along the HC trajectory, except for those satisfying $f(Y_1^*)=f(Y_2^*)=f(Y_3^*)=C$ (which have $\langle f \rangle_t=C$). In general, HC trajectories are not $\rho$- regular and the phase average $\langle f \rangle_{\rho}$ may not exist. \\


\begin{center}
	\includegraphics[scale = 0.45]{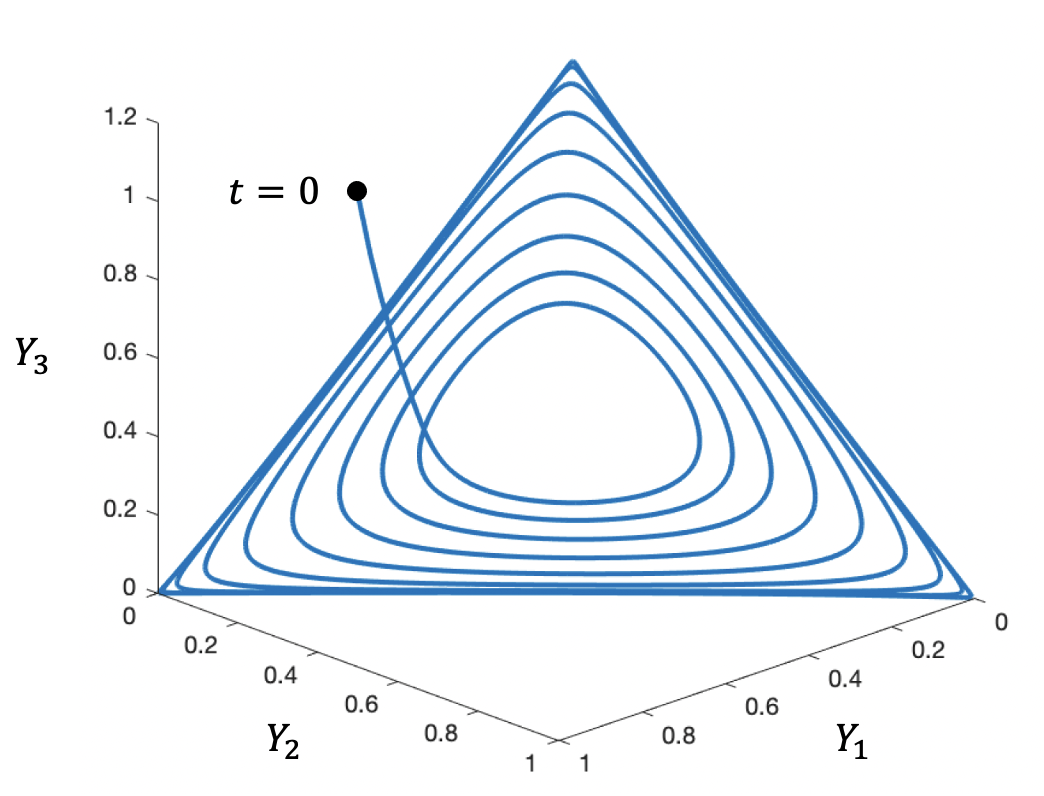}
\end{center}


 
Since the definition of $\rho$-regular depends on the initial conditions, one needs to check each trajectory individually. It would be nice to have a way to check \textit{a collection of initial conditions} about their $\rho$-regularity. This question is related to the ergodic theory (see e.g. \cite{mane_ergodic_2011}, \cite{cornfeld_ergodic_2012}, \cite{walters_introduction_2000}), where we will introduce some terminology (adopted for continuous dynamics) in below. \\

\textbf{Definition 7.5.} Let $\Gamma: S \rightarrow S$ be a mapping. Given a set $A \subseteq S$, we denote the image of $A$ under the action of $\Gamma$ as
\begin{equation} 
	\Gamma A := 
	\{Y: Y=\Gamma (Y_0), \;Y_0\in A\}.
\end{equation}

A probability measure $\omega$ is \textit{invariant under} $\Gamma$, if for every Borel set $A\subseteq S$ we have $\omega(\Gamma A) = \omega(A)$. An \textit{ergodic measure} $\omega$ is an invariant probability measure with no invariant subset of intermediate $\omega$-measure, i.e. there is no $A \subseteq S$ with $0<\omega(A)<1$ such that $Y\in A$ implies $\Gamma (Y)\in A$. \\
 
 
 \textbf{Definition 7.6.} Consider a systems ODE: $Y'(t)=G(Y)$ with $Y(t)$ defined for all $t\geq 0$. For a set $A$ and a point $Y_0$, we define their distance $d(A,Y_0)$ by 
 \begin{equation}
 	d(A,Y_0) := \inf _{Y \in A}
 	\{ \lVert Y - Y_0\lVert_{2}\},
 \end{equation}
 where $\lVert . \lVert_{2}$ represents the Euclidean norm. A closed set $A\subseteq S$ is called an \textit{attractor} if the following conditions are satisfied for every $t$:
 
 \begin{enumerate}[label=(\roman*)]
 	\item $Y(t) \in A \Rightarrow Y(t+\tau) \in A$ for $\tau \geq 0$.  
 	\item Exist an open neighborhood $B \supset A$ such that $Y(t) \in B \Rightarrow Y(t+\tau) \in B$ for $\tau \geq 0$.
 	\item If $Y(t) \in B \Rightarrow \lim_{t\rightarrow\infty} d(A,Y(t)) =0 $. 
 	\item There is no proper subset of $A$ satisfies (i),(ii).
 \end{enumerate}
 
The open set $B$ is called the \textit{basin of attraction} of attractor $A$, denoted as $B(A)$.  \\

\vspace{10 pt}

\textbf{Proposition 7.7.} Consider a systems ODE $\frac{dY}{dt}=G(Y)$, where the solution $Y(t), \; t\in [0,\infty)$ is confined in bounded space $S \subseteq \mathbb{R}^n$. Define 
\begin{equation}
	\begin{aligned}
		\Gamma^{\tau}: \,S &\rightarrow S, \\
					     Y(t) &\mapsto Y(t+\tau).
	\end{aligned}
\end{equation}
If there exists an ergodic measure $\rho$ under $\Gamma^{\tau}$,  and an attractor $A\subseteq S$ with $\rho(A)=1$, then for every $Y(0) \in B(A)$ the trajectory $Y(t)$ is $\rho$-regular. \\

\textbf{Proof.} By the property (ii) of basin of attractor, the open set $B(A)$ is invariant under the mapping of $\Gamma^{\tau}$ and we can consider the restriction of $\Gamma^{\tau}$ on $B(A)$:
 
\begin{equation}
	\begin{aligned}
		\Gamma^{\tau}|_{B(A)}: B(A) &\rightarrow B(A), \\ 
							   Y(t) &\mapsto Y(t+\tau).
	\end{aligned}
\end{equation}

For the mapping $\Gamma^{\tau}|_{B(A)}$, consider the open set $E := B(A) \backslash A$. Since every $Y(t) \in E$ is attracted to $A$, the occurrence frequency measure $\rho(E)$ is zero. Therefore, there is no invariant probability measure on $E$. Therefore, by assumption $\rho$ is the only ergodic measure on $B(A)$. By the property of unique ergodicity (see (\cite{walters_introduction_2000}), Theorem 6.19), $Y(t)$ is $\rho$-regular. $\;\;\blacksquare$ \\

\newpage
\section{Biomass transfer on scalable reaction networks (SRNs)}

Our next goal is to extend the DDE formulation from linear reaction networks (LRNs) to scalable reaction networks (SRNs). We start with the following definition. \\

 
\textbf{Definition 8.1.} In a scalable reaction network (SRN), every reaction flux $J_a(X)$ satisfies the following conditions: 
\begin{enumerate}[label=(\roman*)]
	\item $J_a(X)$ is positive on $\mathbb{R}^n_{>0}$ and continuously differentiable on $\mathbb{R}^n_{\geq0} \backslash \{0\}$.
	\item $J_a(X)$ is upstream-limited, i.e. if $S_{ka}<0$, then $J_a(X) = 0$ whenever $X_k =0$.
	\item $J(cX) = cJ(X)$ for all $c \geq 0$.
\end{enumerate}

In this work, we also require \textit{auxiliary conditions} that allow us to analyze the ODE in Lagrangian view. 
 
\begin{enumerate}[label=(\roman*), start=4]
	\item If a reaction $\phi_a$ has an upstream node in the environment, then exist a system node $x_g$ such that $J[\phi_a] =0$ whenever $X_g = 0$. 
	\item The trajectory $Y(t)$ is $\rho$-\textit{regular} with  an ergodic measure $\rho$ on $\Delta^{n-1}$.
	\item The system has long-term growth rate $\lambda > 0$. 
\end{enumerate}

\textbf{Note.} Flux functions that satisfies condition (i)-(iii) are called \textit{scalable flux functions}. Properties of SRNs have been studied (\cite{lin_origin_2020}). By SRN condition (iii), the original system dynamics $\frac{dX}{dt}= SJ(X) := F(X)$ implies a \textit{rescaled system dynamics} $\frac{dY}{dt} = F(Y)-\mu(Y)Y$ where $Y=X/N $ is the normalized biomass vector living in unit simplex space $\Delta^{n-1}$. \\

In this work, we will only consider the cases where the long-term average and $\rho$ exist. The following Lemma allows us to analyze scalable flux functions under Lagrangian view: \\


\textbf{Lemma 8.2.} Consider a scalable flux functions $J_a(X)$. If $X_p \rightarrow 0$ implies $J_a(X) \rightarrow 0$, then we can expressed $J_a(X)$ as
\begin{equation}
J_a(X) = R_{a,p}(Y) X_p,
\end{equation}
with $R_{a,p}(Y)$ defined and continuous on $\Delta^{n-1}$. \\ 

\textbf{Proof.} We define
\begin{equation}
	\begin{aligned} \label{L8.2A}
		R_{a,p}(Y) &:= 
		\frac{J_a(X)}{X_p} = \frac{J_a(Y)}{Y_p},
		 \qquad &\text{ for } Y_p>0, \\
		&:= \frac{\partial J_a(Y)}{\partial Y_p}\bigg |_{Y_p=0}, 
		 \qquad &\text{ for } Y_p=0. \\
	\end{aligned}
\end{equation}

For $Y_p > 0$, the expression (\ref{L8.2A}) is well-defined. Also, by scalable condition (i), $J_a(X)$ is continuous on $\mathbb{R}^n_{\geq 0}\backslash \{0\}$. Hence, $J_a(Y)$ is continuous when $Y_p > 0$, and this implies the continuity of $R_{a,p}(Y)$ when $Y_p > 0$. \\ 

Next, we show that $R_{a,p}(Y)$ is defined and continuous for the region with $Y_p = 0$. First we notice that 
\begin{equation} \label{L8.2C}
\begin{aligned}
R_{a,p}(Y) \big |_{Y_p=0} 
= \frac{\partial J_a(Y)}{\partial Y_p}\bigg |_{Y_p=0} 
&= \frac{\partial J_a(X)}{\partial X_p}\bigg |_{X_p=0, \,X_p \in \Delta^{n-1}}, 
\end{aligned}
\end{equation}
By scalable condition (i), $J_a(X)$ is continuously differentiable on $\mathbb{R}^n_{\geq 0}\backslash \{0\}$, and hence $R_{a,p}(Y)|_{Y_p=0}$ is well-defined. To show $R_{a,p}(Y)$ is continuous at $Y_p=0$, we calculate the limit as $Y_p \rightarrow 0$. Note that   
\begin{equation} \label{L8.2C}
\begin{aligned}
\lim_{Y_p \rightarrow 0} \frac{J_a(Y)}{Y_p}
= \lim_{Y_p \rightarrow 0} \frac{J_a(Y)-0}{Y_p-0}
= \lim_{Y_p \rightarrow 0} \frac{\partial J_a(Y)}{\partial Y_p}
&= \lim_{X_p \rightarrow 0, \,X_p \in \Delta^{n-1}} 
\frac{\partial J_a(X)}{\partial X_p} \\
&= \frac{\partial J_a(X)}{\partial X_p}\bigg |_{X_p=0, \,X_p \in \Delta^{n-1}},
\end{aligned}
\end{equation}
where the last equality is again by scalable condition (i). This shows $R_{a,p}(Y)$ is continuous at $Y_p=0$. $\;\;\blacksquare$ \\ 


Unlike in LRNs where reactions can only have one upstream node, reactions in SRNs can have multiple upstream nodes. We will use SRN condition (ii), (iv) and the above Lemma to describe the biomass transfer process on an \textit{equivalent SRN}. To illustrate the basic idea, let us consider the following example: \\


\textbf{Example 8.3.} Assume $\mathfrak{m}(x_j)=1$ for all nodes $x_j$ and consider a reaction 
\begin{equation}
	\phi_a: p_1 x_1 + p_2x_2 \rightarrow q_3 x_3 + q_4 x_4.
\end{equation}
We assign a scalable flux function to $\phi_a$ as $J[\phi_a] = \frac{cX_1X_2}{N}$. Now, we can decompose $\phi_a$ into two reactions, with the same flux functions: 
\begin{equation}
	\begin{aligned}
		&\phi_{a,1}:  p_1 x_1 \rightarrow 
		\big(\frac{p_1}{p_1+p_2}\big) q_3 x_3 
		+ \big(\frac{p_1}{p_1+p_2}\big) q_4 x_4, \;\;
		&J[\phi_{a,1}] = (cY_2)X_1 \\
		&\phi_{a,2}:  p_2 x_2 \rightarrow 
		\big(\frac{p_2}{p_1+p_2}\big) q_3 x_3 
		+ \big(\frac{p_2}{p_1+p_2}\big) q_4 x_4, \;\;
		&J[\phi_{a,2}] = (cY_1)X_2 \\
	\end{aligned}
\end{equation}\

The two reactions are both scalable and have single upstream node. It is clear that $J[\phi_a]=J[\phi_{a,1}]=J[\phi_{a,2}]$, and the biomass flux $\phi_a$ is decomposed into biomass fluxes on $\phi_{a,1}$ and $\phi_{a,2}$. Therefore, the ODEs before and after the decomposition are the same, and hence the original and decomposed systems have the same dynamics. \\ 

Note that the purpose to decompose one reactions with multiple upstream node into multiple reactions with single upstream node is for exploring the similarity between LRNs and SRNs. In the following, we denote an SRN as $\mathcal{N}(x, S, J)$ where $S$ is the stoichiometry matrix and $J$ is the collection of scalable flux function.  \\


\textbf{Definition 8.4 (Mono-upstream decomposition).} Consider an SRN satisfying conditions (i)-(vi), denoted as $\mathcal{N}(x, S, J)$. Consider a general reaction

\begin{equation}
	\phi_a: \sum_{i=1}^n c_{ia}x_i + \sum_{j=1}^{n'}c_{ja}'E_j
	\rightarrow
	\sum_{i=1}^n d_{ia}x_i + \sum_{j=1}^{n'}d_{ja}'E_j,
\end{equation}
where  $c_{ia}, c'_{ja}, d_{ia}, d'_{ja}$ are nonnegative integers. Let $up(\phi_a)$, $dw(\phi_a)$ represents upstream and downstream nodes, respectively (see Definition 1.3). We can define another reaction $\phi_a^*$ as the following: \\
\begin{equation} 
\begin{aligned}
	\phi_a^*: 
	&\sum_{x_i \in up(\phi_a)} |d_{ia} - c_{ia}|\, x_i 
	+ \sum_{E_j \in up(\phi_a)} |d_{ja}' - c_{ja}'|\, E_j \\
	\rightarrow
	&\sum_{x_k \in dw(\phi_a)} |d_{ka} - c_{ka}|\, x_k
	+ \sum_{E_{\ell} \in dw(\phi_a)} 
	|d_{\ell a}' - c_{\ell a}'|\,E_{\ell},
\end{aligned}
\end{equation}
One can verify that $\phi_a$ and $\phi_a^*$ yield the same ODE in the deterministic setting. Now, recall the definition of upstream fraction $\nu[x_i, \phi_a^*] \in [0,1]$ in Definition 1.5, which represent the fraction of biomass from upstream node $x_i$ versus the biomass from all upstream nodes in reaction $\phi_a^*$. We can split the reaction $\phi_a^*$ into multiple reactions, where each reaction has exactly one upstream nodes from $\phi_a^*$. Namely, for each node $x_i \in up(\phi_a^*)$, define
\begin{equation}
	\begin{aligned}	
		\phi_a^{x,i}: 
		|d_{ia}-c_{ia}|\,x_i
		\rightarrow
		&\sum_{x_k \in dw(\phi_a)}
		\bigg(
		\nu[x_i,\phi_a^*] \cdot
		|d_{ka}-c_{ka}| 
		\bigg)\,x_k \\
		&+ \sum_{E_\ell \in dw(\phi_a)}
		\bigg(
		\nu[x_i,\phi_a^*] \cdot
		|d'_{\ell a}-c'_{\ell a}| 
		\bigg)\,E_\ell,
	\end{aligned}	 
\end{equation}
and for each node $E_j \in up(\phi_a^*)$, define
\begin{equation}
	\begin{aligned}	
		\phi_a^{E,j}: 
		|d_{ia}-c_{ia}|\,E_j
		\rightarrow
		&\sum_{x_k \in dw(\phi_a)}
		\bigg(
		\nu[E_j,\phi_a^*] \cdot
		|d_{ka}-c_{ka}| 
		\bigg)\,x_k \\
		&+ \sum_{E_\ell \in dw(\phi_a)}
		\bigg(
		\nu[E_j,\phi_a^*] \cdot
		|d'_{ka}-c'_{ka}| 
		\bigg)\,E_\ell.
	\end{aligned}	 
\end{equation}
For all reactions in above, we define their flux functions as  
\begin{equation}
	J[\phi_{a}^{x,i}] = J[\phi_{a}^{E,j}] = J[\phi_a].  
\end{equation}

Perform this decomposition for all reactions, we arrive a new SRN where each reaction has one upstream node and also satisfies condition (i)-(vi), denoted as $\mathcal{N}(x,S^*,J^*)$.  \\


\textbf{Lemma 8.5.} Given an SRN satisfying condition (i)-(vi), the ordinary differential equation can be expressed as 
\begin{equation}
	\frac{dX}{dt} = SJ(X) = M(Y)X
\end{equation}	
where $M$ is an $n$-by-$n$ matrix of functions. Each entry $M_{ij}(Y)$ is a function of $Y$, continuous and bounded on $\Delta^{n-1}$. \\

\textbf{Proof.} For the original SRN, we apply the mono-upstream decomposition in Definition 8.4 and obtain a new SRN $\mathcal{N}(x,S^*,J^*)$. By construction, the ODE of $\mathcal{N}(x,S,J)$ and $\mathcal{N}(x,S^*,J^*)$ are the same. \\

For $\mathcal{N}(x,S^*,J^*)$, each flux function has the form $J_a(X)=R_a(Y)X_k$ for one $X_k$. Suppose there are $m'$ fluxes in this network, define an $m'$-by-$n$ matrix $P$:
\begin{equation}
	\begin{aligned}
		P_{ak}(Y) & = R_a(Y),\;\; 
		\text{if } J_a(X)=R_a(Y)X_k, \\
		&= 0,\;\;
		\text{\;\;\;\;\;\;\;\;otherwise.}
	\end{aligned}
\end{equation}

The definition gives $J(X)=P(Y)X$, where $P$ is a continuous and bounded function of $Y$. Define $M(Y) := S^* P(Y)$, then each entry $M_{ij}$ is also a function of $Y$, bounded and continuous on $\Delta^{n-1}$, and we have
\begin{equation}
\frac{dX}{dt}=SJ=S^*J^*=S^*P(Y)X=M(Y)X.
\end{equation}
$\;\;\blacksquare$ \\





\textbf{Corollary 8.6.} Consider an SRN satisfying conditions (i)-(vi). Assume the trajectory $Y(t)$ converge to a fixed point $Y^* \in \Delta^{n-1}$. Then, the long-term dynamics can be analyzed with an LRN defined by $\mathcal{N}(x,S^*,J^*(Y^*))$ with $\frac{dX}{dt}=M(Y^*)X$. \, Furthermore, the long-term growth rate $\lambda$ can be analyzed with DDE: $\frac{dZ}{dt} = -\beta Z + \int_0^\infty \alpha(\tau)Z(t-\tau)\,dt$, where $\alpha(\tau), \beta$ given by the decomposed network with $M(Y^*)$. \\ 


\textbf{Example 8.7.} In this example we consider SRN with the following reactions, with biomass contents $\mathfrak{m}(x_i) = \mathfrak{m}(E_j)=1$ for all nodes.  \\

\includegraphics[scale = 0.55]{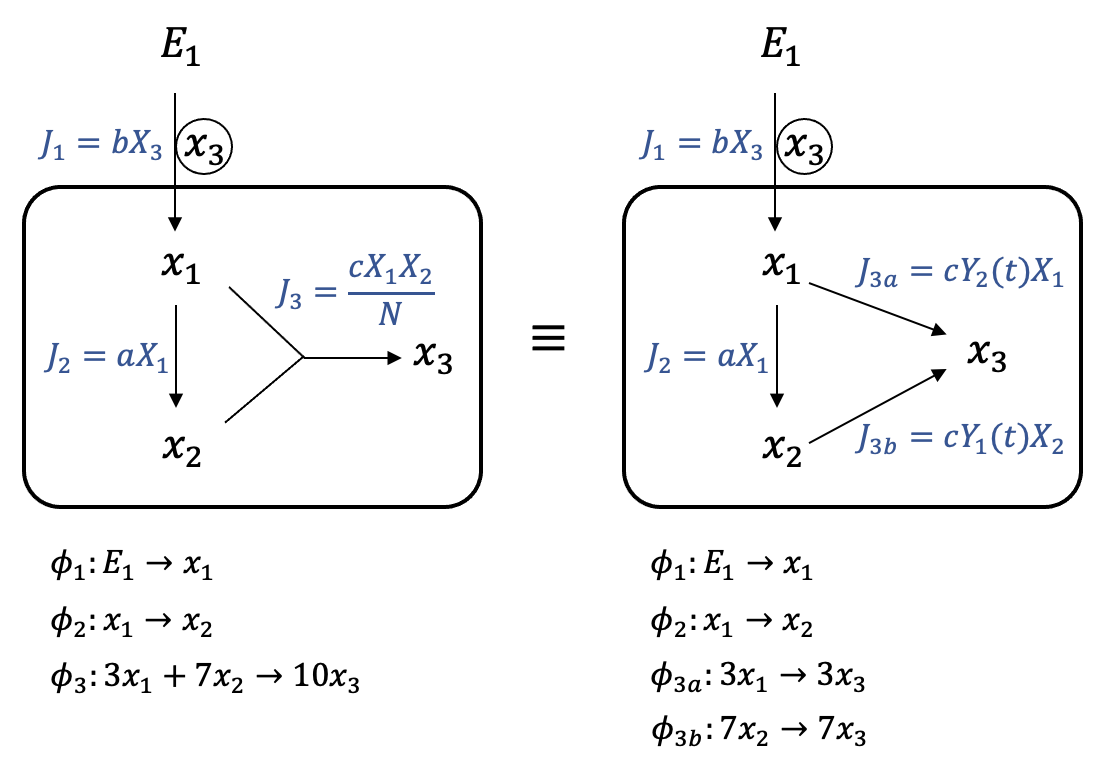}\\

The rescaled system of this SRN is (express $Y_3$ as $1-Y_1-Y_2$):
\begin{equation}
\begin{aligned}
	Y_1'(t) &= -aY_1+b(1-Y_1)(1-Y_1-Y_2)-3cY_1Y_2\\
	Y_2'(t) &= aY_1- bY_2(1-Y_1-Y_2)-7cY_1Y_2 \\
\end{aligned}
\end{equation}

Consider the parameter set $a=b=c=1$. The rescaled system has two fixed points: one saddle at $(0,1)$ and one stable fixed point at $(Y_1^*,Y_2^*)\approx(0.3041,0.1121)$. Every trajectory with $Y(0)>0$ converge to the stable fixed point $Y^*$. The long-term growth rate is 
$$\lambda=\mu(Y^*)=b(1-Y_1^*-Y_2^*)\approx 0.5838$$ 

From Definition 8.4 there is an equivalent SRNs (see above figure, right). Between two SRNs, the reactions are different (since $\phi_3$ is spitted into $\phi_{3a}$ and $\phi_{3b}$), but their ODEs are the same, which is
\begin{equation}
\begin{aligned}
X' &= M(Y)X,\\
M(Y) &= 	
\begin{pmatrix}
	-a-3cY_2 & 0 & b \\
	a & -7cY_1 & 0 \\
	3cY_2 & 7cY_1 & 0 \\
\end{pmatrix}.	
\end{aligned}
\end{equation}
As a verification, we calculate the top eigenvalue $\mu_{max}$ of $M(Y^*)$ with $a=b=c=1$. Substitute $Y^*=(0.3041,0.1121)$ into $M(Y^*)$, we obtained a value $\mu_{max}\approx 0.5838$, which is equal to $\lambda$ as a consistent result. \\

Now, we adopt Lagrangian view and analyze the catalytic kernel $\alpha$. By our assumption, $x_1,x_2,x_3$ have the same biomass content $\mathfrak{m}(x_i)=1$. From the equivalent SRN (see above figure, right), the gatekeeper is $\mathcal{G}=\{x_3\}$. There are two reaction pathways, and at the stable equilibrium $Y^*$ we have the following properties: \\

\begin{center}
\begin{tabular}{l*{5}{|c}r}
$\pi$ & $\kappa_{\pi}$ & $q_{\pi}$ & $\theta_{\pi}$ &
$\mathcal{L}[f_{\pi}](s)$ \\
\hline
$\pi_1: E_1x_1x_3$ 	& $b$ & $ \frac{3cY_2^*}{a'}$ & $1$ & 
$\frac{a'}{s+a'}$ \\
$\pi_2: E_1x_1x_2x_3$ & $b$ & $\frac{a}{a'}$ & $1$ &
$\frac{a'}{s+a'}\frac{7cY_1^*}{s+7cY_1^*}$ \\
\end{tabular} 
\end{center}
with $a':= a+3cY_2^*$. Now, with the parameters $a=b=c=1$, $Y^*=(0.3041,0.1121)$ we calculate the catalytic spectra numerically. Define $A_{\pi}=\kappa_{\pi}\,q_{\pi}\,\theta_{\pi}\,\mathcal{L}[f_{\pi}]$, we have
\begin{equation}
	\begin{aligned}
		A_{\pi_1}(s) &= \frac{3cY_2^*}{a'}\frac{a'}{s+a'}\approx \frac{0.3363}{s+1.3363}. \\
		A_{\pi_2}(s) &= \frac{a}{a'}\frac{a'}{s+a'}\frac{7cY_1^*}{s+7cY_1^*}
		\approx \frac{2.1284}{(s+1.3363)(s+2.1284)}.
	\end{aligned}
\end{equation}
Solve $\tilde{\alpha}(\lambda) = \lambda$ numerically, we have $\lambda \approx 0.5838$, equal to the long-term growth rate calculated by the matrix methods. \\


Corollary 8.6 specifies how to calculate catalytic kernel when the rescaled system converges to a fixed point $Y^*$. In general, for SRNs the rescaled trajectory $Y(t)$ can converge to different types of attractors, including limit cycle, limit torus and even strange attractors. Our goal is to define and calculate the \textit{effective catalytic kernel} for general attractors. \\

In the following, we assume all SRN satisfied condition (i) to (vi) as in Definition 8.1. Recall that in Definition 2.7 and 2.8 we defined amplification rate and arrival functions for general reaction networks. Now, we will look their expression in SRNs. \\


\textbf{Definition 8.8 (State-dependent amplification rate).} Consider a reaction pathway $ \pi \in \mathcal{F}(\mathcal{E},\mathcal{G}) \cup \mathcal{F}(\mathcal{G},\mathcal{G})$ in an SRN. We define $\kappa_{\pi}^{dep}(Y(t'))$ the \textit{state-dependent amplification rate} of reaction pathway $\pi$ as,
\begin{equation} 
\kappa_{\pi}^{dep}(Y(t')) 
:=\kappa_{\pi}(t') 
= \frac{S[u,\omega_0]J[\omega_0](t')}{Z(t')}.
\end{equation}

\textbf{Note.} The quantity $\kappa(t')$ is defined for general RN (see Definition 2.6). However, for SRNs, $\frac{J[\omega_0](t')}{Z(t')}$ is a ratio between two scalable functions at time $t'$. This allows us to define the amplification rate as a function $Y(t')$, where the superscript '$dep$' indicates that $Y(t')$ is the "departure state (biomass at time $t'$)" for the biomass transfer. 
\begin{center}
\includegraphics[scale=0.55]{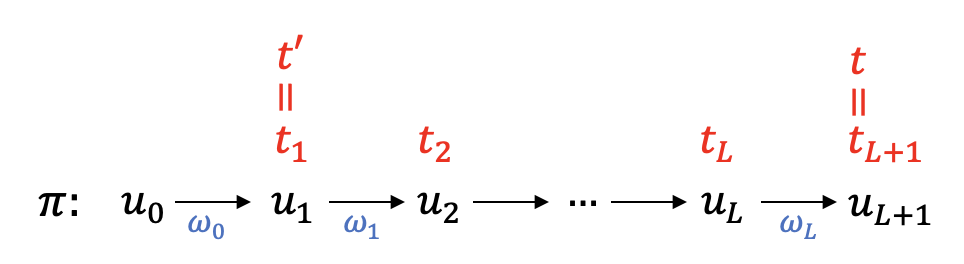} 	
\end{center}


Recall that in LRNs, the arrival function $H_{\pi}(t',t)=H_{\pi}(t-t')$ is a function of $\tau := t-t'$ and hence we can define arrival function $h_{\pi}(\tau)$. For SRNs with non-fixed-point attractors, this is no longer valid, i.e. $H_{\pi}(t',t)$ could depends both departure time $t'$ and arrival time $t$ independently. There is no way to write down a single kernel-like function. However, we could still define the following state-dependent kernel-like functions. We denote  the image of $Y(t)$ by 
\begin{equation}
\text{Img}(Y) := \{Y(t),\; 
t \in [0,\infty)\} \subseteq \Delta^{n-1}.
\end{equation}
Since $\text{Img}(Y)$ is a subset of unit simplex, we can define state-dependent functions on it. \\



\textbf{Definition 8.9 (State-dependent arrival functions).} Consider the arrival function $H_{\pi}(t',t)$ in Proposition 2.7. Since the ODE system is autonomous, the function dependence on $t'$ or $t$ can be expressed as dependence on $Y(t')$ or $Y(t)$. We define \textit{state-dependent arrival functions} as

\begin{equation}
\begin{aligned}
&h_{\pi}^{arr}: \mathbb{R}\times \text{Img}(Y) 
\rightarrow \mathbb{R}, \quad
&h_{\pi}^{arr}(t-t';Y(t)) := H_{\pi}(t',t).  \\
\end{aligned}
\end{equation}
\textbf{Note.} The superscript '$arr$' indicates "arrival state (biomass at time $t$)" for the biomass transfer. \\


\textbf{Proposition 8.10.} Consider an SRN satisfying condition (i)-(vi), and a reaction pathway $\pi \in \mathcal{F}(\mathcal{E},\mathcal{G}) \;\cup\; \mathcal{F}(\mathcal{G},\mathcal{G})$ in the SRN. Define $\tau := t-t'$, and recall the definition of $\alpha^{\bigstar}(\tau; t)$ and $\beta^{\bigstar}(t)$ in Lemma 2.9 for general reaction networks. For SRNs, we have    

\begin{equation} \label{E8.13A}
\begin{aligned}
\alpha^{\bigstar}(\tau; t) =
\sum_{\pi \in \mathcal{F}(\mathcal{E},\mathcal{G}) \;\cup\; \mathcal{F}(\mathcal{G},\mathcal{G})}
\kappa_{\pi}^{dep}(Y(t'))\, h_{\pi}^{arr}(\tau; Y(t)),
\end{aligned}
\end{equation} 

\begin{equation} \label{E8.13B}
\begin{aligned}
\beta^{\bigstar}(t) = (-1)
\sum_{x_g \in \mathcal{G}} \;
\sum_{\phi_c\in out(x_g)}
(1-\xi[\mathcal{G}, \phi_c])\,
S[x_g,\phi_c]\;
\frac{J_c(X(t))}{Z(t)}  > 0.
\end{aligned}
\end{equation}
Since in SRNs both $\alpha^{\bigstar}(\tau; t)$ and $\beta^{\bigstar}(t)$ are functions of $Y$, we define 
\begin{equation}
	\begin{aligned}
		\alpha(\tau; Y(t)):&= 
		\alpha^{\bigstar}(\tau; t), \\
		\beta(Y(t)):&=
		 \beta^{\bigstar}(t). 
	\end{aligned}
\end{equation}
With above definitions, the DDE can be expressed as  
\begin{equation}
\frac{dZ}{dt}(t) = 
-\beta(Y(t))Z(t)
+ \int_0^{t} \alpha(\tau; Y(t))\,Z(t-\tau)\,d\tau
+ C_{ini}(t),
\end{equation}
with $\lim_{t\rightarrow\infty} \frac{C_{ini}(t)}{Z(t)} = 0$. \\

\textbf{Proof.} Equation (\ref{E8.13A}), (\ref{E8.13B}) follows Lemma 2.9. We notice that the state-dependent function $\alpha(\tau; Y(t))$ and $\beta(Y(t))$ in SRNs can be viewed as generalization of $\alpha(\tau)$ and $\beta$ in LRNs. Note that since the scalable reaction network satisfies an autonomous ODE, $\;\frac{dY}{dt} = F(Y)-\mu(Y)Y$, the values of $Y(t)$ and $\tau$ uniquely determine the value of $Y(t+\tau)$. Similarly, for $t\geq \tau$, the values of $Y(t)$ and $\tau$ uniquely determine the value of $Y(t-\tau)$. Therefore, although the formula (\ref{E8.13A}) depends explicitly on $Y(t)$, $Y(t')$ and $\tau$, we can express $\alpha^{\bigstar}$ as a function of $\alpha(\tau; Y(t))$. \\

To show $\lim_{t\rightarrow\infty} \frac{C_{ini}(t)}{Z(t)} = 0$, we express the limit as 
\begin{equation} \label{E8.13}
	\begin{aligned}
		\frac{C_{ini}(t)}{Z(t)} = 
		\frac{1}{Z(t)}\frac{dZ(t)}{dt}
		+ \beta(Y(t))
		- \int_0^{t} \alpha(\tau; Y(t))\,
		\frac{Z(t-\tau)}{Z(t)} \,d\tau.
	\end{aligned}
\end{equation}

Note that the three terms in the right hand side of (\ref{E8.13})  are all scale-invariant under the transformation $X(t) \mapsto cX(t)$ for $c \in \mathbb{R}$. It is also clear that all terms in the right hand side of (\ref{E8.13}) are continuous function of $Y$. Therefore, we can expressed $C_{ini}(t)/Z(t):=f(Y(t))$ as a continuous function of $Y(t)$. Now, by SRN condition (v), the trajectory $Y(t)$ is $\rho$-regular and hence $\langle f(Y)\rangle_{\rho}$ converge to an real number, denoted as $c^*$. By definition, $C_{ini}(t)/Z(t)$ is nonnegative. If $c^*>0$, then since $Z(t)$ is unbounded for large $t$, the value of $C_{ini}(t)$ must also be unbounded for large $t$. However, this contradict to the fact that $\int_0^\infty C_{ini}(t)\,dt \leq N(0)$. Therefore, we must have $\langle C_{ini}(t)/Z(t)\rangle_{\rho} = 0$. This implies $\lim_{t\rightarrow\infty} \frac{C_{ini}(t)}{Z(t)} = 0$ since $C_{ini}(t)/Z(t)$ is nonnegative. $\;\blacksquare$ \\


With the state-dependent arrival function, our next goal is to introduce ergodic averaging for the catalytic kernel. By SRN auxiliary condition (v), the trajectory $Y(t)$ is $\rho$-regular and hence $\langle f(Y(t))\rangle_t = \langle f(Y)\rangle_{\rho}$ for every continuous function $f(Y)$. Our goal is to find "effective" $\alpha,\beta$ that are state-independent, serving as the suitable "averaged quantities". \\

The simplest ergodic averaging is when the attractor of $Y(t)$ is a fixed point $Y^*$. In this case, we can define $\alpha_{eff}(\tau):= \alpha(\tau; Y^*)$ and $\beta_{eff}:= \beta(Y^*)$. When the attractor is not fixed points, choosing the effective kernel and degradation is more complicated. Naively, one may define $\alpha_{eff}(\tau) := \langle \alpha(\tau; Y)\rangle_{\rho}$ as the ergodic averaging. It turns out that this is inadequate. Theorem D gives the adequate average. \\

In the following, we adopt the semigroup notation: for functions $f$ on trajectory $X$ we denote $f(X(t+\tau))$ as $f(\Gamma^{\tau}X)$ where $\Gamma^\tau: \mathbb{R}^n \rightarrow \mathbb{R}^n$ is a semigroup operator mapping $X(t)$ to $X(t+\tau)$. Note that for SRNs, the nonnegative quadrant is both forward- and backward- invariant by SRN condition (ii). Therefore, we can also have $\tau < 0$ for $\Gamma^\tau$. Finally, we express $Z(X) = \sum_{x_g \in \mathcal{G}} X_g$ as function of $X$. \\


\textbf{Theorem C}: Consider an SRN satisfying condition (i)-(vi) with a trajectory having long-term growth rate $\lambda$ and occurrence frequency measure $\rho$. Then, the gatekeeper biomass $Z(t)$ satisfies 

\begin{equation}
\lim_{T\rightarrow\infty}\frac{1}{T} \int_0^T \; 
\frac{Z(t)}{Z(t+\tau)}\;dt
=
\bigg \langle
\frac{Z(t)}{Z(t+\tau)} 
\bigg \rangle_t
=\bigg \langle
\frac{Z(X)}{Z(\Gamma^{\tau}X)}
\bigg \rangle_{\rho} 
= e^{-\lambda \tau},
\end{equation}
for every $\tau \geq 0$. \\

\textbf{Note.} Since we assumed the long-term growth rate is $\lambda$, it is relative easy to show that 

\begin{equation}
\bigg \langle
\frac{Z(X)}{Z(\Gamma^{\tau}X)}
\bigg \rangle_{\rho} 
= U(\tau)\,e^{-\lambda\tau},
\end{equation}
with $\lim_{\tau\rightarrow\infty} \frac{1}{\tau}\log U(\tau) = 0$. However, it takes more effort to show that $U(\tau) = 1$ for all $\tau \in \mathbb{R}$. We will prove this in the next section. \\


\textbf{Theorem D}: Consider an SRN satisfying condition (i)-(vi), with a trajectory having long-term growth rate $\lambda$ and occurrence frequency measure $\rho$.  Define 
\begin{equation}
\begin{aligned}
\alpha_{eff}(\tau) 
&:= e^{\lambda \tau} 
\bigg \langle \alpha(\tau;Y(t)) \frac{Z(t)}{Z(t+\tau)}
\bigg \rangle_{t}
= e^{\lambda \tau} 
\bigg \langle \alpha(\tau;Y) \frac{Z(X)}{Z(\Gamma^{\tau}X)}
\bigg \rangle_\rho \\
\beta_{eff} 
&:=  
\big \langle \beta(Y(t))\big \rangle_t
= \big \langle \beta(Y) 
\big \rangle_\rho
\end{aligned}
\end{equation}
 
Then the long-term growth rate satisfies 
\begin{equation}
\lambda + \beta_{eff} = \tilde{\alpha}_{eff}(\lambda),
\end{equation} 
where $\tilde{\alpha}_{eff}(s)$ is the Laplace transform of $\alpha_{eff}(\tau)$. \\

We will prove Theorem C and D in the next section. Note that when the the attractor of $Y(t)$ is a fixed point $Y^*$, the  gatekeeper biomass $Z$ can be analyzed by the DDE 

\begin{equation} 
\frac{dZ}{dt} = -\beta_{eff}Z
+ \int_0^t \alpha_{eff}(\tau)\,Z(t-\tau)\,d\tau	
+ \gamma(t),
\end{equation}
with $\lim_{t\rightarrow\infty} \frac{\gamma(t)}{Z(t)} = 0$. However, when the attractor of $Y(t)$ is not a fixed point, the trajectory of $Z(t)$ does not converge to the above equation, but depends on additional dynamics on the simplex. To illustrate the idea of effective kernel in non-steady-state, we consider the following concrete example. \\


\newpage

\textbf{Example 8.11}: We consider an SRN with four nodes, among them three nodes ($x_2,x_3,x_4$) repress each other like the classical repressilator (see Figure 5A in the main text). We assume  $\mathfrak{m}(x_k) = 1$ for all node, and the systems ODE follows: \\ 
\begin{equation}\begin{aligned}
&\frac{dX_1}{dt} = J_1-J_2-J_3-J_4, \\	
&\frac{dX_k}{dt} = J_k,\qquad k=2,3,4,\\
&J_1(X) = bX_2, \\
&J_2(x) = \frac{cX_1}{1+KY_3^{\ell}},\qquad
J_3(x) = \frac{cX_1}{1+KY_4^{\ell}},\qquad
J_4(x) = \frac{cX_1}{1+KY_2^{\ell}}. 
\end{aligned} \end{equation}\

This is a scalable reaction network, where node $x_2,x_3,x_4$ represses the synthesis fluxes of each others. Note that $x_2$ is responsible for the system influx and hence the gatekeeper node. Since there is no degradation, we have $\lambda = b\langle Y_2 \rangle > 0 $. Based on the parameter, this network can exhibit either balanced growth or oscillatory growth. We examine the parameter $b=c=1$, $K=1000$ with different values of Hill coefficient $\ell$. Under this parameter regime, the system satisfies SRN conditions (i) to (vi). \\


\begin{center}
\includegraphics[scale=0.4]{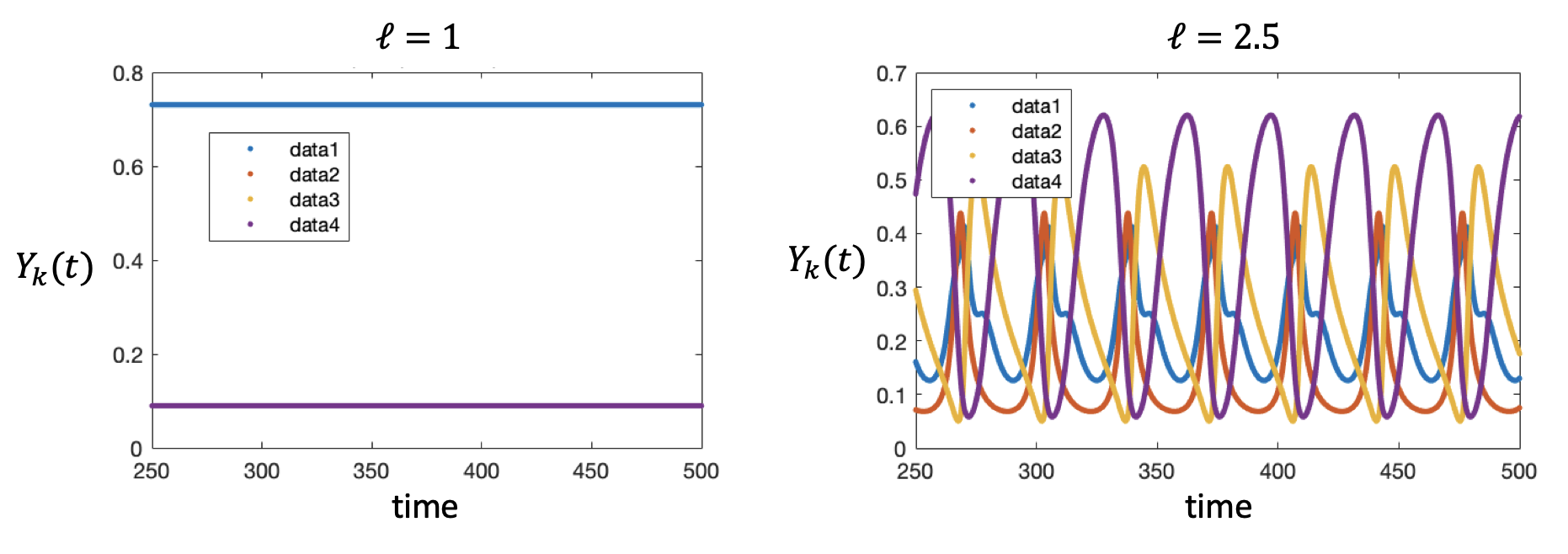} 	
\end{center}


(Case I, $\ell = 1$): For this parameter, the rescaled system $Y(t)$ reaches a fixed point (see figure in above) $Y^*\approx(0.7308,0.0897,0.0897,0.0897)$. We hence have $\lambda = bY_2^* = 0.0897$. To analyze in Lagrangian perspective, we notice that the only reaction pathway in the system is $\pi: E_1x_1x_2$ and the gatekeeper biomass is $Z=X_2$. We have 
\begin{equation}
\kappa_{\pi}=b=1, \;\; 
h_{\pi}(\tau) := \theta_{\pi} \,q_{\pi}\,f_{\pi}(\tau). \\
\end{equation} 

Since there no branch in $\pi$, the transmission efficiency is $\theta_{\pi}=1$. By symmetry of node $x_2,x_3,x_4$ we have $Y_2^*=Y_3^*=Y_4^*$ and hence $J_2=J_3=J_4$ in the long-term. Therefore, the pathway probability is $q_{\pi}=\frac{J_2}{J_2+J_3+J_4} = \frac{1}{3}$. Finally, the conditional waiting time follows
\begin{equation}
f_{\pi}(\tau) = R_{tot} e^{-R_{tot}\tau},\;\;
R_{tot} := \frac{3c}{1+K(Y_2^*)^{\ell}} \approx 0.0331.
\end{equation}
(Note that the factor 3 comes from the fact that three fluxes $J_2,J_3,J_4$ have the same magnitude, and hence the total transition rate is three times of $J_2$). Now we can calculate $\alpha(\tau) = \kappa_{\pi}h_{\pi}(\tau)$ and have the Laplace transform 
\begin{equation}
	\tilde{\alpha}(s) = b q_{\pi}\frac{R_{tot}}{s+R_{tot}}.
\end{equation}
Substitute $s=\lambda=0.0897$, we can verified numerically that the relation $\tilde{\alpha}(\lambda) = \lambda$ is satisfied. \\


(Case II, $\ell = 2.5$): For this parameter, the rescaled system $Y(t)$ converges to a limit cycle. In Eulerian perspective, we could obtain $\lambda$ by integration $\mu(Y) = bY_2$ with the stationary measure on the limit cycle. Numerically, this can be calculated by extract one complete limit cycle of $Y(t)$ and perform time averaging of $\mu(Y(t))$. Numerical integration gives $\lambda \approx 0.1500$. \\

In Lagrangian perspective, For this case, the catalytic kernel $\alpha(\tau;Y(t))$ is state-dependent. To calculate this, we use the formula from Proposition 8.10,
\begin{equation}
\alpha_{\pi}(\tau; Y(t))= 
\kappa_{\pi}^{dep}(Y(t'))\;
h_{\pi}^{arr}(Y(t)).
\end{equation}
In this case, we have 
\begin{equation} 
\begin{aligned}
	\kappa_{\pi}^{dep}(Y(t')) &= b = 1, \\
	h_{\pi}^{arr}(Y(t)) &= \theta_{\pi} \,R_2(t)\,H_{\pi}(t',t), \\
\end{aligned}
\end{equation}
with the following terms 
\begin{equation}
\begin{aligned}
&\theta_{\pi}=1, \;
R_2(t) = \frac{c}{1+K[Y_3(t)]^{\ell}}, \;
R_3(t) = \frac{c}{1+K[Y_4(t)]^{\ell}}, \;
R_4(t) = \frac{c}{1+K[Y_2(t)]^{\ell}}, \; \\
& R_{tot}(t) = R_2(t)+R_3(t)+R_4(t), \\
&H_{\pi}(t',t) = exp \bigg( - \int_{t'}^{t} R_{tot}(z)dz \bigg). 
\end{aligned}
\end{equation}

As described in Theorem C, we calculate the effective catalytic kernel by numerical integration of $\alpha_{eff}(\tau) = e^{\lambda \tau} \bigg \langle \alpha(\tau;Y(t)) \frac{Z(t)}{Z(t+\tau)} \bigg \rangle_t $. In comparison, we also calculate the mean catalytic kernel by $ \alpha_{mean}(\tau) = \big \langle \alpha(\tau;Y) \big \rangle_t $. We performed Laplace transform and obtained $\tilde{\alpha}_{eff}(s)$ and $\tilde{\alpha}_{mean}(s)$ from both kernels, and testing the formula $\tilde{\alpha}(\lambda) = \lambda$. Substitute $\lambda = 0.1500$ into the spectra, we found that $\tilde{\alpha}_{eff}(\lambda) \approx 0.1497$, closed to $\lambda$ within 0.2$\%$ of numerical error. In comparison, $\tilde{\alpha}_{mean}(\lambda) \approx 0.1816$, significantly deviate from $\lambda$. (see Figure 5B in the main text). This shows that when the rescaled system $Y(t)$ converge to non-fixed-point type attractors, effective kernel $\alpha_{eff}$ is the appropriate generalization for the catalytic kernel.  $\;\;\blacksquare$ \\


\newpage
\section{Proofs of Theorem C and D}



In below, we discuss some propositions which will be used in the proof of Theorem C. We denote the omega-limit set of the trajectory $Y(t)$ as $\Omega_{Y} \subseteq \Delta^{n-1}$. From SRN condition (v), it is also a minimal compact support of ergodic measure $\rho$. \\


\textbf{Proposition 9.1.} Consider an SRN satisfying condition (i)-(vi). For each $k \in \mathbb{N}$ and $\tau \geq 0$, the following functions
\begin{equation}
\begin{aligned}
		f_k(Y,\tau) &:= \bigg[\frac{Z(t-\tau)}{Z(t)}
		e^{\lambda\tau} \bigg]^k, 
		\\
		f_k(Y,-\tau) &:= \bigg[\frac{Z(t)}{Z(t-\tau)}
		e^{-\lambda\tau} \bigg]^k.
		\\
\end{aligned}
\end{equation}
are well-defined on $\Omega_Y \times \mathbb{R}$.  \\  

\textbf{Proof.} First we show that the ratio $\frac{Z(t-\tau)}{Z(t)}$ is a function of $Y(t)$ and $\tau$. Notice that  
\begin{equation}
\frac{Z(t-\tau)}{Z(t)}
= 1 - \frac{1}{Z(t)} 
\int_{s=0}^\tau 
Z'(t-s)\,ds,
\end{equation}
where $Z'$ is linear combination of scalable flux functions. Therefore, $f_1(Y,\tau)$ is well-defined. Next, note that $f_k(Y,\tau) = f_1(Y,\tau)^k$ so they are all well-defined as function of $Y(t)$ and $\tau$. $\;\;\blacksquare$ \\


\textbf{Note.} In the following, our calculations involve the time and phase averages of $f_k(Y(t),\tau)$ for fixed $\tau$. For the time average, $Z(t-\tau)$ is only well-defined for $t \geq \tau$ and hence we will interpret $\langle \frac{Z(t-\tau)}{Z(t)}\rangle_t$ as averaging on time interval $[\tau, \infty)$. This implies for fixed $\tau \geq 0$,

\begin{equation}
\begin{aligned}
	\bigg \langle
	\frac{Z(t-\tau)}{Z(t)}
	\bigg \rangle_t
	&=
	\bigg \langle
	\frac{Z(t)}{Z(t+\tau)}
	\bigg \rangle_t, 
	\\
	\bigg \langle
	\frac{Z(t)}{Z(t-\tau)}
	\bigg \rangle_t
	&=
	\bigg \langle
	\frac{Z(t+\tau)}{Z(t)}
	\bigg \rangle_t. 
	\\
\end{aligned}
\end{equation}

\vspace{20 pt}


\textbf{Proposition 9.2.} With the same conditions and notations as in Proposition 9.1, for every $\tau \in \mathbb{R}$ we have 

\begin{enumerate} [label=(\roman*)]
	\item 
	\begin{equation}
	\bigg \langle f_k(Y,\tau) \bigg \rangle_{\rho}
	\leq
	\bigg \langle f_{2k}\big(Y,\frac{\tau}{2} \big) \bigg \rangle_{\rho},
	\end{equation}
	\item
	\begin{equation} 
	\bigg \langle f_k(Y,\tau) \bigg \rangle_{\rho}
	\bigg \langle f_k(Y,-\tau) \bigg \rangle_{\rho}
	\geq 1.
	\end{equation}
\end{enumerate}


\textbf{Proof.} We define $g(t):= Z(t)e^{-\lambda t}$ and express $f_k(Y,\tau)$ as 
\begin{equation}
f_k(Y,\tau) = 
\bigg[\frac{Z(t-\tau)}{Z(t)}e^{\lambda\tau} \bigg]^k 
= \frac{g(t-\tau)^k}{g(t)^k}.
\end{equation}
From SRN condition (v), we can interchange time and phase averages during our calculation. By Cauchy-Schwarz inequality on $L^2(\rho)$ measure, we have
\begin{equation}
\bigg\langle |AB| \bigg\rangle_{\rho}^2
\leq
\bigg\langle |A|^2 \bigg\rangle_{\rho}
\bigg\langle |B|^2 \bigg\rangle_{\rho}
\end{equation}
for every pairs of function $A,B \in L^2(\rho)$. For inequality (i), consider 
\begin{equation}
		A(t) := \frac{g(t-\tau)^k}{g(t-\frac{\tau}{2})^k} \geq 0,\;\;\;
		B(t) := \frac{g(t-\frac{\tau}{2})^k}{g(t)^k} \geq 0	.
\end{equation}
By the translation symmetry of $\tau$, the long-term average of $A^2$ and $B^2$ is the same. Hence, the phase average of $A^2,B^2$ on $\Omega_Y$ is also the same. This gives
\begin{equation}
\bigg\langle |A|^2 \bigg\rangle_{\rho} =
\bigg\langle |B|^2 \bigg\rangle_{\rho} =
\bigg\langle \frac{g(t- \frac{\tau}{2})^{2k}}{g(t)^{2k}} \bigg\rangle_{t}
= \bigg\langle f_{2k}\big(Y,\frac{\tau}{2} \big)  \bigg\rangle_{\rho}.
\end{equation}
On the other hand, $A(t)B(t) = \frac{g(t-\tau)^k}{g(t)^k}$ and hence 
\begin{equation}
\bigg\langle |AB| \bigg\rangle_{\rho} =
\bigg\langle 
\frac{g(t-\tau)^k}{g(t)^k}
\bigg\rangle_{t} 
=  \bigg\langle f_k(Y,\tau) \bigg\rangle_{\rho}.
\end{equation}

Using the Cauchy-Schwarz inequality and taking square root of both sides, we obtain the relation (i). For inequality (ii), consider

\begin{equation}
		A(t) := \sqrt{
		\frac{g(t-\tau)^k}{g(t)^k} }\geq 0,\;\;\;
		B(t) := \sqrt{
		\frac{g(t)^k}{g(t-\tau)^k} }\geq 0,\;\;\;		
\end{equation}
By the translation symmetry of $\tau$, the long-term average of $\frac{g(t)^{k/2}}{g(t-\tau)^{k/2}}$ is the same as $\frac{g(t+\tau)^{k/2}}{g(t)^{k/2}}$. Therefore, 
\begin{equation}
	\bigg \langle |B|^2 \bigg \rangle_{\rho} 
	= \bigg \langle 
	\frac{g(t+\tau)^k}{g(t)^k} 
	\bigg\rangle_{t} 
	= \langle f_k(Y,-\tau) \rangle_{\rho}
\end{equation}
On the other hand, we have $A(t)B(t)=1$ in this case. By Cauchy-Schwarz inequality we obtain the relation (ii). $\;\blacksquare$ \\


\textbf{Proposition 9.3.} Consider an SRN satisfying condition (i)-(vi), then whenever $\frac{dZ}{dt} \leq 0$ we have 
$ \big|\frac{dZ}{dt} \big|\leq CZ $
where $C$ is a positive constant.  \\ 

\textbf{Proof.}  Let $\mathcal{C}$ denote the collection of efflux reactions from gatekeeper nodes, i.e. $\mathcal{C}:=\{\phi_a \,|\, \phi_a \in out(x_g), x_g \in \mathcal{G}\}$. For gatekeeper node, its changing rate $dZ/dt$ is contributed by influxes positively and effluxes negatively. Hence,  
\begin{equation}
\frac{dZ}{dt} 
\geq \sum_{x_g \in \mathcal{G}}
\sum_{\phi_{a}\in \mathcal{C}} 
S[x_g, \phi_a]J[\phi_a].
\end{equation}
Since $S_{ja} < 0$ for $\phi_a \in \mathcal{C}$, this implies when $dZ/dt \leq 0$,
\begin{equation}
\bigg| \frac{dZ}{dt} \bigg| 
\leq 
\bigg|
\sum_{x_g \in \mathcal{G}}
\sum_{\phi_{a}\in \mathcal{C}}  
S[x_g,\phi_a]J[\phi_a]
\bigg|
=
\sum_{x_g \in \mathcal{G}}
\sum_{\phi_a \in \mathcal{C}} 
\bigg|S[x_g,\phi_a]\bigg|\;J[\phi_a](X).
\end{equation}

Now, from the upstream limited conditions of SRNs, each efflux function $\phi_a \in \mathcal{I}$ can be expressed as $J_a(X)=R_a(Y)X_g$ for a continuous and bounded $R_{a}(Y)$ with $x_g \in \mathcal{G}$. Therefore, the last term of the above equation follows 
\begin{equation}
\sum_{x_g \in \mathcal{G}}
\sum_{\phi_a \in \mathcal{I}} 
\bigg|S[x_g,\phi_a]\bigg|\;J[\phi_a](X) 
:=\sum_{x_g \in \mathcal{G}} 
M_g(Y) X_g 
\leq CZ,
\end{equation}
where $M_g(Y)$ is the linear combination of $R_a(Y)$ of effluxes from node $x_g$, and $C>0$ is a constant. $\blacksquare$ \\

\vspace {10 pt}


Given an autonomous dynamical system $Y'(t) = G(Y)$ on $S$ and denote the semigroup operator $\Gamma_Y^t: Y(0) \mapsto Y(t)$. Recall Definition 7.5, a subset $A\subseteq S$ is called \textit{invariant} under the action of $\Gamma^t$, if $\Gamma^tA = A$ for all $t \rightarrow \mathbb{R}$. \\


\textbf{Proposition 9.4}: Consider an SRN satisfying condition (i)-(vi), define $S_Z \subseteq \Delta^{n-1}$ as the region $S_Z: \{Y \in \Delta^{n-1}, \frac{Z}{N} =0. \}$. Let $\rho$ denote the ergodic measure corresponded to the trajectory $Y(t)$, we have $\rho(S_Z)=0$. \\

\textbf{Proof.} Intuitively, the region $\frac{Z}{N}=0$ corresponds to the states with no biomass on gatekeeper nodes. For a growing system, the occurrence of system trajectory $Y(t)$ in this region should be zero. To show this explicitly, we first show that a trajectory start with $Z(0)>0$ cannot reach $Z(t)=0$ at a finite time $t$. Assume the contrary that $Z(0)>0$ and $Z(t^*)=0$, then by continuity of the ODE (condition (i)) there is a time interval $[t',t^*]$ where $dZ/dt \leq 0$. Integrating the system on this interval and using Proposition 9.3, we get 

\begin{equation}
\int_{t'}^{t^*} 
\frac{1}{Z}\bigg| \frac{dZ}{dt}\bigg|dt
\leq 
\int_{t'}^{t^*} C\,dt
= C(t^*-t').
\end{equation}

For the right hand side, $C(t^*-t')$ is a finite number. For the left hand side, $\log \,Z(t')-\log Z(t^*)$ diverges since we assume $Z(t^*)=0$. This contradict to our assumption and hence there is no real number $t^*$ can satisfies the system conditions. \\

We have shown that a trajectory $X(t)$ with $Z(0)>0$ stay for $Z(t)>0$ for any positive $t$. This implies the set $\Delta^{n-1} \backslash S_Z$ is invariant under the action of $\Gamma_Y^t$. By the definition of ergodic measures, $\rho(\Delta^{n-1} \backslash S_Z)$ must be either 0 or 1. This implies $\rho(S_Z)$ is either 0 or 1. \\

Suppose $\rho(S_Z)=1$, this would imply $\lim_{t\rightarrow\infty} Z(t)/N(t) = 0$. By Proposition 2.2(ii), the ratio $Z(t)/N(t) \rightarrow 0$ implies $\lambda \leq 0$ if $\lambda$ exists. This contradict to the SRN condition (vi), which assumes $\lambda>0$. Therefore, we must have $\rho(S_Z) = 0$. \;$\blacksquare$ \\


\textbf{Corollary 9.5}: Consider an SRN satisfying condition (i)-(vi), define the region 
\begin{equation}
S_{Z}^{\,out} := \{Y(t)\in \Delta^{n-1},\; Y(t-\tau) \in S_Z 
\textit{ for some } \tau > 0 
\},
\end{equation}
then we have $\rho(S_{Z}^{\,out})=0$. \\

\textbf{Proof.}: From Proposition 9.4, a trajectory cannot reach $S_Z$ in a finite time, and hence the trajectory $Y(t)$ cannot commute between $\{Z(t)=0\}$ and $\{Z(t)>0\}$ for multiple times. Therefore, the set $S_Z^{\,out}$ is not recurrent and the stationary probability flux from $S_Z$ to simplex interior must be zero. This implies $\rho(S_{Z}^{\,out}) = 0$. $\;\blacksquare$ \\


We will use the next Lemma for differentiation under integral. \\

\textbf{Differentiation Lemma 9.6}. (Theorem 7.28 in \cite{klenke_probability_2013}): Let $V$ be a subspace of $\mathbb{R}^n$, and $I \subset \mathbb{R}$ be an nontrivial open interval. Suppose a function $f(Y,\tau): V\times I \rightarrow \mathbb{R}$ satisfies the following properties: 

\begin{enumerate} [label=(\roman*)]
	\item For every $\tau \in I$ the function $f(Y,\tau) \in L^1(\rho)$. 
	\item For $\rho$-almost all $Y \in V$, $f(Y,\tau)$ is differentiable for $\tau$.
	\item Exist a positive function $H(Y) \in L^1(\rho)$ such that $\frac{\partial f}{\partial \tau} < H(Y)$ for all $\tau \in I$ and $\rho$-almost everywhere.
\end{enumerate}

Then, for any $\tau \in I$ we have $\frac{\partial f}{\partial \tau}\in L^1(\rho)$ and 

\begin{equation}
\frac{d}{d\tau} \int_V f(Y,\tau) \,\rho(dY)
= \int_V \frac{\partial f}{\partial \tau} 
\;\rho(dY) .
\end{equation}\\


\textbf{Proposition 9.7}: Under the same condition of Proposition 9.1 and define 
\begin{equation}
f_k(Y,\tau) := 
\bigg[\frac{Z(t-\tau)}{Z(t)}e^{\lambda\tau} \bigg]^k,
\end{equation}
we have
\begin{equation}
\bigg[
\frac{d}{d\tau}\int_{\Delta^{n-1}}
f_k(Y,\tau)\,\, \rho(dY) 
\bigg ]_{\tau=0} = 0,
\end{equation}
for every $k \in \mathbb{N}$.\\


\textbf{Proof.} We need to check if the condition (i),(ii),(iii) are satisfied for the Differentiation Lemma in above. We choose a fixed number $\varepsilon>0$ and consider the interval $I := [-\varepsilon, \varepsilon]$. \\


For (i), we first show that for $\tau \in I$, 
\begin{equation}
	f_k(Y, \tau)\,e^{-k\lambda\tau} = \frac{Z(t-\tau)^k}{Z(t)^k} \leq 1 + k \varepsilon M^*
\end{equation}
on $\Omega_Y$, where $M^*$ is a constant. Fixed $\tau \in I$ and use Mean Value Theorem, we have 
\begin{equation}
f_k(Y, \tau)\,e^{-k\lambda\tau} = 1 + \frac{k}{Z}\frac{dZ}{dt} (-\tau_m),
\end{equation}
with $\tau_m \in [0,\tau]$. The potential issue is when $Z(t)\rightarrow 0$ the term in the right-hand side could diverge. In the next step, we show that $\frac{1}{Z}\frac{dZ}{dt}$ is bounded on $\Omega_Y$. \\

To proceed, we investigate the trajectory $Y(t)$ when $Z(t) \rightarrow 0$, that is, when $Y(t)$ close to the vicinity of $S_Z$. We claim that for any recurrent trajectory $Y(t)$ we must have $\frac{dZ}{dt} \leq 0$ as $Z \rightarrow 0$. To see this, suppose $Z(0)=0$ with a positive time derivative $\frac{dZ}{dt}|_{t=0} > 0$. Then for $t>0$ the trajectory belongs to $S_{Z}^{out}$, which is transient for the ergodic measure $\rho$ by Proposition 9.4. \\

Now, since $\frac{dZ}{dt} \leq 0$ at the vicinity of $S_Z$, by Proposition 9.3 we have $\frac{1}{Z}\frac{dZ}{dt} \leq C$ in the vicinity of $S_Z$. Away from $S_Z$, we define another region $Z/N \geq \varepsilon$ which is a compact subset of unit simplex and hence $\frac{1}{Z}\frac{dZ}{dt}$ is also bounded by another constant $C'$. Taking $M^*=\max(C,C')$ is sufficient for our need. Now, for every $\tau \in I$ we have
\begin{equation}
f_k(Y,\tau) 
\leq e^{k\lambda \varepsilon} (1+k\varepsilon M^*)
:= C_{max},
\end{equation}
where $C_{max}$ is a constant independent of $\tau$. Since $\Omega_Y$ is compact, this implies $f_k(Y,\tau) \in L^1(\rho)$ for each $\tau \in I$. \\


For (ii), we calculate 
\begin{equation}
\frac{\partial f_k(Y,\tau)}{\partial \tau}	
= k f_k(Y,\tau) 
\bigg \{
\frac{-1}{Z(t-\tau)}\frac{dZ(t-\tau)}{d(t-\tau)}
+\lambda
\bigg \}.
\end{equation}

The above expression is well-defined on $\Delta^{n-1}$ except for the set $S_Z=\{Z/N=0\}$. This set has zero $\rho$-measure by Proposition 9.4, and hence condition (ii) is satisfied. \\


For (iii), note that the term $\frac{1}{Z(\xi)}\frac{dZ(\xi)}{d\xi}|_{\xi=t-t'}$ is a function of $Y$ and is a linear combination of scalable fluxes connected to nodes in $Z$. Therefore, this term is bounded on the unit simplex by a constant $C_2$. We also showed $f_k(Y,\tau)$ is uniformly bounded by $C_1$ in $I$. Therefore, we can choose $H(Y) := kC_1C_2$ which belongs to $L^1(\rho)$ and satisfies the requirement. \\

With conditions (i), (ii), (iii) satisfied, we calculate the derivative at $\tau=0$. Note that
\begin{equation}
\frac{\partial f_k(Y,\tau)}{\partial \tau}	
\bigg|_{\tau=0}
= k f_k(Y,0) 
\bigg \{
\frac{-1}{Z(t)}\frac{dZ(t)}{dt} + \lambda
\bigg \}
= k\lambda -k\frac{d\log Z(t)}{dt},
\end{equation}

Integrating with the measure $\rho$ is equal to the the long-term average of $t$, namely,
\begin{equation}
\bigg[
\frac{d}{d\tau}\int_{\Delta^{n-1}}
f_k(Y,\tau)\, \rho(dY) 
\bigg ]_{\tau=0} 
= k\lambda - k \lim_{T\rightarrow \infty}\frac{1}{T} \log Z(T)
= 0.
\end{equation}
The last equality is from Proposition 2.2. $\;\blacksquare$ \\


%
%
%


\textbf{Theorem C}: Consider an SRN satisfying conditions (i)-(vi), with a trajectory having long-term growth rate $\lambda$ and occurrence frequency measure $\rho$. Then, the gatekeeper biomass $Z(t)$ satisfies 

\begin{equation}
\lim_{T\rightarrow\infty}\frac{1}{T} \int_0^T \; 
\frac{Z(t)}{Z(t+\tau)}\;dt
=
\bigg \langle
\frac{Z(t)}{Z(t+\tau)} 
\bigg \rangle_t
=\bigg \langle
\frac{Z(Y)}{Z(\Gamma^{\tau}Y)}
\bigg \rangle_{\rho} 
= e^{-\lambda \tau},
\end{equation}
for every $\tau \geq 0$. \\

\textbf{Proof of Theorem C.}  We express $Z(t):=g(t)e^{\lambda t}$ and hence $ \frac{Z(t-\tau)}{Z(t)} = \frac{g(t-\tau)}{g(t)}e^{-\lambda \tau} $. Following the notation in Proposition 9.1, we define 

\begin{equation}
f_k(Y,\tau) := 
\bigg ( \frac{g(t-\tau)}{g(t)}\bigg )^k.
\end{equation}

By SRN condition (v), the phase average of $f_k(Y,\tau)$ over probability measure $\rho$ is equal to the time average of $\bigg [\frac{g(t-\tau)}{g(t)}\bigg]^k$, which is denoted by

\begin{equation}
F_k(\tau) 
:= \langle f_k(Y,\tau) \rangle_\rho	
= \bigg \langle \bigg (
\frac{g(t-\tau)}{g(t)}\bigg )^k
\bigg \rangle_t .
\end{equation}

It is clear that $F_k(0) = 1$ for all $k$. The theorem is proved by showing for every $\tau \in \mathbb{R}$,
\begin{equation}
F_1(\tau) := \lim_{T\rightarrow\infty}\frac{1}{T} \int_0^T \; \frac{g(t-\tau)}{g(t)}\,dt  = 1.
\end{equation}

By Proposition 9.2 (i), and equate the time average to the phase average, we have 
\begin{equation}
\bigg \langle 
\bigg(
\frac{g(t-\frac{\tau}{2^k})}{g(t)} 
\bigg) ^{2^k}
\bigg \rangle_t
\leq
\bigg \langle 
\bigg(
\frac{g(t-\frac{\tau}{2^{k+1}})}{g(t)} 
\bigg) ^{2^{k+1}}
\bigg \rangle_t,
\end{equation}
for $k=0,1,2\cdots$. Using this inequality repeatedly, we have 
\begin{equation} \label{D1}
F_1(\tau) \leq\cdots\leq
F_{2^k} \big(\frac{\tau}{2^k} \big)
\leq
F_{2^{k+1}} \big(\frac{\tau}{2^{k+1}} \big) 
\end{equation}
Intuitively, since we have $F_k(0)=1$ for every $k$, the above limit seems to be bounded by $F_{\infty}(0) = 1$ as $k\rightarrow \infty$. However, this argument is not rigorous since it is also possible that $F_{2^k}(\frac{\tau}{2^k})= \infty$ for all $k$ for all $\tau \neq 0$, given that we have not show $F_k(\tau)$ is continuous on $\tau$. \\

To proceed, we calculate the derivative $\frac{dF_k(\tau)}{d\tau}$ at $\tau = 0$. By Proposition 9.7, 

\begin{equation}
\frac{dF_k(\tau)}{d\tau}
\bigg|_{\tau=0}
= \bigg [
\frac{d}{d\tau} \int_{\Delta^{n-1}} f_k(Y,\tau) \rho(dY)
\bigg ]_{\tau = 0}
 = 0.
\end{equation}
This shows that $F_k'(\tau)=0$ at $\tau=0$ for every $k \in \mathbb{N}$. In the meantime, we also shows $F_k(\tau)$ is continuous at $\tau=0$. By definition of derivative, we have 
\begin{equation}
	- \delta_0 < 
	\lim_{\tau\rightarrow 0}
	\frac{F_k(\tau)-F_k(0)}{\tau} = 0 < \delta_0,
\end{equation}
for every $k$ and arbitrary $\delta_0>0$. Since $F_k(0)=1$ for every $k$, we have 
\begin{equation}\label{D2}
	\lim_{\tau\rightarrow 0} 
	|F_k(\tau)-1|  < \delta_0  \tau.
\end{equation}

Next we show for arbitrary $\tau$ the value $F_1(\tau)$ is arbitrary closed to 1. Given any fixed $\delta_0 > 0$, we can choose large enough $k$ such that (i) $\tau^* := \tau /(2^k) < 1$ and (ii) $F_k(\tau^*) < 1+\delta_0$. Using equation (\ref{D1}) and (\ref{D2}), we have 
\begin{equation}
F_1(\tau) < F_{2^k}(\tau^*) < 1+\delta_0\tau^* < 1+\delta_0,
\end{equation}
for arbitrarily small $\delta_0>0$. This implies $F_1(\tau) \leq 1$ for any $\tau \in \mathbb{R}$. Note that this result is valid for both $\tau>0$ and $\tau<0$, and hence we also have $F(-\tau) \leq 1$. Since $F_1(\tau)$ is nonnegative, we conclude that 
\begin{equation}
\begin{aligned}
		0\leq &F_1(\tau) \leq 1, \\ 
		0\leq &F_1(-\tau) \leq 1. \\
\end{aligned}
\end{equation}\

On the other hand, by Proposition 9.2 (ii) we also have
\begin{equation}
	F_1(\tau)F_1(-\tau) \geq 1.
\end{equation}

Together, this implies $F_1(\tau) = 1$ for arbitrary $\tau \in \mathbb{R}$. $\;\;\blacksquare$ \\


%
%
%


\vspace{20 pt}


\textbf{Theorem D}: Consider an SRN satisfying conditions (i)-(vi), with a trajectory having long-term growth rate $\lambda$ and occurrence frequency measure $\rho$. Define 
\begin{equation} \label{ThmD}
\begin{aligned}
\alpha_{eff}(\tau) 
&:= e^{\lambda \tau} 
\bigg \langle \alpha(\tau;Y(t)) \frac{Z(t-\tau)}{Z(t)}
\bigg \rangle_{t\in[\tau,\infty)}
= e^{\lambda \tau} 
\bigg \langle \alpha(\tau;Y) \frac{Z(\Gamma^{-\tau}X))}{Z(X)}
\bigg \rangle_\rho, \\
\beta_{eff} 
&:=  
\big \langle \beta(Y(t))\big \rangle_t
= \big \langle \beta(Y) 
\big \rangle_\rho.
\end{aligned}
\end{equation}
Then, the long-term growth rate satisfies the relation 
\begin{equation}
\lambda + \beta_{eff} = \tilde{\alpha}_{eff}(\lambda).
\end{equation}\ 
 $\;\blacksquare$ \\


\textbf{Proof of Theorem D.} By using Proposition 8.10 and perform time-average from $0$ to $\infty$, we obtain
\begin{equation} \label{Thm-D0}
\begin{aligned}
\bigg\langle 
\frac{1}{Z}\frac{dZ}{dt} 
\bigg\rangle_{t} 
&= \bigg\langle 
\frac{1}{Z(t)}
\bigg(
-\beta(Y(t)) Z(t) + 
\int_{0}^t \alpha(\tau; Y(t))\,Z(t-\tau)\,d\tau
\bigg)
\bigg\rangle_{t}  \\
&= 
-\bigg\langle
\beta(Y(t))
\bigg\rangle_{t} 
+ 
\bigg\langle
\frac{1}{Z(t)}
\int_{0}^t \alpha(\tau; Y(t))\,Z(t-\tau)\,d\tau
\bigg\rangle_{t}  \\
&= 
-\bigg\langle
\beta(Y(t))
\bigg\rangle_{t} 
+ 
\bigg\langle
\int_{0}^t \alpha(\tau; Y(t))\,
\frac{Z(t-\tau)}{Z(t)} \,d\tau
\bigg\rangle_{t}. 
\end{aligned}
\end{equation}

First, we show that $\lambda = \langle \frac{1}{Z}\frac{dZ}{dt}\rangle_t$. Since $\mu_Z := \frac{1}{Z}\frac{dZ}{dt}$ is scale-invariant and hence a function of $Y$, by SRN condition (v), both the time-average and phase-average exist for $\mu_Z$ and are equal, denote as $c_0$. Now, by SRN condition (vi), $\lambda > 0$ and hence from  Proposition 2.2 (iii) we have $\langle\mu_Z\rangle_t = c_0 = \lambda$. Second, we have $\beta_{eff}=\big \langle \beta(Y(t))\big \rangle_t$ by definition. Therefore, the theorem will be proved if 
\begin{equation}
\bigg\langle
\int_{0}^t \alpha(\tau; Y(t))\,
\frac{Z(t-\tau)}{Z(t)} \,d\tau
\bigg\rangle_{t}
= \tilde{\alpha}_{eff}(\lambda).
\end{equation} 
By direct calculation, we have  
\begin{equation}
\tilde{\alpha}_{eff}(\lambda)
= \int_0^\infty
\bigg \langle \alpha(\tau;Y(t)) \frac{Z(t-\tau)}{Z(t)}
\bigg \rangle_{t\in[\tau,\infty)}
d\tau
\end{equation} 
Therefore, our task is to show that the time-averaging integration of $t$ can be interchanged with the $\tau$- integration. We first define the integrand by 
\begin{equation}
	K(\tau,t) := \alpha(\tau; Y(t))\,
	\frac{Z(t-\tau)}{Z(t)},
\end{equation}
and define 
\begin{equation}
	K_1(\tau, t) := \frac{Z(t-\tau)}{Z(t)} \times
	\sum_{\pi' \in \mathcal{F}(\mathcal{E}, \mathcal{G}) \cup \mathcal{F}(\mathcal{G}, \mathcal{G})}
    \kappa_{\pi'}(t-\tau)\, \delta(\tau),
\end{equation}
with $\pi'$ to be the reaction pathways that has only one system nodes, and 
\begin{equation}
\begin{aligned}
    K_2(\tau,t) := \frac{Z(t-\tau)}{Z(t)} \times
    \sum_{\pi'' \in \mathcal{F}(\mathcal{E}, \mathcal{G}) \cup \mathcal{F}(\mathcal{G}, \mathcal{G})}
    \kappa_{\pi''}(t-\tau)\, h_{\pi''}(\tau),
\end{aligned}
\end{equation}
with $\pi''$ to be the reaction pathways that has two or more system nodes. From Proposition 8.10, $K(\tau,t)=K_1(\tau,t)+K_2(\tau,t)$. \\


Now, we show that both $K_1(\tau, t)$ and $K_2(\tau,t)$ can undergoes the interchange of time-average and $\tau$-integration. For $K_1(\tau,t)$, note the Dirac's delta measure $\delta(\tau)$ picks the value of $\tau = 0$, and hence

\begin{equation}
	\begin{aligned}
		\bigg \langle 
		\int_0^t	
		K_1(\tau,t) 
		d\tau
		\bigg \rangle_t
		&= 
		\bigg \langle 
		\frac{Z(t-0)}{Z(t)} \times
		\sum_{\pi' \in \mathcal{F}(\mathcal{E}, \mathcal{G}) \cup \mathcal{F}(\mathcal{G}, \mathcal{G})}
   		 \kappa_{\pi'}(t-0)\,    		 
		\bigg \rangle_t\\
		&= 
		\sum_{\pi' \in \mathcal{F}(\mathcal{E}, \mathcal{G}) \cup \mathcal{F}(\mathcal{G}, \mathcal{G})}
		\bigg \langle
	    \kappa_{\pi'}(t)\,	
		\bigg \rangle_t. \\
	\end{aligned}
\end{equation}
On the other hand, 
\begin{equation}
	\begin{aligned}
		\int_0^ \infty
		\bigg \langle 
		K_1(\tau, t) 
		\bigg \rangle_{t\in[\tau,\infty)}
		d\tau
		&= 
		\int_0^ \infty
		\bigg \langle 
		\frac{Z(t-\tau)}{Z(t)} \times
		\sum_{\pi' \in \mathcal{F}(\mathcal{E}, \mathcal{G}) \cup \mathcal{F}(\mathcal{G}, \mathcal{G})}
   		 \kappa_{\pi'}(t-\tau)\,    		 
		\bigg \rangle_{t\in[\tau,\infty)}
		\delta(\tau)\, d\tau \\
		&= 
		\bigg \langle 
		\frac{Z(t-0)}{Z(t)} \times
		\sum_{\pi' \in \mathcal{F}(\mathcal{E}, \mathcal{G}) \cup \mathcal{F}(\mathcal{G}, \mathcal{G})}
   		 \kappa_{\pi'}(t-0)\,    		 
		\bigg \rangle_{t\in[0,\infty)} \\
		&= 
		\sum_{\pi' \in \mathcal{F}(\mathcal{E}, \mathcal{G}) \cup \mathcal{F}(\mathcal{G}, \mathcal{G})}
		\bigg \langle
	    \kappa_{\pi'}(t)\,	
		\bigg \rangle_t. \\
	\end{aligned}
\end{equation}
Therefore, $K_1(\tau, t)$ allows interchange of time-average and $\tau$-integration. Finally, from Definition 8.8., we have 
\begin{equation}
	\sum_{\pi' \in \mathcal{F}(\mathcal{E}, \mathcal{G}) \cup \mathcal{F}(\mathcal{G}, \mathcal{G})}
		\bigg \langle
	    \kappa_{\pi'}(t)\,	
		\bigg \rangle_t \\
	= \sum_{\pi' \in \mathcal{F}(\mathcal{E}, \mathcal{G}) \cup \mathcal{F}(\mathcal{G}, \mathcal{G})}
		\bigg \langle
	    \kappa_{\pi'}^{dep}(Y)\,	
		\bigg \rangle_\rho.
\end{equation}
The existence of the above summation is guaranteed by and by SRN condition (v), and with the fact that the number of reaction pathway $\pi' \in \mathcal{F}(\mathcal{E}, \mathcal{G}) \cup \mathcal{F}(\mathcal{G}, \mathcal{G})$ with only one system node is finite. \\


For $K_2(\tau,t)$, we need to compare the two integrals in below: 
\begin{equation}
	\bigg \langle 
		\int_0^t	 K_2(\tau,t) d\tau
	\bigg \rangle_t
	\text{ v.s. } 
	\int_0 ^\infty
	\bigg \langle
	K_2(\tau,t)
	\bigg \rangle_{t \in [\tau, \infty)}
	d\tau.
\end{equation}

Notice that $K_2$ is a nonnegative function defined in the region $\Omega: \{(\tau,t) \in \mathbb{R}_{\geq 0}^2: \tau \leq t \}$. We need to calculate the time-average and $\tau$-integration as improper double integral. For convenience, we extend the definition of $K_2$ by defining $K_2(\tau,t) = 0$ in the region of $\tau>t$. In this way, our integration can be extended to $\mathbb{R}_{\geq 0}^2$. The first integral now satisfies
\begin{equation}
	\bigg \langle 
		\int_0^t	 K_2(\tau,t) d\tau
	\bigg \rangle_t
	= \bigg \langle 
		\int_0^\infty K_2(\tau,t) d\tau
	\bigg \rangle_t.
\end{equation}
	
For the second integral, we notice that for fixed $\tau, T$ with $T>\tau$, 
\begin{equation}
\begin{aligned}
\bigg \langle 
K_2(\tau,t) 	
\bigg \rangle_{t \in [\tau,T] }
&= \frac{1}{T-\tau}
\int_\tau ^T
K_2(\tau,t) 	
\,dt \\
&= \frac{1}{T-\tau}
\int_0 ^T
K_2(\tau,t) 	
\,dt \\
&= \frac{T}{T-\tau}
\bigg \langle 
K_2(\tau,t) 	
\bigg \rangle_{t \in [0,T] }.
\end{aligned}
\end{equation}
Therefore, for every fixed $\tau$, taking $T\rightarrow\infty$ yields
\begin{equation}
	\bigg \langle
	K_2(\tau,t)
	\bigg \rangle_{t \in [\tau, \infty)}
	= \lim_{T\rightarrow\infty}
	\bigg \langle 
	K_2(\tau,t) 	
	\bigg \rangle_{t \in [\tau,T] }
	= \bigg \langle 
	K_2(\tau,t) 	
	\bigg \rangle_{t \in [0,\infty) }.
\end{equation}
Hence, the second integral follows the equality 
\begin{equation}
	\int_0 ^\infty
	\bigg \langle
	K_2(\tau,t)
	\bigg \rangle_{t \in [\tau, \infty)}
	d\tau
	=
	\int_0 ^\infty
	\bigg \langle
	K_2(\tau,t)
	\bigg \rangle_{t \in [0,\infty) }
	d\tau
\end{equation}
By construction, $K_2(\tau,t)$ is nonnegative and continuous on $\tau$ and $t$, hence it is a nonnegative, measurable function on $\mathbb{R}_{\geq 0}^2$. This allows us to use Tonelli's Theorem for interchange the sequential time-average and $\tau$-integration. Specifically, we can truncate the double integral in the bounded region $[0,T]^2$ and taking the large-$T$ limit. This gives
\begin{equation}
\begin{aligned}
	\bigg \langle 
		\int_0^t	 K_2(\tau,t) d\tau
	\bigg \rangle_t
	&= \bigg \langle 
		\int_0^\infty	
		 K_2(\tau,t) d\tau
	\bigg \rangle_t \\
	&= 
	\lim_{T\rightarrow\infty}
	\frac{1}{T}
	\int_0^T	 
	\bigg(
	\int_0^T
	K_2(\tau,t)
	\,d\tau
	\bigg)
	dt \\
	&= \lim_{T\rightarrow\infty}
	\frac{1}{T}
	\int_0^T	 
	\bigg(
	\int_0^T
	K_2(\tau,t)
	\,dt
	\bigg)
	d\tau \\
	&= \int_0^\infty	 
	\bigg \langle
	K_2(\tau,t)
	\bigg \rangle_{t\in[0,\infty)}
	d\tau \\
	&= \int_0^\infty	 
	\bigg \langle
	K_2(\tau,t)
	\bigg \rangle_{t \in[\tau,\infty)}
	d\tau. 
\end{aligned}
\end{equation}
Therefore, the function $K_2(\tau,t)$ also allows interchange of time-average and $\tau$-integration. Combining the results from $K_1$ and $K_2$, we have 
\begin{equation}
\bigg\langle
\int_{0}^t K(\tau,t) \,d\tau
\bigg\rangle_{t}
= \int_0^\infty
\bigg \langle K(\tau,t)
\bigg \rangle_{t\in[\tau,\infty)}
d\tau
= \tilde{\alpha}_{eff}(\lambda)
= \lambda + \beta_{eff}.
\end{equation} 
This completes our proof for Theorem D. $\;\;\blacksquare$ \\


\newpage


\small
\printbibliography

\end{document}